\documentclass[12pt]{article}
\input epsf.sty
\input axodraw.sty
\newcommand{\be}{\begin{equation}}
\newcommand{\ee}{\end{equation}}
\newcommand{\ba}{\begin{eqnarray}}
\newcommand{\ea}{\end{eqnarray}}
\newcommand{\bi}{\begin{itemize}}
\newcommand{\ei}{\end{itemize}}
\newcommand{\rmi}[1]{{\mbox{\scriptsize #1}}}
\newcommand{\nr}[1]{(\ref{#1})}
\newcommand{\tr}{{\rm Tr\,}}
\newcommand{\re}{\mathop{\rm Re}}
\newcommand{\Hc}{{\rm H.c.\ }}
\newcommand{\im}{\mathop{\rm Im}}
\newcommand{\nn}{\nonumber \\}
\newcommand{\fr}[2]{{\frac{#1}{#2}}}
\newcommand{\msbar}{\overline{\mbox{\rm MS}}}

\newcommand{\bfx}{{\bf x}}
\newcommand{\bfy}{{\bf y}}
\newcommand{\bfi}{{\bf i}}

\newcommand{\<}{\langle} 
\renewcommand{\>}{\rangle}  

\newcommand{\bmu}{\bar{\mu}}

\newcommand{\Att}{|\hat A_t|^2}
\newcommand{\muu}{|\hat\mu|^2}
\newcommand{\sbb}{\sin\!2\beta\,}
\newcommand{\cbb}{\cos\!2\beta\,}
\newcommand{\tb}{\tan\beta}
\newcommand{\cf}{\cos\phi}
\newcommand{\uy}{U(1)\ }
\newcommand{\pf}{\frac{T^2}{(4\pi)^2}}
 
\newcommand{\eq}{Eq.~}
\newcommand{\eqs}{Eqs.~}
\newcommand{\fig}{Fig.~}
\newcommand{\figs}{Figs.~}
\newcommand{\se}{Sec.~}

\def\lsi{\raise0.3ex\hbox{$<$\kern-0.75em\raise-1.1ex\hbox{$\sim$}}}
\def\gsi{\raise0.3ex\hbox{$>$\kern-0.75em\raise-1.1ex\hbox{$\sim$}}}
\newcommand{\lsim}{\mathop{\lsi}}
\newcommand{\gsim}{\mathop{\gsi}}

\def\la{\label}
\makeatletter \@addtoreset{equation}{section} \makeatother
\renewcommand{\theequation}{\arabic{section}.\arabic{equation}}
\makeatletter
\renewcommand\section{\@startsection {section}{1}{\z@}%
                                   {-5.5ex \@plus -1ex \@minus -.2ex}
                                   {2.3ex \@plus.2ex}%
                                   {\normalfont\large\bfseries}}
\renewcommand\subsection{\@startsection{subsection}{2}{\z@}%
                                     {-3.25ex\@plus -1ex \@minus -.2ex}%
                                     {1.5ex \@plus .2ex}%
                                     {\normalfont\normalsize\bfseries}}
\renewcommand\thesection {\@arabic\c@section}
\renewcommand\thesubsection   {\thesection.\@arabic\c@subsection}
\renewcommand{\@seccntformat}[1]{%
\csname the#1\endcsname.\hspace{1.0em}}
\makeatother

\begin{document}
 
\begin{titlepage}
\begin{flushright}
CERN-TH/2000-226\\
NORDITA-2000/80HE\\
hep-lat/0009025\\
\end{flushright}
\begin{centering}
\vfill
 
{\bf TWO HIGGS DOUBLET DYNAMICS AT THE ELECTROWEAK\\ 
PHASE TRANSITION: A NON-PERTURBATIVE STUDY}

\vspace{0.8cm}
 
M.~Laine$^{\rm a,b}$, 
K.~Rummukainen$^{\rm c,d}$ \\ 

\vspace{0.3cm}
{\em $^{\rm a}$Theory Division, CERN, CH-1211 Geneva 23,
Switzerland\\ }
\vspace{0.3cm}
{\em $^{\rm b}$Dept.\ of Physics,
P.O.Box 9, FIN-00014 Univ.\ of Helsinki, Finland\\ }
\vspace{0.3cm}
{\em $^{\rm c}$NORDITA, Blegdamsvej 17,
DK-2100 Copenhagen \O, Denmark\\ }
\vspace{0.3cm}
{\em $^{\rm d}$Helsinki Inst.\ of Physics,
P.O.Box 9, FIN-00014 Univ.\ of Helsinki, Finland\\ }

\vspace*{0.7cm}
 
\end{centering}
 
\noindent
Using a three-dimensional (3d) effective field theory and
non-perturbative lattice simulations, we study the MSSM electroweak
phase transition with two dynamical Higgs doublets.  We first carry
out a general analysis of spontaneous CP violation in 3d two Higgs
doublet models, finding that this part of the parameter space is well
separated from that corresponding to the physical MSSM.  We then
choose physical parameter values with explicit CP violation and a light 
right-handed stop, and determine the strength of the phase transition. 
We find a transition somewhat stronger than in 2-loop perturbation theory, 
leading to the conclusion that from the point of view of the non-equilibrium 
constraint, MSSM electroweak baryogenesis can be allowed even for a 
Higgs mass $m_H\approx 115$ GeV. We also find that small values of 
the mass parameter $m_A$ ($\lsim 120$ GeV), which would relax the 
experimental constraint on $m_H$, do not weaken the transition 
noticeably for a light enough stop. Finally we determine the 
properties of the phase boundary.
\vfill
\noindent
 

\vspace*{1cm}
 
\noindent
CERN-TH/2000-226\\
NORDITA-2000/80HE\\
November 2000
 
\vfill

\end{titlepage}
 

\section{Introduction}

The observed existence of a baryon asymmetry in 
our Universe is direct evidence 
for physics beyond the Minimal Standard Model. Indeed, the
electroweak theory contains anomalous baryon number violation 
which is rapid enough to be in thermal equilibrium at 
temperatures $T\gsim 100$ GeV~\cite{krs,asy,gdm,db}, so that any pre-existing
asymmetry is washed out. (Unless there is an asymmetry in 
baryon minus lepton number, $B-L$, which would also require physics beyond
the Minimal Standard Model; for an overview on recent scenarios
based on this, see~\cite{bp}). 
As the Universe then cools down, it turns
out that there is no electroweak phase transition at all for 
Higgs masses $m_H \gsim 72$ GeV~\cite{isthere,gis,cfh}, or 
$m_H \gsim 90$ GeV in the presence of primordial magnetic fields~\cite{bext}:
the cosmological evolution is just smooth and continuous.
Taking the experimental lower bound $m_H \gsim 110$ GeV into 
account~\cite{lep} (the factual bound is even a few GeV higher by now), 
this would mean that the baryon symmetric 
high temperature state simply
freezes in, in contradiction with observation. 

It is quite interesting that already one of the simplest 
extensions of the Minimal Standard Model, 
the Minimal Supersymmetric Standard Model (MSSM), offers
a definite and experimentally testable way 
of changing this conclusion. One can uniquely identify 
a bosonic degree of freedom, the right-handed stop, 
which can be ``light'' and dynamical at the phase transition
point without violating experimental constraints at zero
temperature, and interacts strongly enough with the Higgs
to strengthen the phase transition 
significantly~\cite{cqw1}--\cite{mlo3}. 
There are also new sources of CP violation available which 
could potentially have a favourable effect~\cite{non-eq}--\cite{risa}. 
Many details of the electroweak phase 
transition in this region have recently been 
studied~\cite{hsch}--\cite{cmqsw}.   

In this paper, we address several issues related to the 
electroweak phase transition in the MSSM. 
The first is CP violation
in the background configuration 
related to the fact that there are two Higgs doublets. 
The second is the strength
of the phase transition at an experimentally viable parameter point
corresponding to a Higgs mass of about 105 GeV, 
not excluded for the MSSM with smallish $m_A$.
(We also consider other values close to these, notably
$m_H$ up to $\sim$ 115 GeV.) 
The third is the structure of the phase boundary, or
bubble wall, at the physical parameter point. 

In particular, 
as to the first of these issues, 
we will pay some attention to the phenomenon of 
spontaneous CP violation. Spontaneous CP violation can 
in principle take place in any two Higgs doublet model~\cite{lee},
but for the physical MSSM parameters it cannot be 
realized at $T=0$~\cite{nm,apo}.  However, 
it has been suggested that it might be more easily realized 
at finite temperatures~\cite{cp,emq}, or even only in the 
phase boundary between the symmetric and broken
phases~\cite{cpr,fkot} (in which case it is 
sometimes called ``transitional'' CP violation). 
The existence of spontaneous CP violation
would mean that even small explicit phases can have a large physical 
effect, and such a situation 
within the phase boundary
would conceivably be useful for 
electroweak baryogenesis~\cite{mstv}--\cite{ckv}.

The method we use to study all the three questions
is the construction of an effective 3d field
theory with the method of dimensional reduction, and its 
non-perturbative analysis with simple 
lattice simulations. The dimensional
reduction step was already carried out in~\cite{cpown}, and 
here we concentrate on the non-perturbative part.

The plan of this paper is the following. 
In \se\ref{theories} we review briefly the form of the 3d effective field
theory for the MSSM, arising after dimensional reduction. 
In \se\ref{lattice} we discuss the lattice
implementation of this theory --- this section should 
be skipped by those not interested. In \se\ref{phasediag}
we carry out a general analysis of the phase diagram of the
3d theory, with particular view on spontaneous CP violation. 
In \se\ref{strength} we focus on a physical choice of 
parameters and determine the strength of the phase 
transition, while in \se\ref{phasewall} we determine the properties 
of the phase boundary appearing in the physical transition, 
checking also for the possibility of transitional CP violation. 
We summarise and discuss implications in \se\ref{conclusions}. 


\section{The effective theory}
\la{theories}

\subsection{The action} 
\la{sec:ac}

At finite temperatures around the electroweak 
phase transition, the thermodynamics 
of the MSSM can be represented by a 3d effective field theory 
containing two SU(2) Higgs doublets $H_1,H_2$ 
and one SU(3) stop triplet $U$~\cite{cpown}.
The action is of the
most general gauge invariant form,
\ba
{\cal L}_\rmi{3d} & = & 
\fr12 \tr G_{ij}^2 + (D_i^s U)^\dagger (D_i^s U) + 
m_U^2(T) U^\dagger U  + \lambda_U (U^\dagger U )^2 \nn
& + & 
\gamma_1 U^\dagger U H_1^\dagger H_1 + 
\gamma_2 U^\dagger U H_2^\dagger H_2
+ \Bigl[\gamma_{12} U^\dagger U H_1^\dagger \tilde H_2+\Hc\Bigr] \nn
& + & 
\fr12 \tr F_{ij}^2 
+ (D_i^w H_1)^\dagger(D_i^w H_1) + 
(D_i^w H_2)^\dagger(D_i^w H_2) \nn
& + &  m_1^2(T) H_1^\dagger H_1 +  m_2^2(T) H_2^\dagger H_2 
+ \Bigl[ m_{12}^2(T) H_1^\dagger \tilde H_2 +\Hc  \Bigr]  \nn
& + &  \lambda_1 (H_1^\dagger H_1)^2 + 
 \lambda_2 (H_2^\dagger H_2)^2 + 
 \lambda_3 H_1^\dagger H_1 H_2^\dagger H_2 + 
 \lambda_4 H_1^\dagger \tilde H_2 \tilde H_2^\dagger H_1 \nn
& + & \Bigl[ \lambda_5 (H_1^\dagger \tilde H_2)^2 + 
 \lambda_6 H_1^\dagger H_1 H_1^\dagger \tilde H_2 +
 \lambda_7 H_2^\dagger H_2 H_1^\dagger \tilde H_2 + \Hc\Bigr]. 
\la{action}
\ea
Here $D_i^s, D_i^w$ are the spatial
SU(3) and SU(2) covariant derivatives, 
$G_{ij}, F_{ij}$ the corresponding field strength tensors, 
and $\tilde H_2 = i \sigma_2 H_2^*$. We denote the SU(3) and
SU(2) gauge couplings by $g_s,g_w$. We have neglected the
dynamical effects of the U(1) subgroup since they are expected
to be small~\cite{su2u1} (although some aspects of the system
with a dynamical U(1) remain to be understood~\cite{allor}). 
Note also that compared with the MSSM, 
$U$ denotes the complex conjugate of the original 
right-handed stop triplet. The couplings in \eq\nr{action}
can be expressed in terms of the physical parameters of 
the MSSM and the temperature $T$, as will be specified below.

For future reference, let us recall that if one is only interested in 
the strength of the phase transition, the effective theory 
in \eq\nr{action} is even unnecessarily complicated. Indeed, it
is easy to understand~\cite{generic,ckold,loold,mssmown,bjls,cpown} 
(see also \ref{inthiggs})
that if the two Higgs doublet mass matrix is diagonalized, one
of the eigendirections is always heavy, and can be perturbatively
integrated out. This results in an effective theory with a 
single SU(2) Higgs doublet, and the right-handed stop. We 
studied that effective theory with lattice simulations in~\cite{mssmsim}.
The reason we keep here both Higgs doublets is that we measure
a number of observables which are specific to the existence
of both fields, and cannot be addressed with the simpler theory.

The couplings appearing in the 3d theory in \eq\nr{action}
have the dimension GeV, and the fields have the dimension
GeV$^{1/2}$. For later convenience, 
we will parameterise the couplings by introducing
what from the point of view of the 3d theory is an arbitrary
scale, the temperature $T$. We then 
scale all the couplings to a dimensionless form by dividing
with $T$, and all the fields into a dimensionless form by 
dividing with $T^{1/2}$:
\be
g^2 \equiv \frac{g_\rmi{3d}^2}{T}, \quad
\lambda_i \equiv \frac{\lambda_{i,\rmi{3d}}}{T}, \quad
\gamma_i \equiv \frac{\gamma_{i,\rmi{3d}}}{T}, \quad
\frac{H_i}{T} \equiv \frac{H_{i,\rmi{3d}}}{T^{1/2}}, \quad 
\frac{U}{T} \equiv \frac{U_{\rmi{3d}}}{T^{1/2}}.
\ee
Expressed in terms of the newly defined coupling 
constants and fields, the action goes over into 
\be
S = \int\! d^3x\, {\cal L}_\rmi{3d,orig} \to 
\frac{1}{T} \int\! d^3x\, {\cal L}_\rmi{3d},
\ee
where the dimensionality of the new ${\cal L}_\rmi{3d}$ is GeV$^4$
as usually in 4d. We shall use this formulation throughout the paper, 
with $T$ taken to be the physical temperature. 

\subsection{Approximate physical values of couplings} 

Expressions for the parameters
in \eq\nr{action}, corresponding to 
the MSSM, have been derived in~\cite{cpown};
for a summary, see Appendix~A.7 there. 
We cite the precise values used in Secs.~\ref{strength}, \ref{phasewall}
later on, but let us recall the general orders of magnitude already here. 

It is important to keep in mind a basic 
difference between the effective theories corresponding 
to the Standard Model and the MSSM.
In the former, the physical Higgs mass
is determined by the effective
quartic scalar coupling, $\lambda\sim (g_w^2/8)(m_H^2/m_W^2)$, while 
the temperature is determined by the 3d scalar mass parameter, 
$m_\rmi{3d}^2(T)\sim -m_H^2/2 + 0.3 T^2$. 
In the MSSM, on the contrary, 
the quartic Higgs couplings are fairly
inert, $\sim g_w^2/8$, and are affected by
the zero-temperature Higgs mass (i.e., by $\tan\beta$) 
only through small radiative corrections.
Thus {\em both} the physical Higgs spectrum 
and the temperature reside now dominantly in the
effective mass parameters. 
The rough generic orders of magnitude 
for the quartic couplings in the MSSM 
can therefore be cited~\cite{cpown}, 
independent of the Higgs mass and temperature:
\ba
& & 
\lambda_1\sim  0.07, \quad
\lambda_2\sim  0.13, \quad
\lambda_3\sim  0.08, \quad
\lambda_4\sim -0.22, \la{mssmc} \\
& &  
|\lambda_5|\lsim 0.002, \quad 
|\lambda_6|\lsim  0.01, \quad
|\lambda_7|\lsim  0.02, \la{mssmc2} \\
& & 
\lambda_U \sim 0.19, \quad
g_w^2 \sim 0.42, \quad 
g_s^2 \sim 1.1. 
\ea
In $\lambda_1,\lambda_3,\lambda_4$, we have included 
only the tree-level terms, but in $\lambda_2$
also the dominant 1-loop terms proportional to the top Yukawa
coupling to the fourth power,  $h_t^4$,
which are absent in $\lambda_1,\lambda_3,\lambda_4$
(we recall that $h_t\approx 1.0$). In order
to get the estimates for $\lambda_5...\lambda_7$, we have 
taken into account that when the right-handed stop is light, 
the squark mixing parameters cannot be too large compared
with the left-handed stop mass, because of 
the stability of the theory (see below). 

As to the three couplings $\gamma_i$ in \eq\nr{action}, they can 
be reparameterised in terms of the three couplings 
$h_t,\hat A_t, \hat\mu$ as
\be
\gamma_1 = -h_t^2 |\hat\mu|^2, \quad
\gamma_2 = h_t^2 (1-|\hat A_t|^2), \quad
\gamma_{12} = h_t^2 \hat A_t^* \hat \mu^*, \la{gpar}
\ee
where $A_t, \mu$ are the mixing parameters in the 3rd 
generation squark mass matrix, $m_Q$ is the 
corresponding left-handed squark mass parameter, 
and $\hat A_t \approx A_t/m_Q$, $\hat \mu \approx \mu/m_Q$. 
There are of course radiative corrections to these relations, 
but we can also view  them as a more general reparameterization, 
since $\hat A_t, \hat\mu$ are essentially free parameters. 
We mostly assume $\hat A_t, \hat\mu \lsim 0.3$, again in order
not to destabilize the theory~(see, e.g., \cite{cpown}), 
and also since large values tend to weaken the
phase transition (see, e.g., \cite{e,mssmown,cm}).

Consider finally the mass parameters. Working in the limit 
$\tilde m_U^2 \equiv - m_U^2 \ll (2 \pi T)^2 \ll m_Q^2$, 
where $m_U^2$ is the right-handed stop mass parameter, 
we have at leading order
\ba
m_U^2(T) & \approx  & -\tilde m_U^2 + 
\biggl( \fr23 g_s^2 + \fr13 h_t^2 -\fr16 h_t^2 (\Att+\muu)
\biggr) T^2, \\ 
m_1^2(T) & \approx &  \fr12 m_A^2 - \fr12 (m_A^2+m_Z^2)\cbb+ 
\biggl(\fr38 g_w^2 -\fr14 h_t^2 \muu  
\biggr) T^2, \la{seq1st} \\
m_2^2(T) & \approx &  \fr12 m_A^2 + \fr12 (m_A^2+m_Z^2)\cbb + 
\biggl(\fr38 g_w^2+\fr12 h_t^2 -\fr14 h_t^2 \Att
\biggr) T^2, \la{seq2nd} \\
m_{12}^2(T) & \approx &  -\fr12 m_A^2\sbb + 
\fr14 h_t^2 \hat A_t^*\hat\mu^* T^2. \la{seq3rd} 
\ea
Here $m_A, \tb$ are the usual MSSM input parameters.
We cite these expressions because they lead
to some generic properties relevant for our discussion below, in 
particular that the trace of the two Higgs doublet mass matrix,
$m_1^2(T)+m_2^2(T)$, is always positive in the region relevant for us.


\section{Lattice formulation}
\la{lattice}

For future reference, we recall next briefly how the theory in 
\eq\nr{action} can be discretized.

\subsection{Lattice action}
\la{lattact}

\newcommand{\bH}{{\rm\tilde H}}
\newcommand{\tH}{\tilde H}
\newcommand{\tU}{\tilde U}
\newcommand{\HH}{\tH^\dagger\tH}
\newcommand{\UU}{\tU^\dagger\tU}

We discretize the action in \eq\nr{action} in the standard way. 
The finite lattice spacing $a$ enters as $\beta_w=4/ (a g_w^2 T)$,
$\beta_s=6/ (a g_s^2 T)$, and the lattice volume 
is denoted by $N_1 N_2 N_3$. 
The gauge field terms $(1/2) \tr G_{ij}^2 + (1/2) \tr F_{ij}^2$
are treated with the usual Wilson formulation, 
as in~\cite{mssmsim}. 
The scalar fields are rescaled into a dimensionless form by
$H_i \to \hat H_i \equiv H_i/T$, 
$U\to \hat U \equiv U/T$, and then, e.g., 
\ba
S
& = & \frac{1}{T} \int\! d^3x\, \Bigl[
(D_i^s U)^\dagger (D_i^s U) + 
m_U^2(T) U^\dagger U  + \lambda_U (U^\dagger U )^2 
 \Bigr] \nn 
& \to &  \hspace*{1cm} 
\sum_\bfx \Bigl[ 
- 2 a T \sum_i 
\re \hat U^\dagger_{\bfx} U^s_{{\bfx},i} \hat U_{\bfx+\bfi} \nn
& & \hspace*{1.3cm} + 
\Bigl( 6 a T + (aT)^3 (m_{U,B}^2/T^2) \Bigr) 
\hat U^\dagger_{\bfx} \hat U_{\bfx}  
+ \lambda_U (aT)^3 
(\hat U^\dagger_{\bfx} \hat U_{\bfx})^2 
\Bigr]. \hspace*{0.5cm} \la{lattaction}
\ea
Here $U^s_{{\bfx},i}$ is the SU(3) link matrix
at point $\bfx$ in direction $i$,
and the bare lattice mass parameter 
$m_{U,B}^2$ is given in~\ref{cts}.

Let us note that 
by computing $m_{U,B}^2$ 
we have fully renormalized the theory~\cite{contlatt}, but we have
not implemented ${\cal O}(a)$ improvement~\cite{moore_a} here. 
This is because according to Sec.~B.4 of the latter ref.\ in~\cite{moore_a},
one would need $\beta_w > 30$ to be comfortably in the regime where 
improvement works, and we are not able to go that close to the continuum 
limit, with the lattice sizes we can manage in practice. 
Thus we expect a more general lattice spacing dependence
(and use correspondingly a more general fitting ansatz for the
continuum extrapolation). 

\subsection{Update algorithm}
\la{algo}

The lattice simulations of the theory in \eq\nr{action} are quite
demanding, due to the large range of mass scales near the
first order transition temperature (we recall that many mass 
scales have already been removed by the analytic dimensional 
reduction computation, but a number of dynamical scales are still 
left over, particularly since we want to keep both Higgs doublets in the 
effective theory). The Monte Carlo update algorithms employed
are nevertheless qualitatively similar to the
ones used in ref.~\cite{mssmsim} for simulating MSSM with only
one SU(2) Higgs doublet. 

Our update algorithm consists of a mixture of overrelaxation, heat
bath and Metropolis updates for both the Higgs and gauge fields.  For
all of the three Higgses
($H_1,H_2,U$), we use the ``Cartesian overrelaxation''
presented in~\cite{nonpert,mssmsim}.  The overrelaxation update
is much more efficient in evolving the fields than the diffusive heat
bath and Metropolis updates; however, in order to ensure ergodicity
one has to mix diffusive update steps with overrelaxation.  We use a
compound update cycle which consists of 5 overrelaxation sweeps
through the lattice, followed by one heat bath/Metropolis sweep.  For
details, we refer to Sec.~6 of \cite{mssmsim}.

For the simulations to be reliable and
to allow for an extrapolation both to
the infinite volume and continuum limits, the lattice spacing $a$ and
the lattice size $L = Na$ should satisfy two inequalities,
\be
  a \ll \xi_{\rm min}; ~~~~ \xi_{\rm max} \ll Na\,,
  \la{ineq}
\ee
where $\xi_{\rm min}$ and $\xi_{\rm max}$ are the smallest and the
largest physical correlation lengths of the system.  In the present
system the mass scales near the transition can vary by roughly an
order of magnitude (see~\se\ref{sec:masses}).  This makes it
necessary to use relatively large lattice sizes $N$, but even then it
is not easy to satisfy the inequalities in \eq\nr{ineq} with wide margins.
This emphasizes the importance of a careful check of both the
infinite volume and the continuum extrapolations.

The fact that the transition is 
strongly of the first order, 
makes the inequalities in \eq\nr{ineq}
somewhat easier to satisfy than for instance
in the case of a second order transition,  
since $\xi_\rmi{max}$ remains finite. 
However, the transition is now so strong that the
system does not spontaneously tunnel from one metastable phase to
another, especially in large volumes.  On the other hand, probing the
whole tunnelling phase space region is required in order to determine
precisely the critical temperature, the order parameter discontinuities,
and the surface tension.  Thus, to allow for frequent tunnellings, we
use the multicanonical simulation method with automatic weight function
calculation, discussed in~\cite{mssmsim}.

\subsection{Observables}
\la{se:obs}

The simplest observables to be considered are
the gauge invariant bilinears
\be
O_1 = H_1^\dagger H_1, \quad
O_2 = H_2^\dagger H_2, \quad
R = \re H_1^\dagger \tilde H_2, \quad
I = \im H_1^\dagger \tilde H_2. \la{ops}
\ee
The operator $I$ is 
odd under complex conjugation, and constitutes thus
an order parameter for spontaneous C violation (see below). 
In addition to the global expectation values of 
these operators, we also consider 
the corresponding correlation lengths, obtained from
the 2-point functions. 

In perturbative analyses, one often uses
the SU(2) and U(1) symmetries of the theory
to parameterise the two Higgs doublets
for instance as 
\be
H_1 = \frac{v_1}{\sqrt{2}}
\left( 
\begin{array}{l}
1 \\
0
\end{array}
\right), \quad
\tilde H_2 = \frac{v_2}{\sqrt{2}}
\left( 
\begin{array}{l}
\cos\theta e^{i\phi} \\
\sin\theta
\end{array}
\right), \la{prms}
\ee
and $\tan\beta = v_2/v_1$. When used
beyond tree-level and in connection with, say, 
some covariant gauge, the values of the quantities 
$v_1,v_2,\theta,\phi$ in the broken phase
are gauge dependent,
and thus there is no unique relation 
to the values of the operators in \eq\nr{ops}. 
However, at the phase transition point
($T=T_c$) we may define a gauge and scale 
independent generalization of the perturbative parameters
for instance by
\be
\fr12 v_i^2\equiv \< H_i^\dagger H_i\> - 
\left.\< H_i^\dagger H_i\>\right|_\rmi{symmetric}, \quad
\tan^2\beta = \frac{v_2^2}{v_1^2}, \la{lattdef}
\ee
where $i=1,2$ and the latter expectation value is taken
in the homogeneous ``symmetric'' high-temperature phase. 
We could also write $\phi \equiv \arctan(I/R)$, or 
$\phi \equiv \arctan[(I-I_\rmi{symmetric})/(R-R_\rmi{symmetric})]$,
but such ratios are numerically very unstable, 
and we do not use them here. Rather it is the perturbative
values of $v_1,v_2,\theta,\phi$ which should be converted 
into gauge invariant observables
such as those in \eq\nr{ops}.

\subsection{Mean field estimates}

Finally, let us note that due to the relatively heavy cost
of simulating the action in \eq\nr{action}
on the lattice, many of the preliminary parameter
scans have been made using relatively small lattice sizes.
To at least partly account for the finite volume effects in 
the comparison with perturbation theory, we transform
the perturbative results for the quantities in \eq\nr{prms}
into finite volume ``mean field'' estimates 
for the operators in \eq\nr{ops} as follows.

A mean field estimate can be obtained
by taking the fluctuations into account in the effective
potential, and performing then the integral over the 
zero-modes of the fields, parameterised
by $v_1,v_2,\theta,\phi$ as in \eq\nr{prms}. 
In addition, we take into account the fluctuations of $U$, 
and parameterise $U=(1/\sqrt{2})(\chi,0,0)^T$.
The integration measure goes over into
\ba
dH_1 \, dH_2 \, dU  
& = &   
C dv \, d\chi \,d\beta\,  d\theta\, d\phi \, 
v^7 \chi^5 \! \cos^3\!\beta \sin^3\!\beta \cos\!\theta\sin\!\theta,
\ea
where $C$ is a constant. The action can be written as
\be
S \approx \frac{V}{T} 
V_\rmi{eff} (v,\beta,\theta,\phi,\chi),
\ee
where $V=N_1N_2N_3 a^3$ is the volume, 
and the mean field estimates are then obtained  from
\be
\langle O_i \rangle \approx Z^{-1} \int\! dH_1 \, dH_2 \, dU \, O_i\exp(-S),
\la{mfest}
\ee
where the $O_i$ are written using \eq\nr{prms}.


\section{Spontaneous CP violation, and transitional}
\la{phasediag}

We now move to the first physics topic. 
In this section we consider the case of no explicit CP violation
(i.e.,\ all the parameters in \eq\nr{action} are assumed real), 
in order to carry out into completion the analysis outlined in~\cite{cpown}. 
The motivation is that if spontaneous CP violation would exist in the 
system where CP phases are put to zero, then one could expect large 
physical effects once small explicit phases are turned on. 

\subsection{Symmetries and phases}

Let us start by reviewing briefly the overall setup. 
In the space of general couplings, 
the theory in \eq\nr{action} has several ``phases''. 
First of all, there are the usual phases related to 
``gauge symmetry breaking'': the SU(3) gauge symmetry
can be broken by an expectation value of $U$, 
and the SU(2) symmetry
by expectation values of $H_1,H_2$. 
In the physical case the parameters had better be 
such that one does not end up in the
phase with a broken SU(3) symmetry, 
since one would be stuck there forever~\cite{cms}. 

In addition to the gauge symmetries, 
there are also global symmetries in the system
of \eq\nr{action}. There is one continuous U(1)
symmetry corresponding to the hypercharge, 
and then there are the usual discrete symmetries C, P. 
The time translation symmetry T is not directly visible any more
in the 3d effective theory, but it dictates what kind of 
operators can arise in the dimensional reduction step~\cite{parity}.

As to the continuous U(1) symmetry, let us first recall how things
are in the single Higgs doublet SU(2) theory, 
${\cal L} = (D^w_i H)^\dagger (D^w_i H)+ ...\;$. 
In this case the global U(1) symmetry is $H\to \exp(i\alpha)H$. 
This global symmetry cannot get broken, however, since 
any configurations
$H$, $\exp(i\alpha)H$ can be SU(2) gauge transformed
to the same configuration (e.g.\ to the unitary gauge). 
Thus the system is \uy invariant 
independent of the expectation value of $H$, and there is always  
a massless excitation in the gauged version of the theory. 

Things are different if there are two Higgs doublets,  
${\cal L} = \sum_{j=1,2} (D^w_i H_j)^\dagger (D^w_i H_j)+...\;$. 
Taking into account the general form of 
the two Higgs doublet potential, \eq\nr{action}, 
there again remains a global symmetry 
$H_1 \to\exp(i \alpha) H_1, H_2 \to \exp(-i \alpha) H_2$.
However, now this symmetry can get broken: 
if $H_1,H_2$ are not proportional to each other, 
it is not possible to unwind simultaneously the
angle from $\exp(i\alpha)H_1,\exp(-i\alpha)H_2$
by an SU(2) gauge transformation. For such expectation
values of $H_1,H_2$, physics is not U(1) invariant. 
In the gauged version of the theory, the photon 
becomes massive. 
In terms of the parametrization in \eq\nr{prms},
the breaking of the \uy symmetry
corresponds to $|\sin\theta|>0$, 
or $|\cos\theta|<1$.

As to the discrete symmetries, 
for real parameters this theory is even  
under both of the discrete symmetries C, P. 
The C transformation corresponds to 
\be
H_i \to H_i^*.
\ee 
While parity is not expected to be 
spontaneously broken in this theory~\cite{Ambj3,su2u1}, 
the C symmetry can be, thus violating also CP~\cite{lee}.
The breaking of C
is signalled by a non-vanishing expectation value 
of the local gauge invariant order parameter
$I$ in \eq\nr{ops}.
In terms of the parametrization in \eq\nr{prms},
C symmetry corresponds to  
the invariance of the theory 
under $\phi\to -\phi$, and the breaking of C is 
signalled by $|\!\cos\phi\,|<1$.

\subsection{The phase diagram in 1-loop perturbation theory}
\la{se:pdin1loop}

We now want to find out the parameter values for 
which the (global)
phases discussed in the previous section are realized. 
We first do this in perturbation theory,
using the (gauge specific) variables in \eq\nr{prms}.
We work in the Landau gauge. 
We always assume $m_U^2(T) \gsim 0$, 
in order to avoid the dangerous charge and colour breaking minimum~\cite{cms}.
For real $\hat A_t,\hat \mu$, 
the 1-loop effective potential is then
\ba
V(v_1,v_2,\theta,\phi) \!\! & = & \!\! 
\fr12 m_1^2(T) v_1^2+
\fr12 m_2^2(T) v_2^2 + 
m_{12}^2(T) v_1 v_2 \cos\theta\cos\phi \nn
& + & \!\!
\fr14 \lambda_1 v_1^4 + \fr14 \lambda_2 v_2^4 + \fr14 \lambda_3 v_1^2 v_2^2
+ \fr14 \lambda_4 v_1^2 v_2^2 \cos^2\theta \nn
& + & \!\!
\fr12 \lambda_5 v_1^2 v_2^2 \cos^2\theta \cos 2\phi + 
\fr12 (\lambda_6 v_1^2 + \lambda_7 v_2^2) v_1 v_2 \cos\theta\cos\phi \nn
& - & \!\! \frac{T}{16\pi} g_w^3 (v_1^2+v_2^2)^{\fr32}
-\frac{T}{2\pi} ( A + B \cos\theta \cos\phi )^{\fr32}  
- \frac{T}{12\pi}\sum_{i=1}^8 (m_{S,i}^2)^{\fr32}. \hspace*{0.5cm}
\la{v1loop} \la{s1loop} \la{Atmu1loop}
\ea
Here $m_{S,i}^2$ are the real eigenvalues of the 8$\times$8 scalar
mass matrix, obtained after making a shift according to \eq\nr{prms}
in \eq\nr{action}, and
\be
A  =  m_U^2(T) + \fr12 h_t^2 v_2^2 - \fr12 h_t^2 
(|\hat\mu|^2 v_1^2 +  |\hat A_t|^2 v_2^2), \quad
B = h_t^2 \hat A_t \hat \mu \, v_1 v_2.  \la{AB}
\ee

We now note that 
the dominant 1-loop effects in this effective potential
are the terms from the vector bosons and from the stops,
while the 1-loop scalar effects
from $m_{S,i}^2$ on the last line in \eq\nr{v1loop}
are small. This is because the scalar self-couplings $\lambda_i$
are never large in the MSSM, being $\lsim g_w^2/8$ (see above). 
Furthermore, we may set $\hat A_t=\hat\mu=0$ in \eq\nr{AB}
for the moment, which allows for a simple analytic treatment. 
We then work out a complete parametrization 
for the part of the parameter space
leading to spontaneous C~violation. 
The same could be done for the phase with broken U(1), but as
it turns out to lie even farther away from the MSSM, we do not
elaborate on it here. The details of the discussion concerning
the C violating phase are presented in~\ref{comppar}, 
and we address now the results only. 

To be explicit, we fix the couplings 
$\lambda_1...\lambda_4$ to the values given in \eq\nr{mssmc}. 
We then make a full scan of the remaining parameter space according
to \eqs\nr{masses1}--\nr{masses3}, \nr{lam6}--\nr{lam5max},  
without any additional restrictions on 
$\lambda_5,\lambda_6,\lambda_7, m_1^2(T),m_2^2(T),m_{12}^2(T)$.
As to the expectation value $v/T$, we take 
the realistic MSSM into account by recalling that 
there one does not get values as large as $v/T\sim 3$
at the electroweak phase transition, and in 
any case for such vevs dimensional reduction and the 
construction of the effective theory
in \eq\nr{action} start to lose their accuracy. 
Thus we assume $v/T\lsim 3$.

\begin{figure}[p]

\centerline{
\epsfxsize=6cm\epsfbox{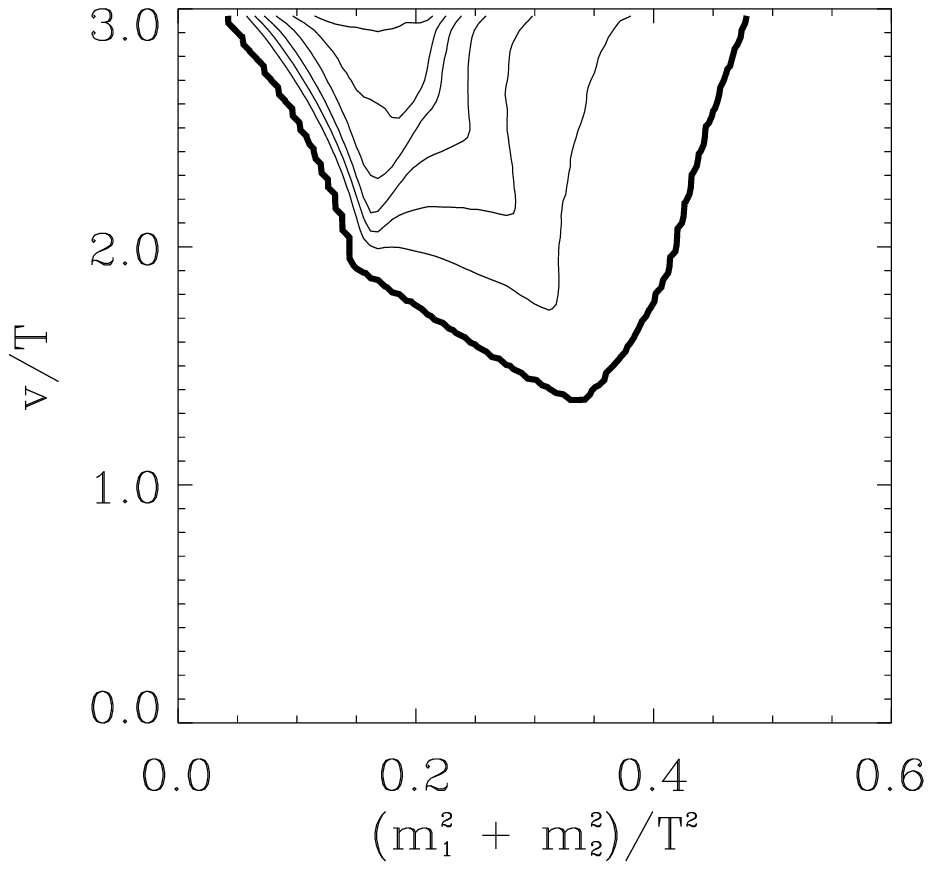}%
\epsfxsize=6cm\epsfbox{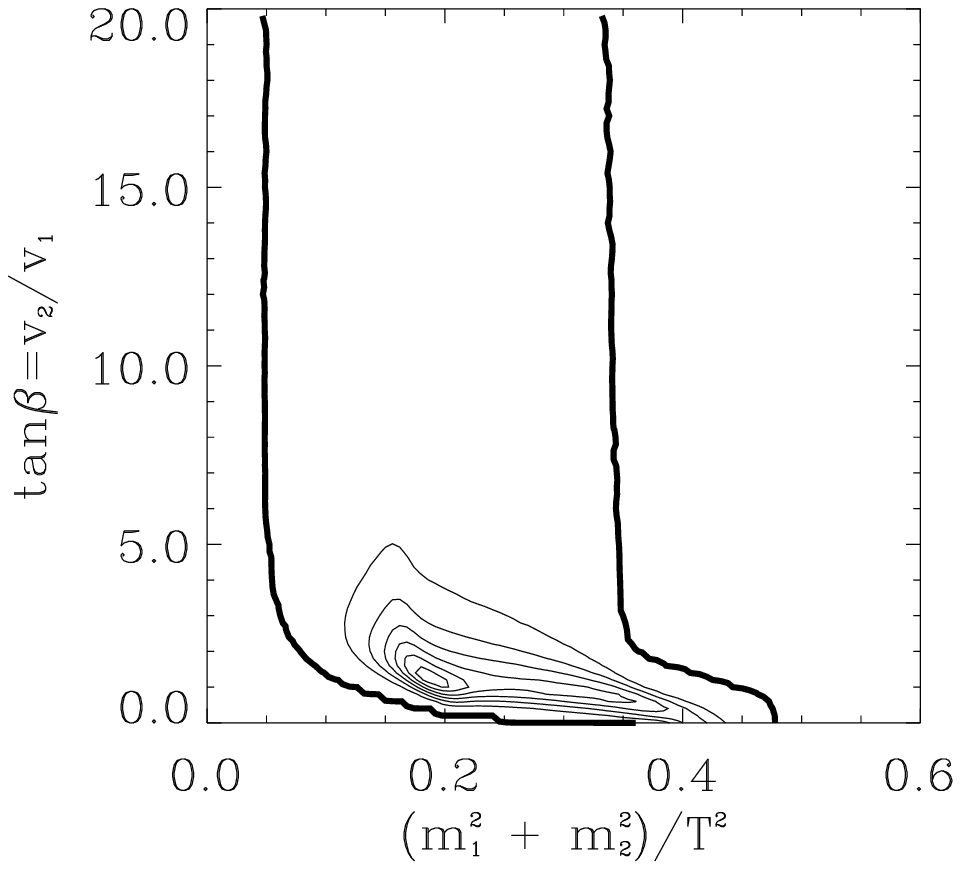}}
\centerline{
\epsfxsize=6cm\epsfbox{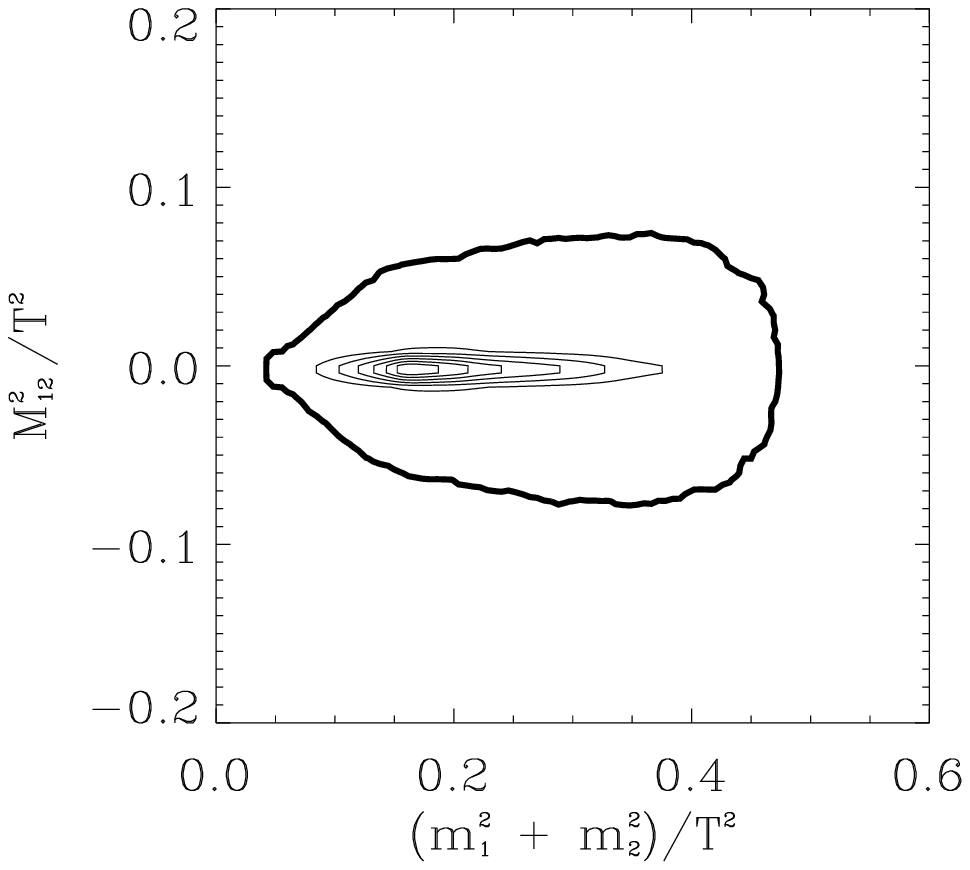}%
\epsfxsize=6cm\epsfbox{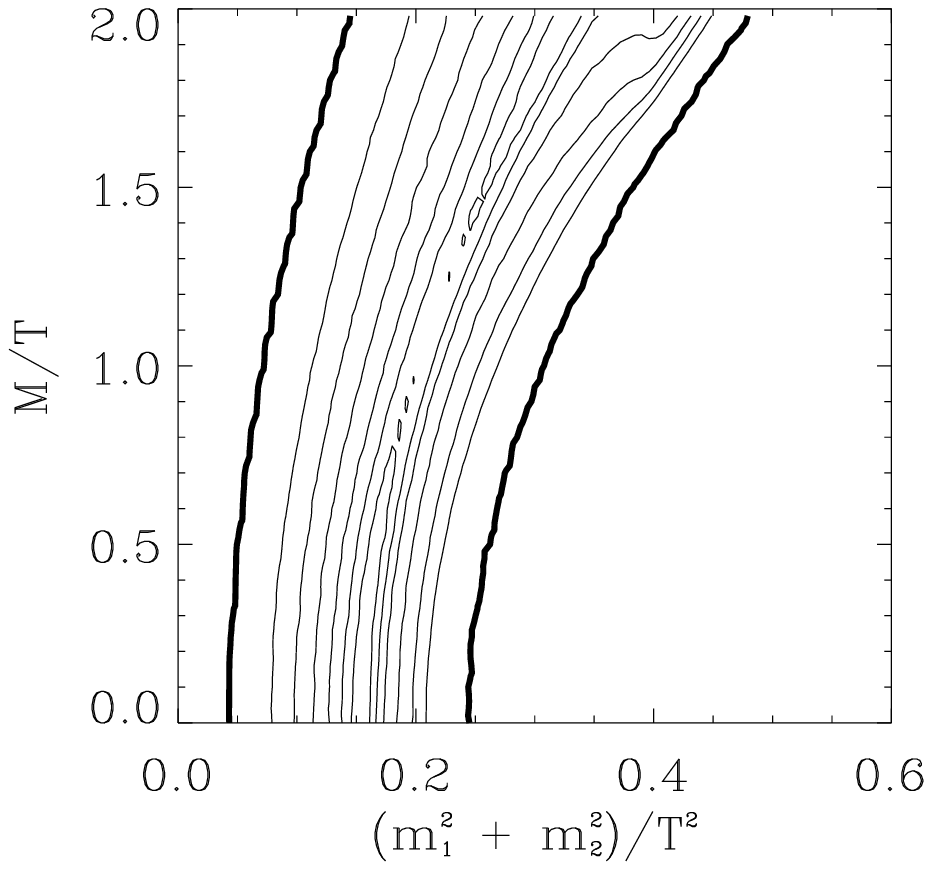}}
\centerline{
\epsfxsize=6cm\epsfbox{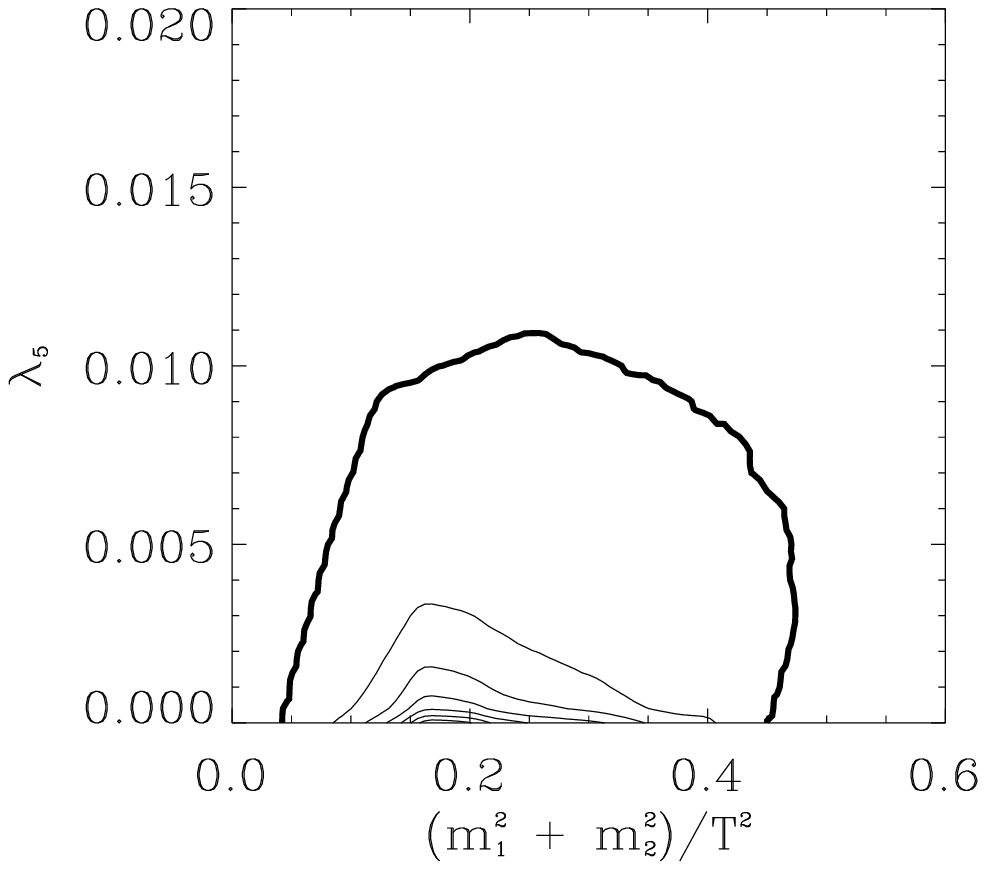}%
\epsfxsize=6cm\epsfbox{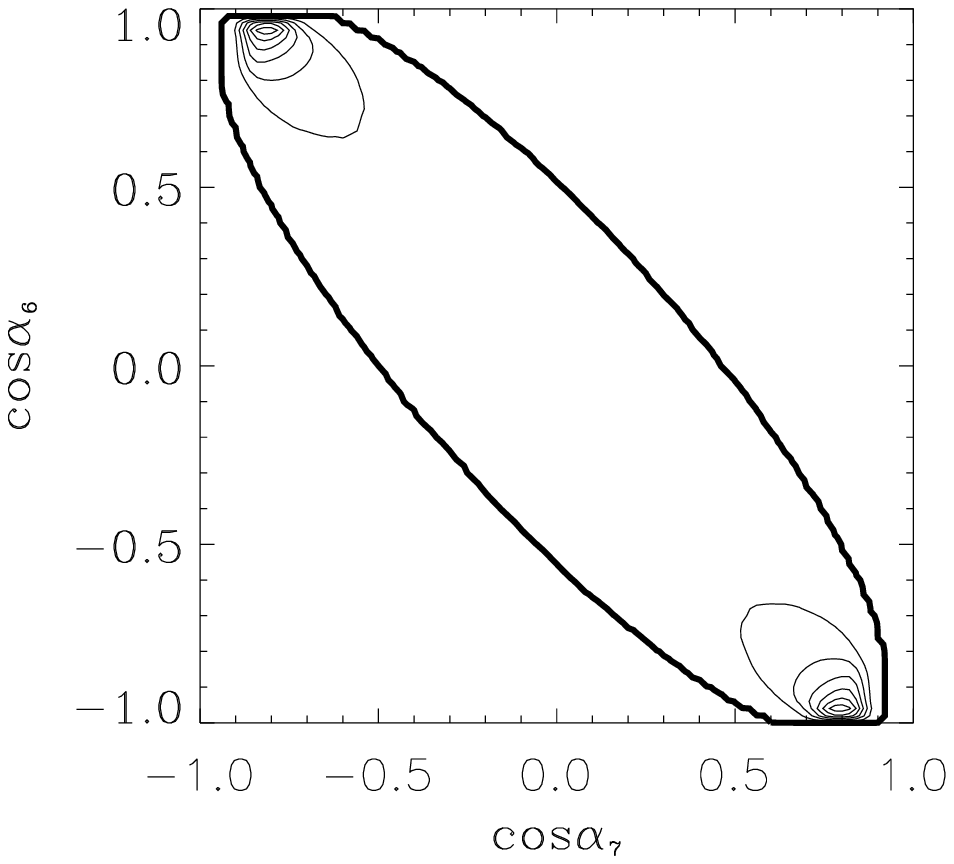}}

\caption[a]{Projections of the region of spontaneous 
C violation ($|\cos\phi|<1$) onto various planes.
Here $M_{12}^2 = m_{12}^2(T) + (1/2)(\lambda_6 v_1^2 + 
\lambda_7 v_2^2)$, $M= \sqrt{2} m_U(T)/h_t$, 
$\cos\alpha_6 = \lambda_6 / (2\sqrt{\lambda_1\lambda_5})$,
$\cos\alpha_7 = \lambda_7 / (2\sqrt{\lambda_2\lambda_5})$,
cf.\ \eqs\nr{M122}, \nr{GHMdef}, \nr{lam6}, \nr{lam7}.
Thin contours (equally spaced, arbitrary normalization)
indicate roughly the thickness of the allowed region of
the parameter space in the orthogonal directions.}
\la{fig:GG}
\end{figure}

The results of the scan are shown in \fig\ref{fig:GG}. 
We show the projections of the parameter space onto different
axes, in most cases with $[m_1^2(T)+m_2^2(T)]/T^2$ as the $x$-axis.
We have only shown the part $[m_1^2(T)+m_2^2(T)]/T^2>0$, as this is
the case relevant for the MSSM, see \eqs\nr{seq1st}, \nr{seq2nd}
and the discussion below them.

Let us make a few observations on the results. 
First of all, for $\lambda_1...\lambda_4$ as given,  
the ``bounded from below'' constraint in \eq\nr{constraint0}  
forbids values $\lambda_5 \gsim 0.026$, and the fact 
that we only allow values $v/T\le 3$ leads to a further 
restriction visible in \fig\ref{fig:GG}, cf.\ \eq\nr{lam5max}. 
(Furthermore, note from \eq\nr{mssmc2} that only small
values are produced by dimensional reduction.) Second, 
$M_{12}^2\equiv m_{12}^2+(1/2)(\lambda_6 v_1^2+\lambda_7 v_2^2)$ 
is very small; this is in order to satisfy \eq\nr{CPviol},
for the small values of $\lambda_5$ that arise. Finally, 
the $x$-axis of the figures, $[m_1^2(T)+m_2^2(T)]/T^2$, is also
restricted to quite small values. This is again
ultimately due
to the smallness of $\lambda_5$, as it was argued
in~\cite{cpown} that one has to satisfy
\be
m_1^2(T)+m_2^2(T) \lsim 2 \lambda_5 v^2. \la{CPmin2}
\ee 
This is strictly speaking quantitatively true only at tree-level, 
but we can observe from \fig\ref{fig:GG}
that there is no 
order of magnitude change due to 1-loop effects. 

As we discuss in~Appendix~\ref{se:Atmu}, the 
inclusions of 1-loop effects from the scalar self-couplings
(contained in $m_{S,i}^2$)
and from allowing $\hat A_t,\hat\mu\neq 0$ in \eq\nr{AB}
do not change these results in an essential way. 
Let us reiterate that the $\lambda_i$'s are always 
small in the MSSM, so that one practically never enters the 
region relevant for the Standard Model
and many other systems, where scalar fluctuations
related to $\lambda_i$'s change the predictions of 
perturbation theory in a qualitative way
(see, e.g.,~\cite{isthere,hart,su3adj}). 

\subsection{The non-perturbative phase diagram}

We now wish to explicitly check how the perturbative
predictions above are changed by non-perturbative effects.
The general experience from  
3d gauge+Higgs systems is that the properties of 
phase transitions are badly described by perturbation 
theory if the transition is weak so that the smallest
mass scale $m_\rmi{min}$ appearing is 
small, $g_w^2 T/(\pi m_\rmi{min})\gsim 1$~\cite{nonpert}.
This happens typically for large scalar self-couplings.
On the other hand, the changes in the locations of the 
phase transitions are expected to be small even in that case:
in the critical temperature $T_c$, non-perturbative effects
arise only at next-to-next-to-leading order~\cite{pa}, 
so that parametrically 
$\delta [m_1^2(T_c)+m_2^2(T_c)]/T_c^2 \sim \# g_w^4/(4\pi)^2$.
We would now like to verify how well this is true
numerically and, in particular,
whether C violation could be more favoured
than was perturbatively estimated.
(For example, could it take place at
somewhat larger $[m_1^2(T)+m_2^2(T)]/T^2$?) 

In order to carry out this check, we shall increase $[m_1^2(T)+m_2^2(T)]/T^2$
starting from the phase where C is broken, crossing the 
phase transition. We parameterise the starting point 
as in \eqs\nr{masses1}--\nr{masses3}, and add then 
a new dimensionless parameter $y$ to $m_1^2(T)/T^2, m_2^2(T)/T^2$:
\be
\frac{m_i^2(T)}{T^2} \to \frac{m_i^2(T)}{T^2} + y, \quad
i = 1,2. \la{newmasses2}
\ee
For simplicity we shall also denote what used to be $v/T$ 
in \eqs\nr{masses1}--\nr{masses3} by $\nu$. 
Increasing $y$ corresponds to increasing
the temperature in the 4d language, 
while $\nu,\tb,\cf$ are used to parameterise  
$m_1^2(T)/T^2,m_2^2(T)/T^2,m_{12}^2(T)/T^2$ 
at the reference point $y=0$.

The parameter space of the theory is quite large, so 
we have to make some choices.
In the following, we choose
a relatively small value $\lambda_5=0.001$
because this is natural from 
the point of view of dimensional reduction in the MSSM 
(cf.\ \eq\nr{mssmc}), and since a small value leads more
easily to relatively small $v/T$, not much larger than unity 
(cf.\ \eq\nr{lam5max}), 
which is also what we expect around the electroweak phase 
transition. We do not expect our results to change 
qualitatively with variations of $\lambda_5$. 

\begin{figure}[t]

\vspace*{0.0cm}

\centerline{\epsfxsize=6cm\hspace*{0cm}\epsfbox{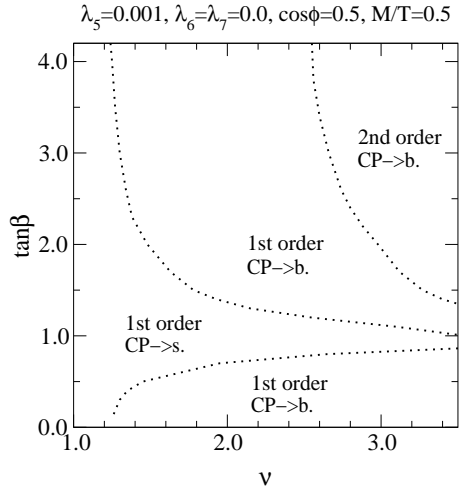}}

\vspace*{0.0cm}

\caption[a]{The 1-loop phase diagram for the transition
from the C violating phase 
to the usual broken and symmetric 
phases as $y$ is varied, for fixed
$\lambda_5=0.001, \lambda_6 = \lambda_7 = 0$ 
($\alpha_6=\alpha_7=\pi/2$),  
$\cos\phi=0.5$, 
$M/T=0.5$ ($m_U(T)/T = 0.35$),
$\hat A_t = \hat\mu = 0$. The 
transition from the C violating phase can lead 
either directly to the symmetric phase (``s.'') or to the usual 
broken phase (``b.''), and it can be either of 1st or 2nd order.} 
\la{fig:tbnu}
\end{figure}

We will also fix $\cos\phi=0.5$, which
does not have a significant effect. Moreover, 
we choose here $\alpha_6=\pi/2$, $\alpha_7=\pi/2$, 
setting $\lambda_6,\lambda_7$ to zero
(according to \eqs\nr{lam6}, \nr{lam7}, 
$\lambda_6 = 2\sqrt{\lambda_1\lambda_5}\cos\alpha_6, 
\lambda_7 = 2\sqrt{\lambda_2\lambda_5}\cos\alpha_7$). 
If one would take other values, one needs to have
an anti-correlation in the signs of $\lambda_6,\lambda_7$, to get 
$M_{12}^2 = m_{12}^2(T)+\fr12 \lambda_6 v_1^2 + \fr12 \lambda_7 v_2^2$ close
to zero; see \fig\ref{fig:GG}. However, we again do not expect
new qualitative effects from relaxing this assumption. 
As to the stop sector, we somewhat arbitrarily
choose $m_U^2(T)/T^2 \approx 0.13$, 
but vary $\hat A_t, \hat\mu$ within $|\hat A_t|, |\hat \mu| \lsim 0.3$. 
We denote $M = \sqrt{2} m_U(T)/h_t$ (cf.\ \eq\nr{GHMdef})
so that $M/T = 0.5$.

These choices fix the quartic couplings of the theory as well as
the stop sector, but still leave the diagonal entries of 
the SU(2) scalar mass matrix, parameterised by $\nu,\tb$
in \eqs\nr{masses1}, \nr{mm2H}, open. 
The perturbative phase diagram in this space, 
based on the full 1-loop effective potential
in \eq\nr{v1loop},  
is shown in \fig\ref{fig:tbnu}.  
We have chosen a number 
of points from this diagram
for further non-perturbative study, such  
that all different qualitative types of transitions are represented. 

\begin{figure}[t]

\centerline{\epsfxsize=4.5cm\hspace*{0cm}\epsfbox{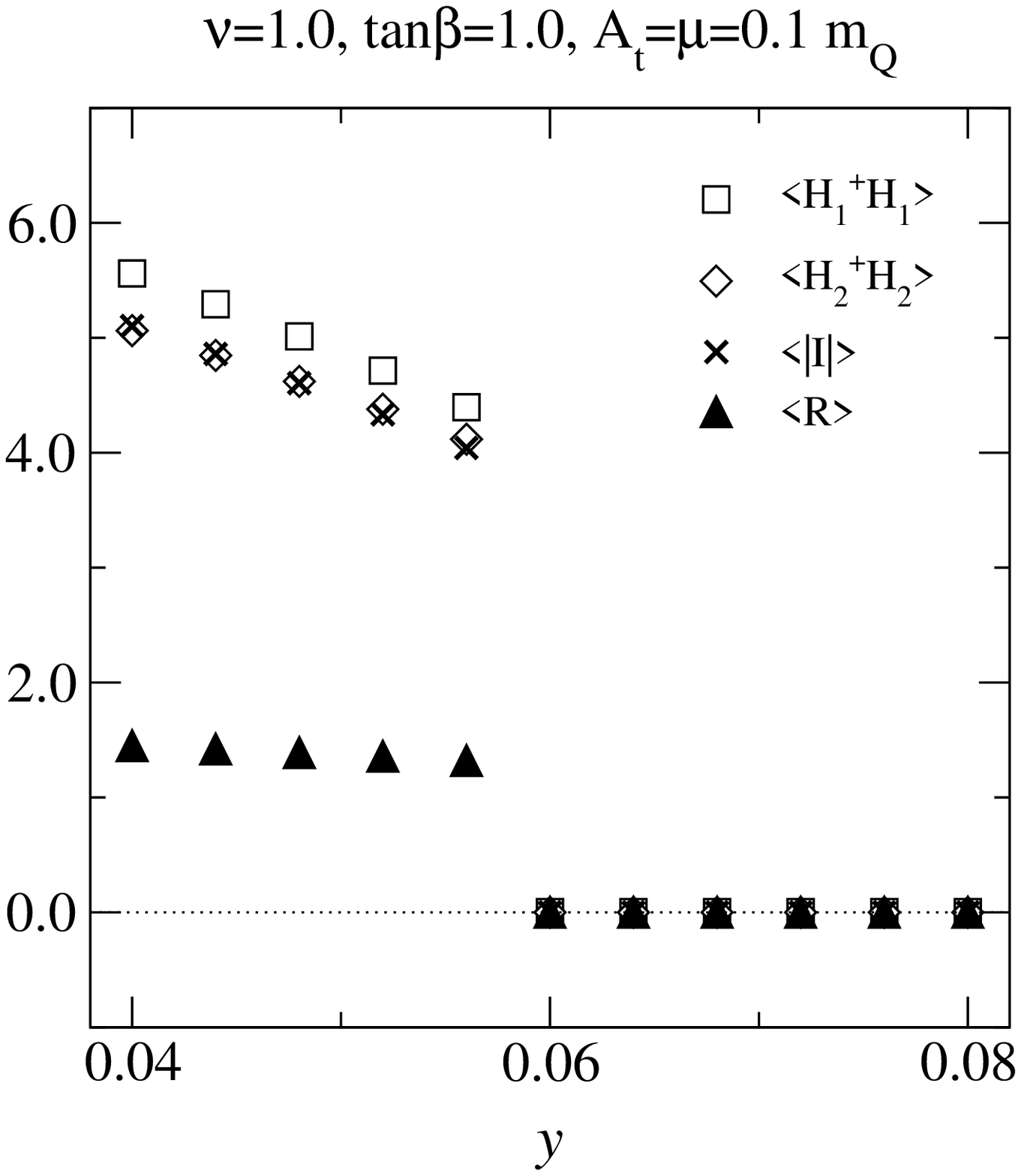}%
\epsfxsize=4.5cm\hspace*{0.7cm}\epsfbox{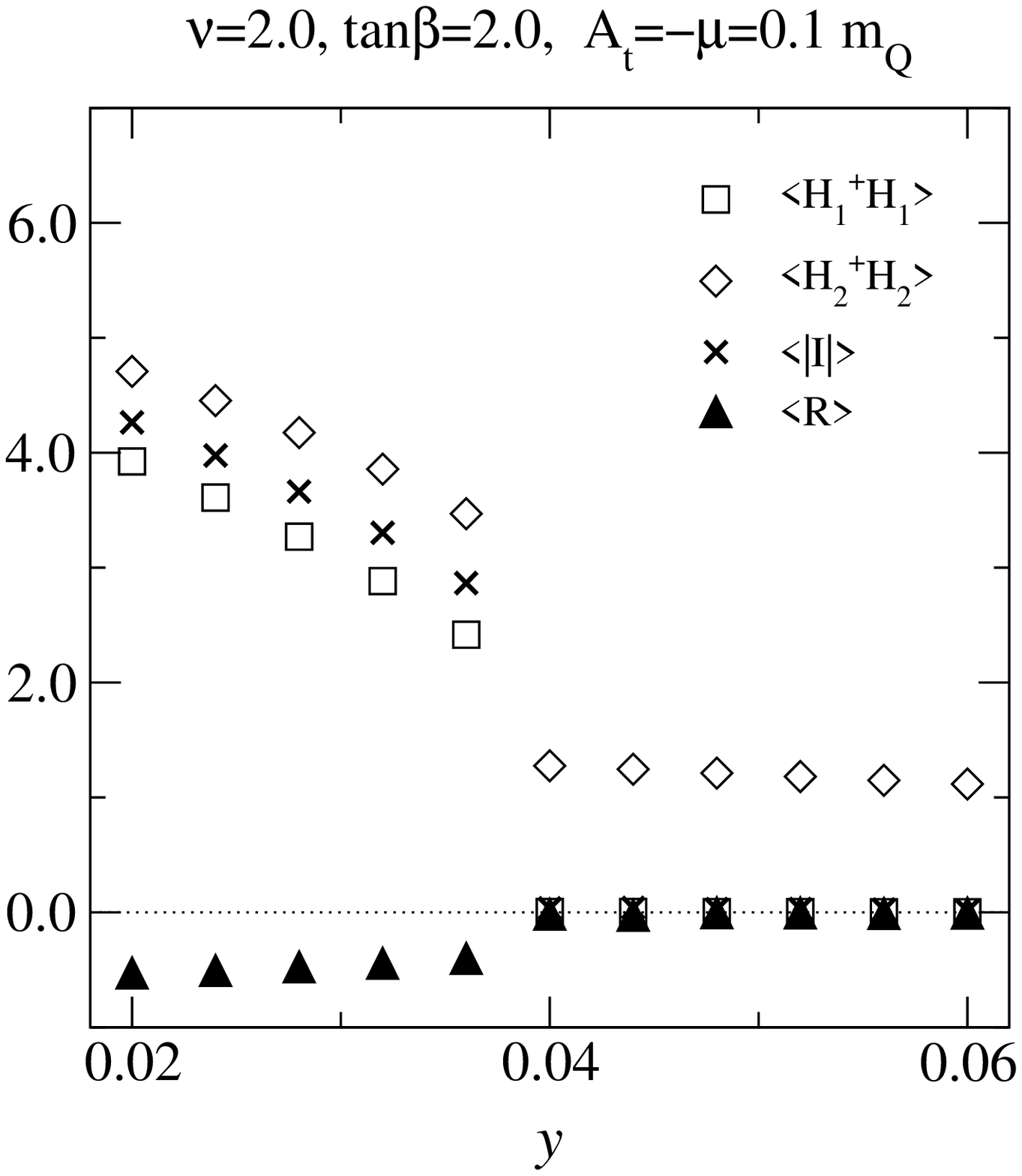}%
\epsfxsize=4.6cm\hspace*{0.7cm}\epsfbox{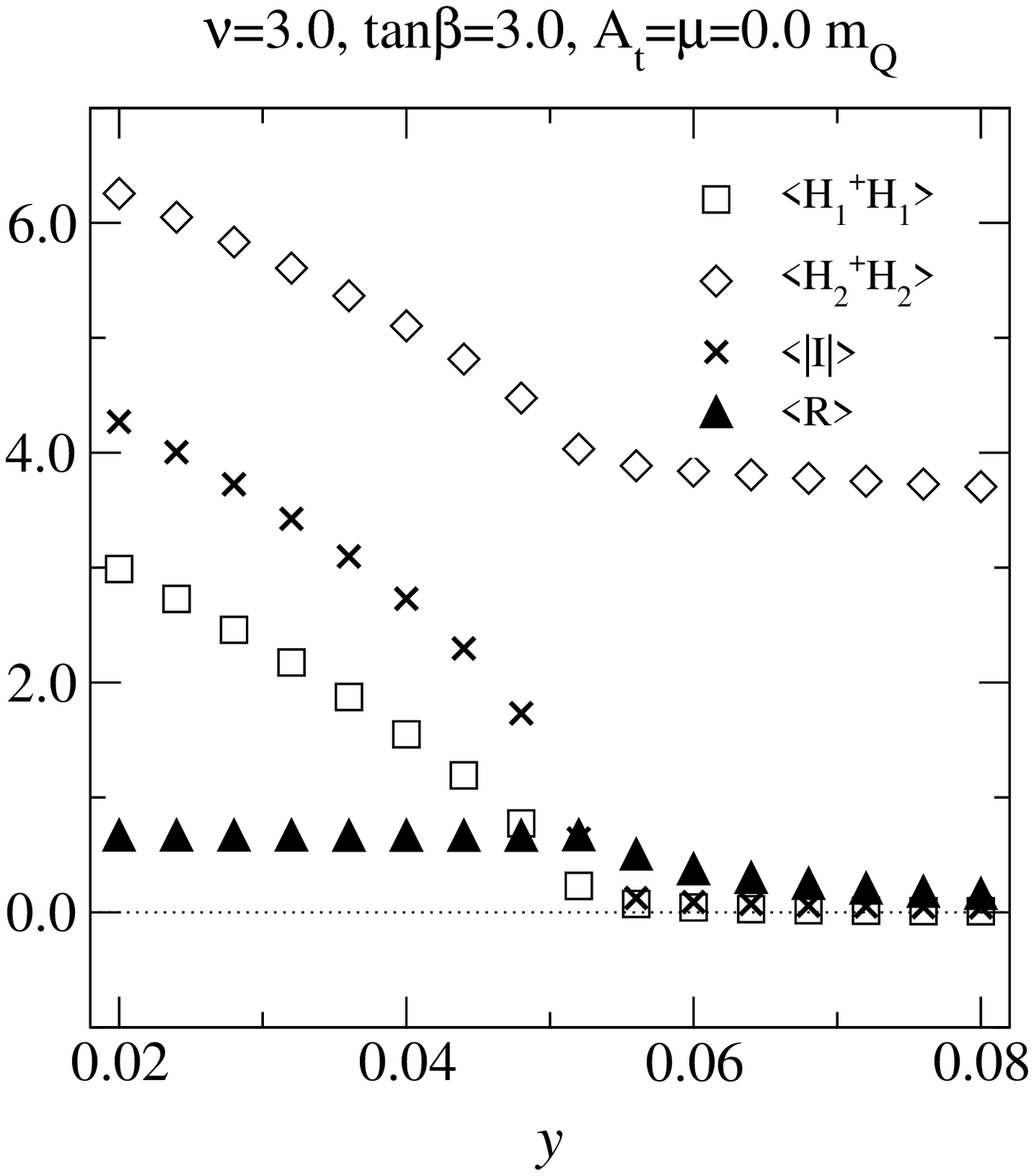}}

\vspace*{0.7cm}

\centerline{\epsfxsize=4.5cm\epsfbox{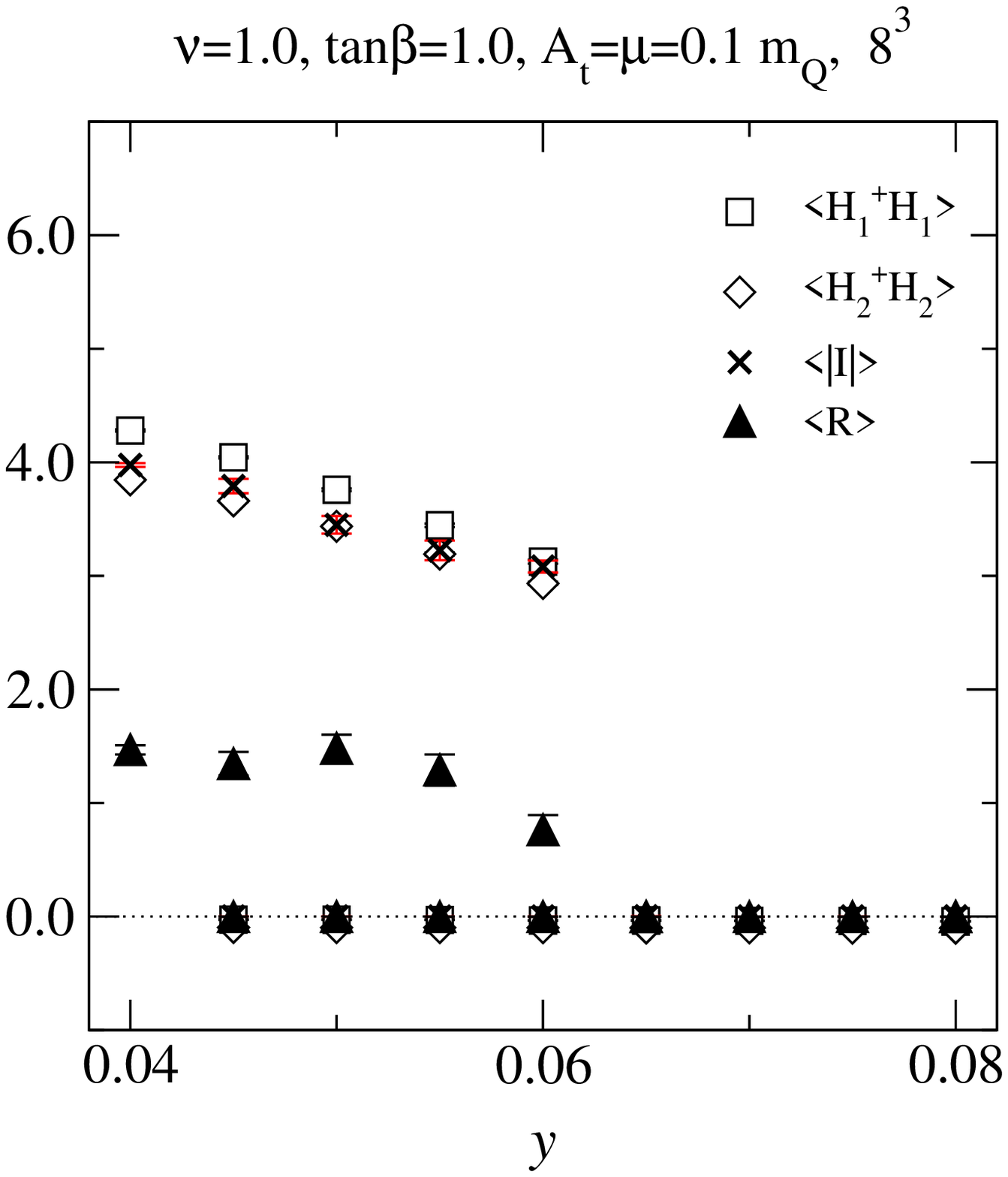}%
\epsfxsize=4.5cm\hspace*{0.7cm}\epsfbox{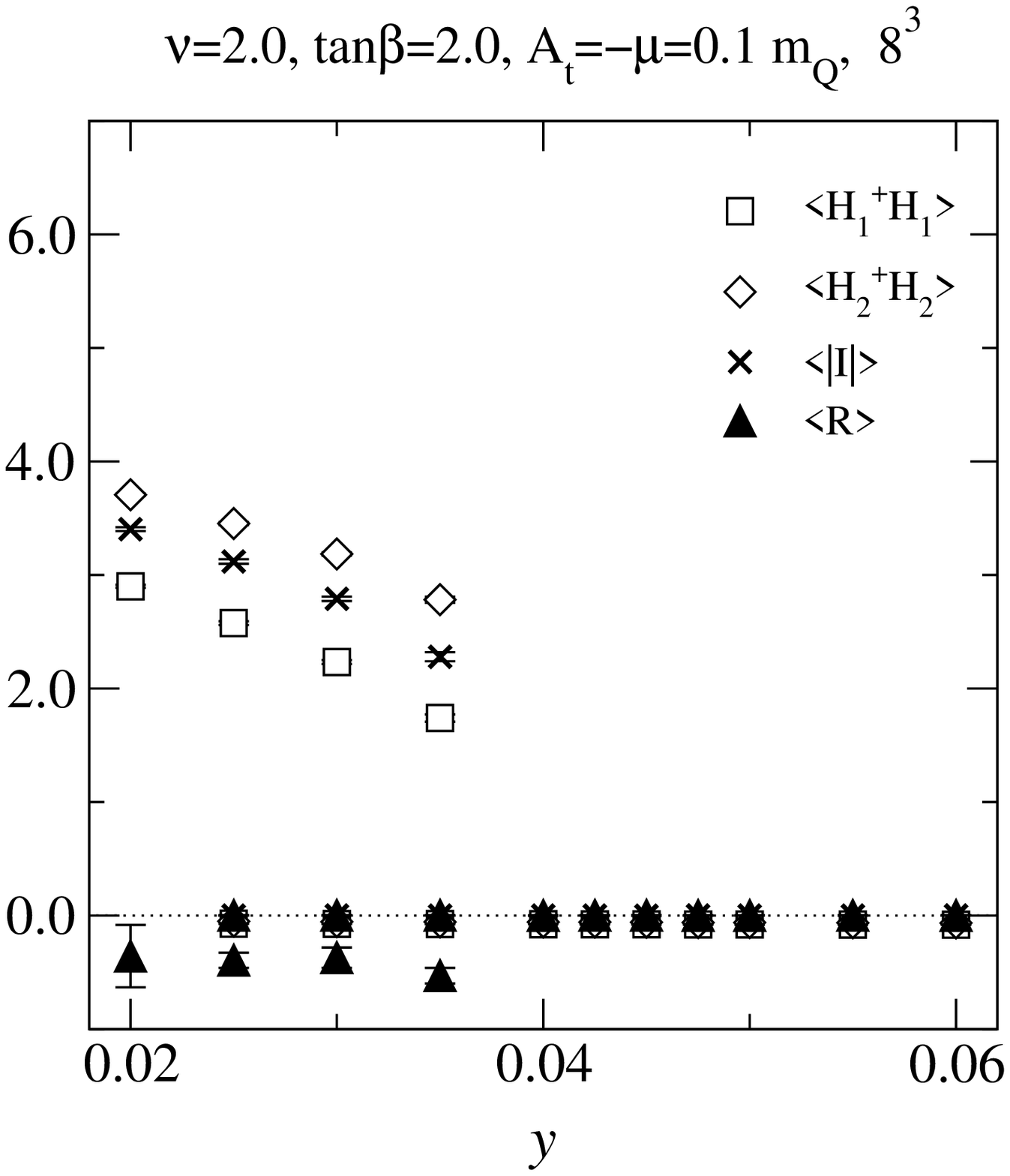}%
\epsfxsize=4.7cm\hspace*{0.7cm}\epsfbox{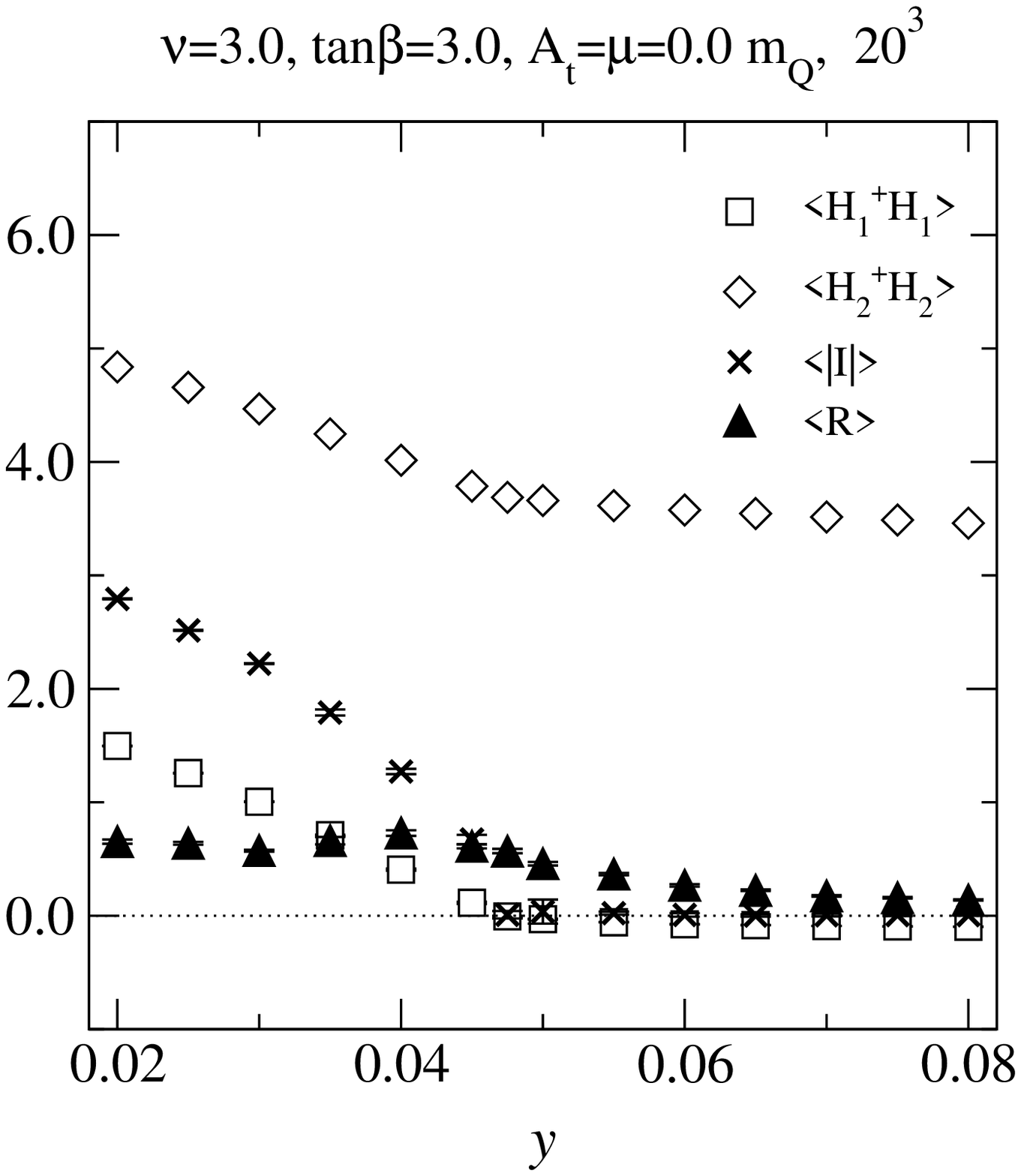}}

\vspace*{0cm}

\caption[a]{Top: Mean field estimates for the $y$-dependence
of different expectation values, corresponding to the physical volume
obtained with $\beta_w=8$, $N_1N_2N_3=20^3$.  For $I$ the absolute
value of the volume average is taken.  Bottom: Lattice results at $\beta_w=8$.
For the first two sets the transition is strongly of the first order, and
a small volume has been used; in spite of this, some $y$-values 
allow for separate measurements in two different metastable phases. 
For the middle set, the transition goes on the lattice directly
to the C conserving symmetric phase, rather than to the C conserving
broken phase seen in the perturbative plot.}
\la{fig:mf}
\end{figure}

In \fig\ref{fig:mf}, we show the mean field estimates 
for the behaviour of the operators in \eq\nr{ops} for
a few representative choices of parameters.  
These estimates follow from \eq\nr{mfest}, with 
the 1-loop effective potential from \eq\nr{v1loop}, supplemented
by a non-vanishing value of $\chi$
as needed in \eq\nr{mfest}, whereby the 
contribution on the last line in \eq\nr{s1loop}
goes over into a sum over the eigenvalues of a 9$\times$9 matrix,
with the radial $U$ direction coupled to the SU(2) scalars.  
(The Goldstone modes of $U$ appear 
separately as, e.g.,\  in~\cite{bjls}.)

The mean field estimates can be compared with the lattice results, 
shown also in \fig\ref{fig:mf} for the same parameter values. 
We observe that, as expected, the behaviours are quite 
close to each other.
We then estimate that compared with \fig\ref{fig:GG}, 
the largest value allowing for spontaneous C violation 
can change at most by 
\be
\delta \frac{m_1^2(T)+m_2^2(T)}{T^2} \approx 2\delta y_c \lsim 0.2, \la{dmm}
\ee
where the upper bound is quite conservative.

\subsection{Implications for the physical MSSM}
\la{comparison}

Let us now compare \fig\ref{fig:GG} supplemented by \eq\nr{dmm}, 
with the part of the parameter space allowed by the MSSM. One
could make many comparisons, but we focus here on
$m_{12}^2(T)/T^2$ and, in particular,  
$[m_1^2(T)+m_2^2(T)]/T^2$, cf.\ \eq\nr{CPmin2}.

First of all, we recall from \fig\ref{fig:GG} that 
one needs small values of $M_{12}^2/T^2$ and therefore, 
due to the smallness of $\lambda_6,\lambda_7$, 
small values of $m_{12}^2(T)/T^2$ (cf.\ \eq\nr{M122}). 
It can be observed from \eq\nr{seq3rd} that in the MSSM
this is easier to satisfy at finite temperatures 
than at zero temperature, due to a temperature correction 
which can cancel the $T=0$ part in $m_{12}^2(T)$ for $A_t\mu > 0$.
Furthermore, even if the experimental lower limit on $m_A$
appears to be rather high, $\sim 90$ GeV~\cite{lep}, one can at least
partially compensate
for this by taking a large $\tb$, since only the combination
$m_A^2 \tb /(1+\tan^2\beta)$ appears in \eq\nr{seq3rd}.

However, the constraint on $[m_1^2(T)+m_2^2(T)]/T^2$ works 
in the opposite direction. To get spontaneous C violation 
$m_1^2(T)+m_2^2(T)$ should be small (\fig\ref{fig:GG}), 
its order of magnitude given by $\lambda_5 v^2$ (\eq\nr{CPmin2}).
Finite temperature does not help with this
constraint at all: $v^2$ gets smaller, 
while $m_1^2(T)+m_2^2(T)$ gets larger. 

To be more precise, we obtain 
using \eqs\nr{seq1st}, \nr{seq2nd} that in the MSSM,
\be
m_1^2(T) + m_2^2(T) \approx m_A^2 +
0.5 T^2 + 0.25 T^2 (1-\Att-\muu). \la{sum}
\ee
Let us now reiterate that
in the limit
of large $m_Q^2$ and small $m_U^2$ that we are working at, 
$1-\Att-\muu$ should in general be positive for the theory to be consistent 
from the point of view of boundedness and
electroweak vacuum stability~\cite{cpown}.
Taking into account also the
experimental lower limit on the CP odd Higgs mass parameter
$m_A\gsim 90$ GeV~\cite{lep}, we then get that in the MSSM,  
\be
[m_1^2(T)+m_2^2(T)]/T^2 \gsim 1.3. 
\ee
This holds
for all temperatures below $100$ GeV, i.e., also in the
broken electroweak Higgs phase. 

This result can be contrasted with~\fig\ref{fig:GG}. 
We observe that there is 
no overlap, since $[m_1^2(T)+m_2^2(T)]/T^2 \lsim 0.5$ always.
A non-perturbative change of the order in \eq\nr{dmm}
clearly cannot bridge the gap.

\begin{figure}[t]

\vspace*{0.5cm}

\hspace*{2cm}
\parbox[c]{4.4cm}{
\begin{picture}(60,40)(0,0)

\SetWidth{1.5}
\SetScale{0.9}
\DashLine(0,5)(60,5){5}
\CArc(30,20)(15,0,360)
\GCirc(30,5){3}{0}
\Text(0,14)[c]{$H$}
\Text(40,36)[l]{$U$}

\end{picture}}
\parbox[c]{2.5cm}{
\begin{picture}(60,40)(0,0)

\SetWidth{1.5}
\SetScale{0.9}
\Line(0,40)(20,20)
\DashLine(0,0)(20,20){5}
\DashLine(20,20)(40,0){5}
\Line(20,20)(40,40)
\GCirc(20,20){3}{0}
\Text(55,20)[c]{$=$}

\end{picture}}
\parbox[c]{2.5cm}{
\begin{picture}(60,40)(0,0)

\SetWidth{1.5}
\SetScale{0.9}
\Line(0,40)(20,20)
\DashLine(0,0)(20,20){5}
\DashLine(20,20)(40,0){5}
\Line(20,20)(40,40)
\Text(55,20)[c]{$+$}

\end{picture}}
\parbox[c]{2.5cm}{
\begin{picture}(60,40)(0,0)

\SetWidth{1.5}
\SetScale{0.9}
\Line(0,40)(20,20)
\DashLine(0,0)(20,20){5}
\DashLine(40,20)(60,0){5}
\Line(40,20)(60,40)
\Line(20,21)(40,21)
\Line(20,20)(40,20)
\Line(20,19)(40,19)
\Text(30,30)[c]{$Q$}

\end{picture}}

\vspace*{0.5cm}

\caption[a]{The tadpole graph leading to \eq\nr{f_full}. 
The effective local quartic vertex shown, 
the second line in \eq\nr{action} with $\gamma_i$'s 
from \eq\nr{gpar}, is a good approximation 
as long as $m_U^2 \ll m_Q^2$.}
\la{fig:graph}
\end{figure}
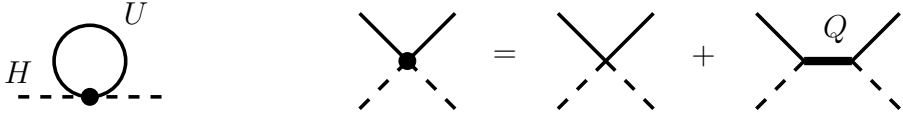

It can be observed from \fig\ref{fig:GG} that increasing $m_U(T)$
seems to allow for larger values of $[m_1^2(T)+m_2^2(T)]/T^2$.
However, this effect is not enough to change our conclusions. 
In fact, for large $m_U(T)$ the field $U$ can be integrated
out, as we review in~\ref{intstop}, and the result is a
theory of the form in \eq\nr{action} but without $U$. 
In this theory, there is an upper limit on $[m_1^2(T)+m_2^2(T)]/T^2$
leading to spontaneous C violation
as well, numerically $[m_1^2(T)+m_2^2(T)]/T^2 \lsim 0.4$
for $v/T\lsim 3.0$~\cite{2hdcp}. 
The effect observed in \fig\ref{fig:GG} is equivalent to the fact 
seen in \eqs\nr{dmm1mm2a}, \nr{dmm1mm2b} that the effective 
$[m_1^2(T)+m_2^2(T)]/T^2$ after integrating out $U$ tends to 
decrease by increasing $m_U(T)$. More precisely,  the last term in
\eq\nr{sum}, $(h_t^2 T^2/4)  (1-\Att-\muu)$, 
is multiplied by the factor 
\be
f = 1-\frac{3}{\pi} \frac{m_U(T)}{T}.
\ee 
However, this decrease does not continue forever, 
but the formula breaks down for large enough $m_U(T)$
when the high temperature expansion is no longer applicable,  
and $f$ then goes over into the tadpole integral shown 
in \fig\ref{fig:graph} (provided that
still $m_U^2 \ll m_Q^2$), 
\be
f
\to
\left. \frac{6}{\pi^2} 
\int_0^\infty dx \frac{x^2}{\sqrt{x^2+y^2}}
\frac{1}{\exp({\sqrt{x^2+y^2}})-1} \right|_{y=m_{U}/T}. \la{f_full}
\ee
This is always positive, 
so that $[m_1^2(T)+m_2^2(T)]/T^2$ does not decrease 
below $m_A^2/T^2 + 0.5 \gsim 1.3$ 
even if $m_U(T)\gsim T$, and our previous
conclusion continues to hold. 

In summary, spontaneous CP violation seems to be excluded
in the MSSM both at $T=0$  and at finite temperatures around
the electroweak phase transition. 

\subsection{Transitional CP violation}

Let us finally come to the issue of ``transitional CP violation''.
There have been suggestions that even if not in the symmetric or broken 
phase, spontaneous C violation could take place  
within the phase boundary between the symmetric and broken
phases~\cite{cpr,fkot}. However, these suggestions could not
be confirmed by later analyses 
for physical parameter values 
(particularly realistic $m_A$)~\cite{pj}. 
Basically, the point is 
that as our discussion above indicated,
spontaneous C violation is always more likely
at large vevs, cf.\ \eqs\nr{CPviol}, \nr{CPmin2}. 
Thus it seems unlikely that C would be 
violated in the phase boundary, if it is not
violated in the broken phase. Below, we will inspect the 
issue numerically at a physical phase boundary corresponding 
to the MSSM, and find the same conclusion.
Let us note that, on the contrary, 
transitional C violation could take place in, say, 
the NMSSM~\cite{pj,hsch}.


\section{The strength of the physical phase transition}
\la{strength}

\subsection{Parameter values}
\la{se:param}

We now move to the electroweak phase transition 
in the physical MSSM. 
Let us start by choosing parameter values. 
We have previously carried out simulations corresponding to 
$\tan\beta=3, m_H=95$ GeV, a large $m_A\gsim 200$ GeV, 
and a light right-handed stop~\cite{mssmsim}.
We found a transition which was somewhat stronger than in 2-loop
perturbation theory and certainly strong enough for baryogenesis. 
Recently, 4d finite temperature lattice simulations have also 
been carried out in a scalar theory with MSSM type couplings, 
at $m_H\approx 45$ GeV~\cite{4dmssm}. There the transition 
is very strong, and it was found to agree well with 
perturbation theory. 

We now want to take a larger $\tan\beta=12$
than in~\cite{mssmsim}, corresponding
to $m_H\approx 105$ GeV for a left-handed 
squark mass parameter $m_Q \approx 1$ TeV and a light
right-handed stop. We also take a smaller
$m_A\approx 120$ GeV for two reasons. First, because then the experimental 
Higgs mass lower bound is somewhat relaxed~\cite{lep}. 
(Recently it has been suggested that 
the experimental Higgs mass lower bound may be
further relaxed for such $m_A$ 
because of large explicit CP violation 
in the system, as we will have~\cite{pw}--\cite{kw}). 
Second, because a smaller 
$m_A$ makes the heavy Higgs doublet eigen\-direction somewhat more
dynamical, since 
effects related to it are suppressed by 
$\sim g_w^2 T/(m_A^2+ 0.5T^2)^{\fr12}$, see~\ref{inthiggs}. 
This means that the situation could  
be favourable for ``dynamical'' CP violation~\cite{cpown}
(i.e.,\, a somewhat non-trivial profile for CP odd observables
within the phase boundary, even if not actual spontaneous CP violation),
as well as for having a non-trivial $\tb$-profile~\cite{mqs,cm,pj}, 
which might affect the actual 
baryon asymmetry produced~\cite{non-eq}--\cite{risa}.

We also introduce a non-vanishing
squark mixing parameter $A_t\approx 200$ GeV, 
as well as gaugino and Higgsino mass parameters $M_2$, $\mu$, 
with $M_2 \sim |\mu|\approx 200$ GeV. The strength of the phase
transition depends little on these 
parameters~\cite{cqw1}--\cite{mlo3}. 
In addition,  to observe 
the CP violating effects more clearly, we choose a maximal
explicit phase for the $\mu$-parameter, $\mu = i|\mu|$.
The first and second generation squarks and sleptons are assumed heavy,
whereby even such a large phase is not in conflict with 
electric dipole moment constraints~\cite{edm}--\cite{ckp}.

Finally and most importantly, we take a negative right-handed
squark mass parameter, $\tilde m_U^2 \equiv -m_U^2 > 0$. Most 
of the time we work at $\tilde m_U = 65$ GeV (see below). To summarise, 
we thus have
\ba
& &  
g_w^2\approx 0.42,\quad 
g_s^2\approx 1.1, \quad
h_t \approx 1.0, \\
& & 
m_Q \approx1 \mbox{ TeV}, \quad
\tilde m_U  \approx 65 \mbox{ GeV}, \quad
m_A \approx120 \mbox{ GeV}, \quad
\tb \approx12,  \\
& &  
A_t\approx |\mu|\approx M_2\approx 200 \mbox{ GeV},
\quad \mu  \approx i |\mu|,
\ea
where on the first row the couplings are assumed
to be evaluated at a scale $\sim 2\pi T$.
These parameters correspond to a lightest 
physical Higgs mass of about 105 GeV, and a lightest 
stop mass of about 155 GeV, with an uncertainty
of a few GeV. 

Applying then the formulas in Appendix A.7 of~\cite{cpown}
(and fixing $T\sim 95$ GeV, 
as suggested by 1-loop perturbation theory, inside logarithms and elsewhere
where its effects are subdominant, in order to simplify the results), 
we obtain the following effective
couplings for the theory in \eq\nr{action}:
\ba
& & 
m_1^2(T) \approx 18380 \mbox{ GeV}^2 + 0.1218 T^2, \la{54} \\
& & 
m_2^2(T) \approx -3980 \mbox{ GeV}^2 + 0.6218 T^2, \\
& & m_{12}^2(T) \approx (-1190 - i 100) \mbox{ GeV}^2 
+ i 0.0030 T^2, \la{m12} \\
& & 
m_U^2(T) \approx -4225 \mbox{ GeV}^2 + 0.8534 T^2, \la{57} \\
& & 
\gamma_1 \approx -0.044, \quad
\gamma_2 \approx  0.96, \quad
\gamma_{12} \approx -i 0.037, \quad
\lambda_U \approx 0.197, \\
& & 
\lambda_1 \approx 0.0652, \quad
\lambda_2 \approx 0.1188, \quad
\lambda_3 \approx 0.0673, \quad
\lambda_4 \approx -0.1948, \\
& & 
\lambda_5 \approx -0.00019, \quad
\lambda_6 \approx i 0.0017, \quad
\lambda_7 \approx i 0.0022. \la{3dprms}
\ea
We should stress that these numbers have of course some 
perturbative errors, but this is not essential for our main 
statements. Indeed, we will compare 3d perturbation theory
with 3d lattice simulations, and precisely the same parameters
in \eqs\nr{54}--\nr{3dprms} are chosen in both cases. This will allow us to 
unambiguously find out whether there are non-perturbative effects 
in the system. Such non-perturbative effects will then
remain very similar even if the 4d parameter values are changed 
slightly, or if the reduction computation leading to 
\eqs\nr{54}--\nr{3dprms} 
is carried out more precisely. 

Finally, let us mention a technical point. 
We treat the mass parameters in \eqs\nr{54}--\nr{57}
as those at the scale $\bmu=T$ inside the 3d theory
(to be more precise: we choose the $\Lambda$-parameters
discussed in Appendix~\ref{Lam} to be $\Lambda \sim 1.0T$).
In order to remove the ambiguity from this choice, a complete
2-loop dimensional reduction computation would be needed
for the mass parameters~\cite{generic}. Unfortunately, 
such computations have been carried out only in special cases.
One is the Standard Model~\cite{generic},
where it turns out that
numerically $\Lambda\sim 7T$. 
In~\cite{mssmsim} it was argued that this should be expected 
also for the MSSM. An explicit computation was then carried out 
in~\cite{mlo2} for a small $m_Q\sim 300$ GeV, 
and the actual scale was found to be of order $\sim 2T$ for the
diagonalized Higgs mass parameter (see \ref{inthiggs} for the definition), 
$\sim 7T$ for the stop mass parameter. However, the scales 
depend on the other parameters of the theory. While 
the way to carry out the computation for large $m_Q$,
the case relevant here, 
has also recently been clarified~\cite{ll}, explicit
results for the full MSSM are still missing,
so we cannot simply take over old values.  

Fortunately, it turns out that this ambiguity 
is of very little significance for our results. Indeed, 
we have tested the effect of changing from $\Lambda\sim T$ to  
$\Lambda \sim 7T$ with 2-loop perturbation theory inside the 3d theory. 
We find that the critical temperature, as well as the value of
$\tilde m_U$ corresponding to the ``triple point''
(see \fig\ref{fig:phasediag}) change by a few GeV, 
but apart from this shift the values of $v/T$ at the 
transition remain almost the same. Thus the ambiguity is
completely inconsequential, if we {\em normalise our parameter 
values with respect to the triple point}.  

\subsection{Phase transition in perturbation theory}
\la{sec:pt}

\begin{figure}[t]

\centerline{\epsfxsize=6cm\hspace*{0cm}\epsfbox{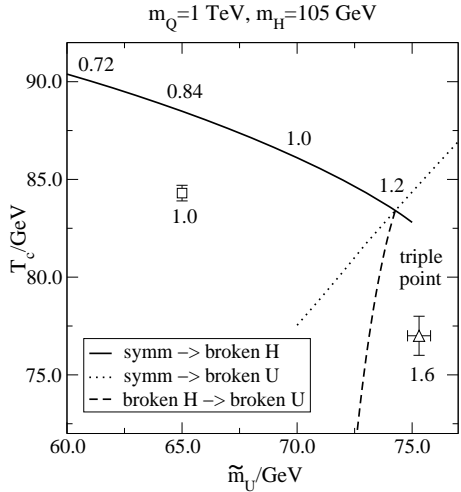}}

\caption[a]{The perturbative 2-loop phase diagram for $m_Q=1$ TeV.
The values of $v/T$ at the transition point are also shown. Lattice
results are displayed at $\tilde m_U = 65$ GeV (square), and 
at the triple point (triangle). The lattice triple point errorbars
include only statistical errors (see the text), and are thus 
an underestimate.}
\la{fig:phasediag}
\end{figure}

\begin{figure}[t]

\centerline{\epsfxsize=6cm\hspace*{0cm}\epsfbox{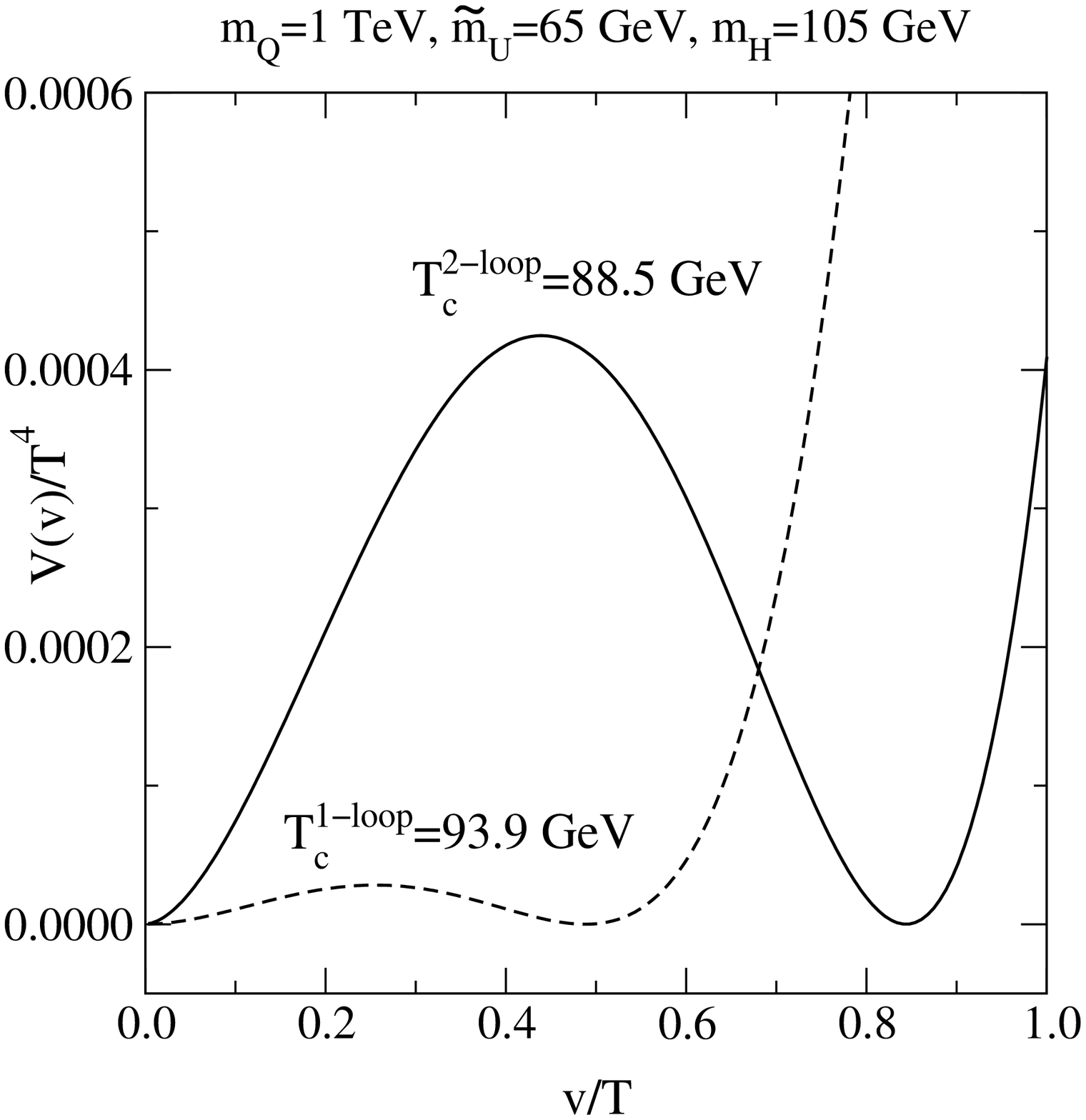}%
\epsfxsize=6cm\hspace*{1cm}\epsfbox{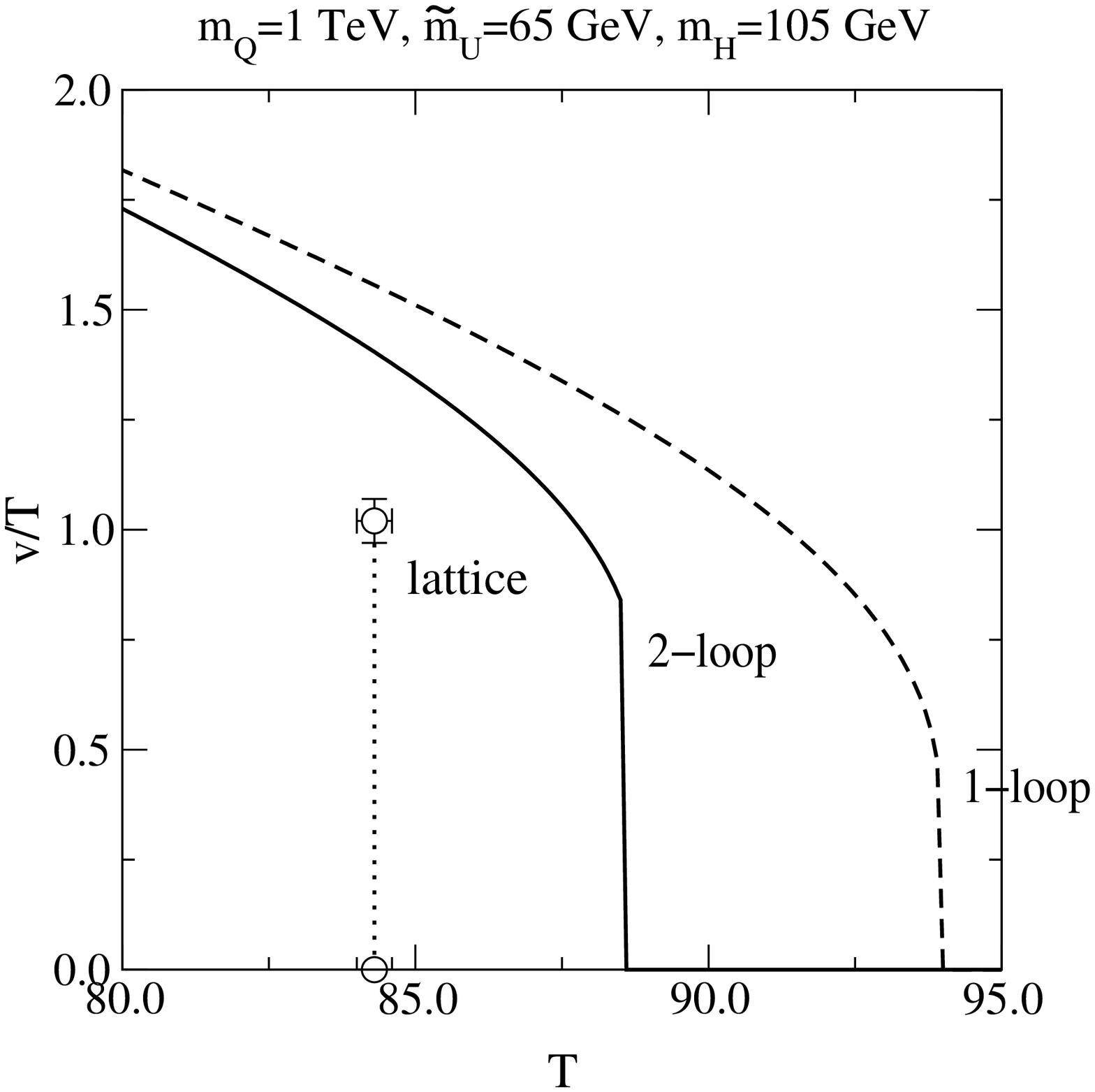}}

\caption[a]{The properties of the phase transition in 1-loop and 2-loop
perturbation theory, as well as on the lattice. Left: the perturbative
effective potentials at the corresponding critical temperatures. Right:
the vacuum expectation values as a function of $T$. The
lattice value refers to $\left. v/T\right|_\rmi{lattice} = 
(2 \Delta \sum_i \hat H_i^\dagger \hat H_i)^{\fr12}$, 
cf.\ \eqs\nr{lattaction}, \nr{lattdef}. }  
\la{fig:broken}
\end{figure}

For the parameter values in \eqs\nr{54}--\nr{3dprms}, the theory 
in \eq\nr{action} has a first order electroweak phase transition. 
Let us first discuss its properties 
in perturbation theory. In the following, we 
use the Landau gauge and the scale parameter $\bmu=T$, 
as is conventional in the literature. 

As we mentioned in \se\ref{sec:ac}, for studying the 
strength of the phase transition the theory in \eq\nr{action}
can be simplified by integrating out 
a linear combination of the two Higgs 
doublets. This is not only a convenience but also a way
of increasing the accuracy of perturbation theory:
large effects related to a heavy excitation 
get resummed. We discuss the details of the integration
out in~\ref{inthiggs}.
After the integration out, we can directly employ the
2-loop effective potential computed in~\cite{mssmsim}. 
We may note that at 1-loop level we have also the 
effective potential in the full theory available, 
see \eq\nr{v1loop}, and in practice
we find quite similar results 
as by using the diagonalized effective theory 
(to 1-loop accuracy). 

In \fig\ref{fig:phasediag} we show the phase diagram
as a function of $\tilde m_U,T$. The three familiar 
types of transitions~\cite{bjls} are well visible. 
However, with our parameters, we observe
that in fact a 2-stage transition as proposed in~\cite{bjls} is excluded, 
unlike with the parameter choice 
which we employed in~\cite{mssmsim}.
Instead one has to worry about whether the metastability of the broken
Higgs phase is sufficiently strong~\cite{mcdo}, 
if $\tilde m_U$ is very close to the triple point. 
However, we can here apply the logic of~\cite{cms} in the 
opposite direction, and state that one would never tunnel into the broken
$U$ direction even if possible in principle. (In the case of~\cite{cms},
the dashed line in \fig\ref{fig:phasediag} was tilted in the other 
direction, and the statement was that if one ends up in the phase
with broken $U$, one can never get from there to the usual electroweak
phase, even if that would be the thermodynamically stable phase.) 
Still, values $\tilde m_U\lsim 70$ GeV would seem safest. 
We choose here $\tilde m_U = 65$ GeV for further study.

As discussed in \se\ref{se:obs},   
the strength of the phase transition is
in perturbation theory usually addressed
in terms of $v/T$ rather than $H_1^\dagger H_1 + H_2^\dagger H_2$, 
or $h^\dagger h$ in the diagonalized 
theory of \ref{inthiggs}. 
The relation is then obtained from \eq\nr{prms} or, 
if one converts from lattice to perturbation theory, from \eq\nr{lattdef}.
We show the 1-loop and 2-loop results for $v/T$ in \fig\ref{fig:broken}.
We observe the familiar feature that the transition is significantly
stronger at 2-loop level~\cite{e}--\cite{mlo2}, 
which is related to the fact that the critical temperature $T_c$ is lower.

Finally, let us recall that because a certain combination  
of the Higgs doublets is heavy, the transition
takes essentially place in one ``light'' direction
in the field space much like in the Standard Model
(see also \fig\ref{fig:ReImb16} below), 
and the sphaleron energy is really
determined by the value of $H_1^\dagger H_1 + H_2^\dagger H_2$ 
in the broken phase. 
Note, in particular, that even though possible in 
principle~\cite{grhi}, we would not expect the sphaleron 
bounds to be modified by the CP violation apparent in the 
couplings in \eqs\nr{54}--\nr{3dprms}, because the Higgs
direction orthogonal to the light eigenmode $h$ is indeed so 
heavy that effects related to it are strongly suppressed;
see~\ref{inthiggs}. 

\subsection{Lattice simulations}\la{lattsim}

\begin{table}[htb]
\centerline{\begin{tabular}{c|l}
\hline
 $\beta_w$ & volumes \\
\hline
 8         & $12^3$ ~~ $16^3$ ~~ $24^3$ \\
12         & $16^3$ ~~ $24^3$ ~~ $32^3$ \\
16         & $24^3$ ~~ $32^3$           \\
24         & $32^3$ ~~ $40^3$ ~~ $56^3$ \\
\hline
\end{tabular}}
\caption[a]{Lattice spacings $\beta_w = 4/(g_w^2 a T)$ and lattice
volumes, used for simulations at the transition temperature.  All 
the simulations here are multicanonical.} 
\la{tab:vols}
\end{table}

In order to inspect the reliability of
the perturbative estimates discussed above, we have
carried out lattice simulations at $\tilde m_U = 65$\,GeV.  
First, a series of simulations was performed
in order to determine the transition temperature
$T_c$, the value of $v/T$ in the broken phase, the latent heat and the
surface tension.  The lattice sizes and lattice spacings are shown in
Table~\ref{tab:vols}.  For each lattice listed we performed 50 000 --
360 000 compound iterations (5 $\times$ overrelaxation + 1 $\times$
heat bath).  All of the simulations in Table~\ref{tab:vols} are
multicanonical, i.e. the probability distribution has been modified to
enhance the tunnelling between the broken and symmetric phases.  For
technical details, we refer to \se\ref{algo}
and to ref.~\cite{mssmsim}, which includes a
detailed description of the application of the multicanonical method
to the MSSM (albeit with only one SU(2) Higgs doublet).

At this point in parameter space the transition is relatively strong,
as can be seen from the probability distributions of $H_1^\dagger
H_1 + H_2^\dagger H_2$ in Fig.~\ref{fig:hg}.  The 
distributions measured
have here been reweighted to a temperature which gives equal
probabilistic weights to the symmetric and broken phases (``equal
weight histograms'').

\begin{figure}[t]

\centerline{\epsfxsize=6cm\epsfbox{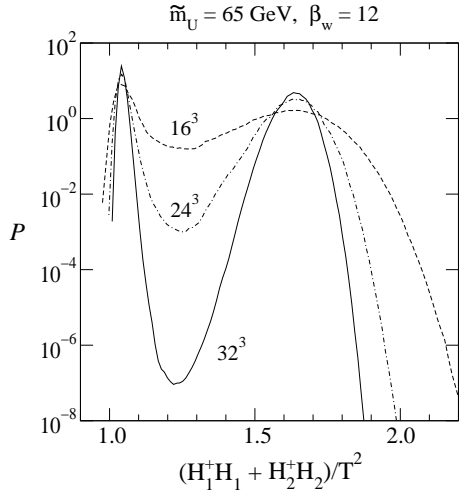}}

\caption[a]{The ``equal weight'' probability distributions of 
$(H_1^\dagger H_1 + H_2^\dagger H_2)/T^2$ (without 
the subtractions in \eq\nr{Higgsub}), 
measured from $\beta_w \equiv 4/(g_w^2 a T) = 12$ lattices.}
\la{fig:hg} 
\end{figure}

\begin{figure}[t]

\centerline{\epsfxsize=6cm\epsfbox{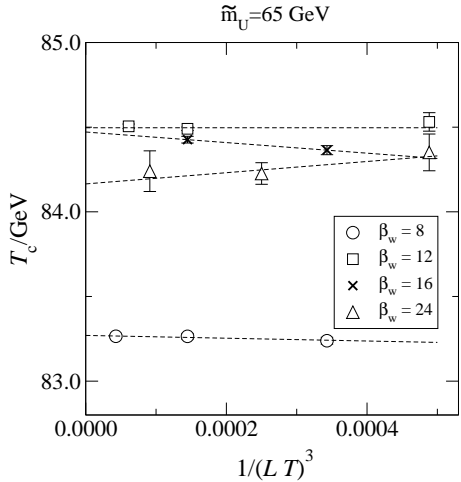}}

\caption[a]{The transition temperatures at different $\beta_w$'s 
and volumes, plotted against inverse volume.
The transition temperatures have been determined with the ``equal weight''
criterion.}
\la{fig:tcrit}

\end{figure}

\subsection{Results for various physical observables}

\paragraph{Critical temperature:}

Tuning the temperature so that the ``order parameter'' probability
distributions have equal weights in symmetric and broken phases
(see Fig.~\ref{fig:hg}) gives a good definition for the transition
temperature at finite volumes.  We use this definition in what
follows.  Other definitions can be found in \cite{mssmsim}; at the
infinite volume limit all of these give identical results
within statistical errors.

In \fig\ref{fig:tcrit} the equal weight temperatures are shown
for all lattices in Table~\ref{tab:vols}.  For each lattice spacing we
perform an infinite volume extrapolation linear in $1/(\mbox{volume})$.
As seen in \fig\ref{fig:tcrit}, the slopes of the fits appear to behave
somewhat unsystematically.  This is caused by the statistical
errors of the individual measurements: the differences of the
transition temperatures at different volumes are small enough that 
a constant (in $1/V$) fit at each of the lattice spacings could have
been used.

On the other hand, we note a clear lattice spacing dependence
between $\beta_w = 8$ and the other sets of data.  
Obviously here the lattice spacing $a = 4/(\beta_w g_w^2 T) 
\approx 1.2/T$ is so large that the subleading corrections
start to be significant.  Thus, we use
only the infinite volume extrapolations
at $\beta_w = 12$, 16 and 24 to estimate  
the continuum limit value, with the result
\be
  T_c = 84.3 \pm 0.3 \, \mbox{GeV}. 
\ee
This can be compared with the 2-loop perturbative value 88.5\,GeV
(see \figs\ref{fig:phasediag}, \ref{fig:broken}).
Indeed, the difference between the perturbative and the non-perturbative
results is much larger than the difference between the results
from different lattice spacings. This behaviour agrees 
qualitatively with our previous study~\cite{mssmsim}.

\begin{figure}[t]

\centerline{\epsfxsize=12cm\epsfbox{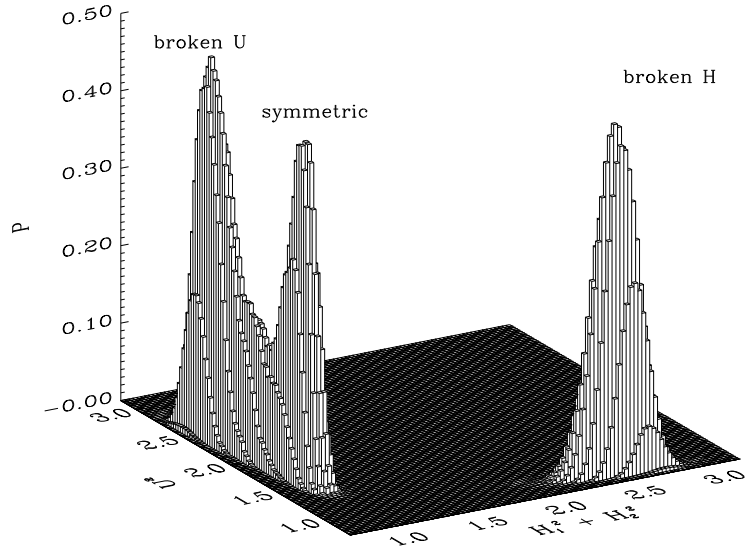}}
\caption[a]{The joint probability distribution
of $\hat H_1^\dagger \hat H_1 + 
\hat H_2^\dagger \hat H_2$ and 
$\hat U^\dagger \hat U$ at the triple
point, measured from a $\beta_w=12$, $12^3$ lattice.  The transition
symmetric$\leftrightarrow$broken $U$ is weak, as can be seen
from the small suppression between the peaks.  This means that 
it is relatively easy to end up in the phase with broken $U$, 
if one goes too close to the triple point. On the other hand,
there is a suppression of $\sim e^{-20}$ between the symmetric
and broken $H$ peaks, signalling a strong transition. }
\la{fig:triplatt}
\end{figure}

\paragraph{Triple point:}

In the perturbative phase diagram in \fig\ref{fig:phasediag}, there is
a ``triple point'' at $\tilde m_U \approx 74.3$ GeV, $T_c\approx 83.4$
GeV, with $v/T\approx 1.24$. We have determined the triple point
location also with lattice simulations at small volumes, using
$\beta_w=12$, volume $12^3$ (see \fig\ref{fig:triplatt}), and
$\beta_w=16$, volume $16^3$.  It is technically difficult and very
time-consuming to perform multicanonical simulations at the
triple point, and we did not attempt to do an infinite volume
extrapolation here.  However, the results obtained from the two
lattices agree reasonably well with each other, and we can make an
estimate for the triple point location, $T_{\rm triple} \approx
77 \pm 1$\,GeV, $\tilde m_{U,\rm triple} \approx 75.3 \pm 0.5$\,GeV, with an
expectation value (see below for its determination) $v/T \approx 1.65
\pm 0.10 $.  We emphasize that these values are just rough estimates; the
lack of an infinite volume extrapolation can be significant here,
since the SU(3) $U$-field is very light at this point.  Nevertheless, a
comparison with the perturbative values is shown in \fig\ref{fig:phasediag}.
The deviation from the perturbative triple point matches the 
behaviour at $\tilde m_U = 65$\,GeV, where we have
a much better control of the systematics.

\paragraph{Order parameter discontinuities:}

Going back to $\tilde m_U = 65$ GeV, 
the discontinuity $\Delta v/T = [2 \sum_i \Delta H_i^\dagger H_i/T^2]^{1/2}$ 
can be quite precisely measured from the
probability distributions (\fig\ref{fig:hg}), by determining the
positions of the symmetric and broken phase peaks separately.  The
results for all the volumes are shown in the left panel of
\fig\ref{fig:delta}.  We see that for each lattice spacing, 
$v/T$ remains practically constant over the whole range of volumes.
Nevertheless, we again make an extrapolation linear in $1/V$ to the
infinite volume limit.

\begin{figure}[t]

\centerline{\epsfxsize=6cm\hspace*{0cm}\epsfbox{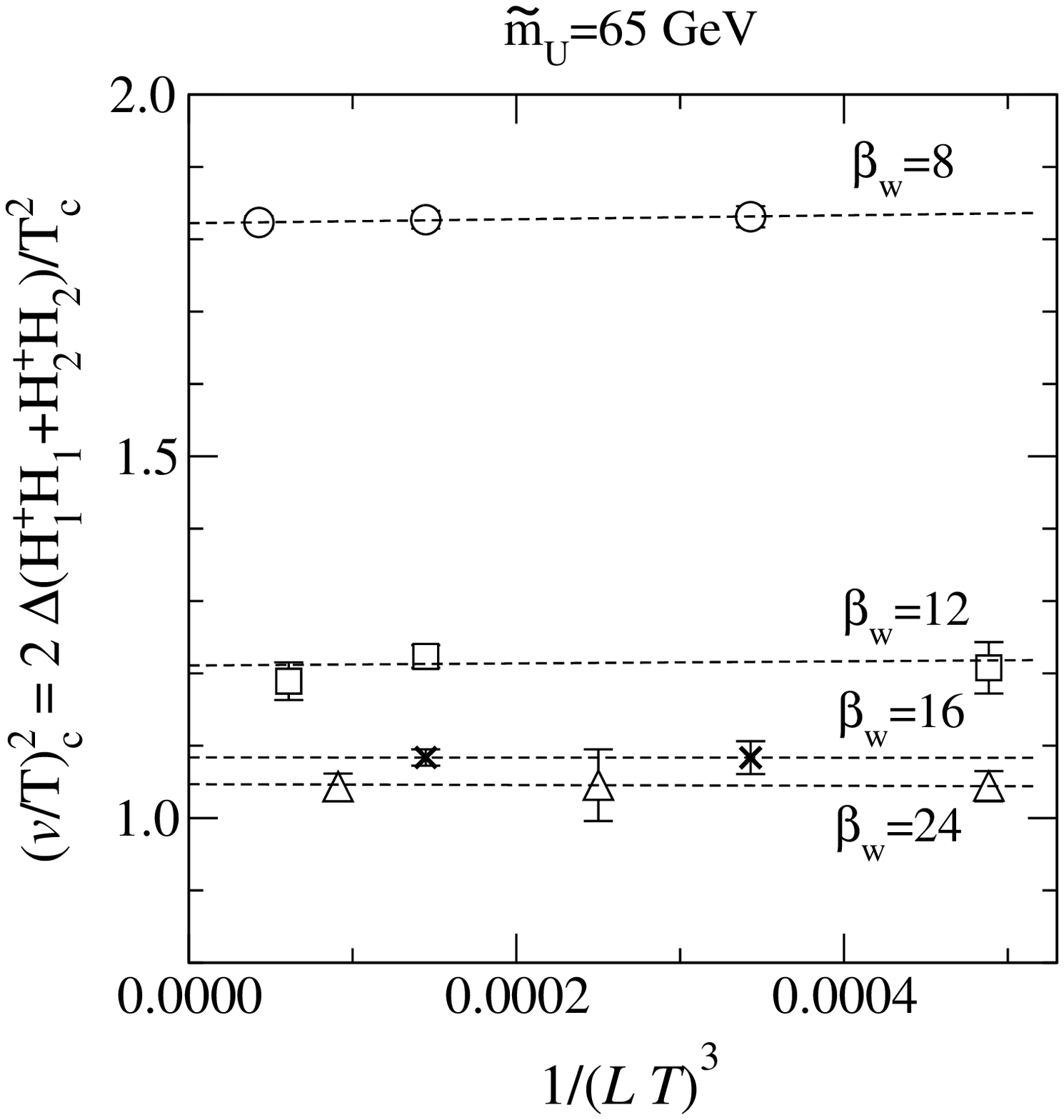}~~%
\epsfxsize=6cm\hspace*{1cm}\epsfbox{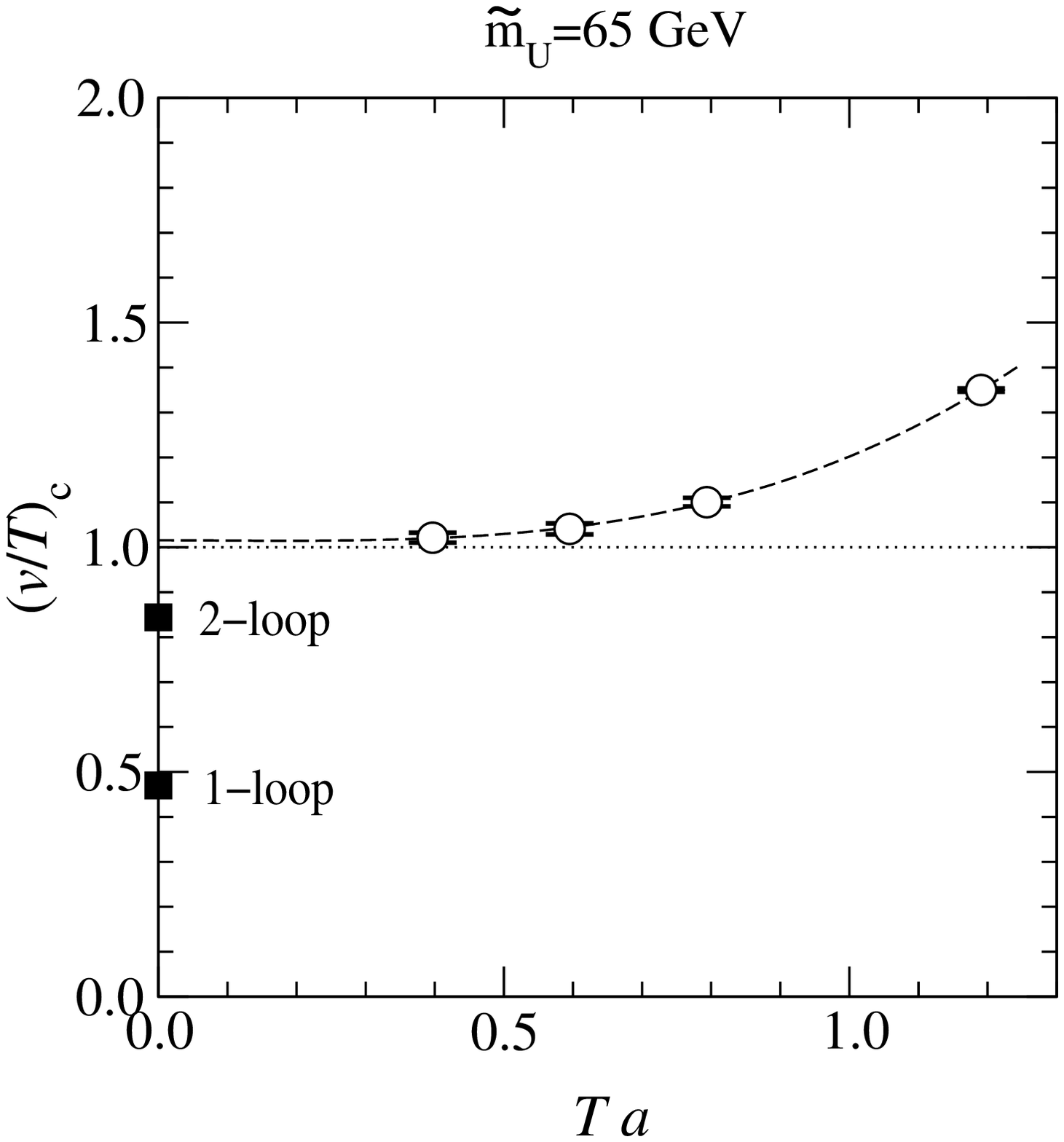}}

\caption[a]{Left: the finite volume dependence of $(v/T)_c^2$, together
with infinite volume extrapolations. 
Right: the continuum extrapolation of $(v/T)_c$.
Note that in units of $T^{-1}$, we are able to operate on relatively
large lattice spacings $a$, since the most ultraviolet fluctuations
of the original 4d theory have been removed by dimensional reduction.}
\la{fig:delta}
\end{figure}

The lattice spacing dependence is relatively pronounced, and the
discontinuity $\Delta v$ becomes smaller as the continuum limit is
approached.  In the right panel of \fig\ref{fig:delta} we show the
infinite volume extrapolations of $v/T$ against the lattice spacing.
The data is extrapolated to the continuum limit using a polynomial fit
of the form
$c_0 + c_2 (aT)^2 + c_3 (aT)^3$.  A priori, there is no
reason why a term linear in $a$ should not be there; however, when it
is included in the fits, $v/T$ invariably becomes larger as $a$ is
decreased at small $a$.  While in principle possible, this kind of a
behaviour does not seem very plausible, especially in view of the
fast apparent convergence of the data as $a$ is decreased in
\fig\ref{fig:delta}.  Thus, we effectively force the term linear in
$aT$ to be $\ge 0$ in the extrapolation.  This behaviour persists if
we drop the largest lattice spacing $aT = 1.2$ from the analysis; in
this case we obtain a good fit with the ansatz $c_0 + c_2 (aT)^2$,
with almost the same value and errors for $c_0$.  Thus, in
the continuum limit, we quote our result as
\be
  (v/T)_c = 1.02 \pm 0.05 \,,
\ee
where the error includes both the statistical errors and the ambiguity
due to different extrapolations.
This can be compared with the 1-loop 
perturbative result $(v/T)_c \approx 0.47$, and the 
2-loop perturbative result $(v/T)_c \approx 0.85$.
See \figs\ref{fig:broken}, \ref{fig:delta} for comparisons.

In addition to the discontinuities at the transition point, 
we have also determined the Higgs condensates at temperatures
close to $T_c$. In the first panel of
\fig\ref{fig:masses} we show how the bilinears $H_2^\dagger H_2$, $R=\re
H_1^\dagger \tilde H_2$, $I=\im H_1^\dagger \tilde H_2$ behave as we go
through the transition.  Since in our simulations $\tan\beta = 12$ is
large, $H_1^\dagger H_1$ is practically constant on the scale of the
plot. We note that $I$
remains almost constant, too, 
and has at most only a very small jump at the
transition; it is non-zero overall because of the small imaginary parts
in the effective parameters, 
particularly $m_{12}^2(T)$, in \eqs\nr{54}--\nr{3dprms}.

\paragraph{Latent heat:}

The latent heat of the transition is closely related to the
order parameter
discontinuities of the 3d theory.  In general, the
latent heat is determined by
\be
   L = -T_c \bigg(\fr{d(\Delta f)}{dT}\bigg)_{T=T_c}\,,
\ee
where $\Delta f = f_\rmi{symmetric} - f_\rmi{broken}$ 
is the free energy density difference
between the symmetric and broken phases, and the derivative is
to be taken along the metastable branches.
More concretely, $L$ measures the discontinuity in the energy density.

In our 3d theory
with the parametrization in \eqs\nr{54}--\nr{3dprms},
explicit temperature dependence appears 
only in the Higgs mass parameters.  
Thus, following~\cite{contlatt,nonpert}, 
the latent heat becomes
\be
  L  =   T_c^3 \Delta \left\langle
 U^\dagger U \fr{d}{d T} \fr{m_U^2(T)}{T^2}  +
 \sum_{i=1}^2 H_i^\dagger H_i  
 \fr{d}{d T}  \fr{m_i^2(T)}{T^2} +  
  \bigg(H_1^\dagger \tilde H_2
  \fr{d}{d T} \fr{m_{12}^2(T)}{T^2}
  + \mbox{H.c.} \bigg)
\right\rangle, 
\la{latent}
\ee
where $\Delta \langle \cdot \rangle = 
\langle \cdot \rangle_{\rm broken} - 
\langle \cdot \rangle_{\rm symmetric}$.

We observe that the latent heat behaves numerically very much like the
discontinuity in $v/T$, but with slightly larger statistical errors,
since more condensates are involved.  
We do not show a separate figure. 
It should be noted that the
$U^\dagger U$-term in \eq\nr{latent} is significant, despite 
the fact that the
$U$-field remains ``unbroken'' at all temperatures at this $\tilde
m_U$.  This is because $\langle U^\dagger U\rangle$ is
somewhat smaller in the phase where the SU(2) Higgs fields $H_i$ are
``broken''.

We use similar infinite volume and continuum extrapolations for the
latent heat as for the condensate $v/T$ in \fig\ref{fig:delta}.
The final result (at our specific parameter point) becomes
\be
  L/T_c^4 = 0.42 \pm 0.03\,,
\ee
which is substantially larger than the perturbative 2-loop 
value $L/T_c^4\approx 0.26$. The difference is quite large, but it is 
again qualitatively similar to the one observed in~\cite{mssmsim}.

\begin{figure}[t]

\centerline{\epsfxsize=6cm\hspace*{0cm}\epsfbox{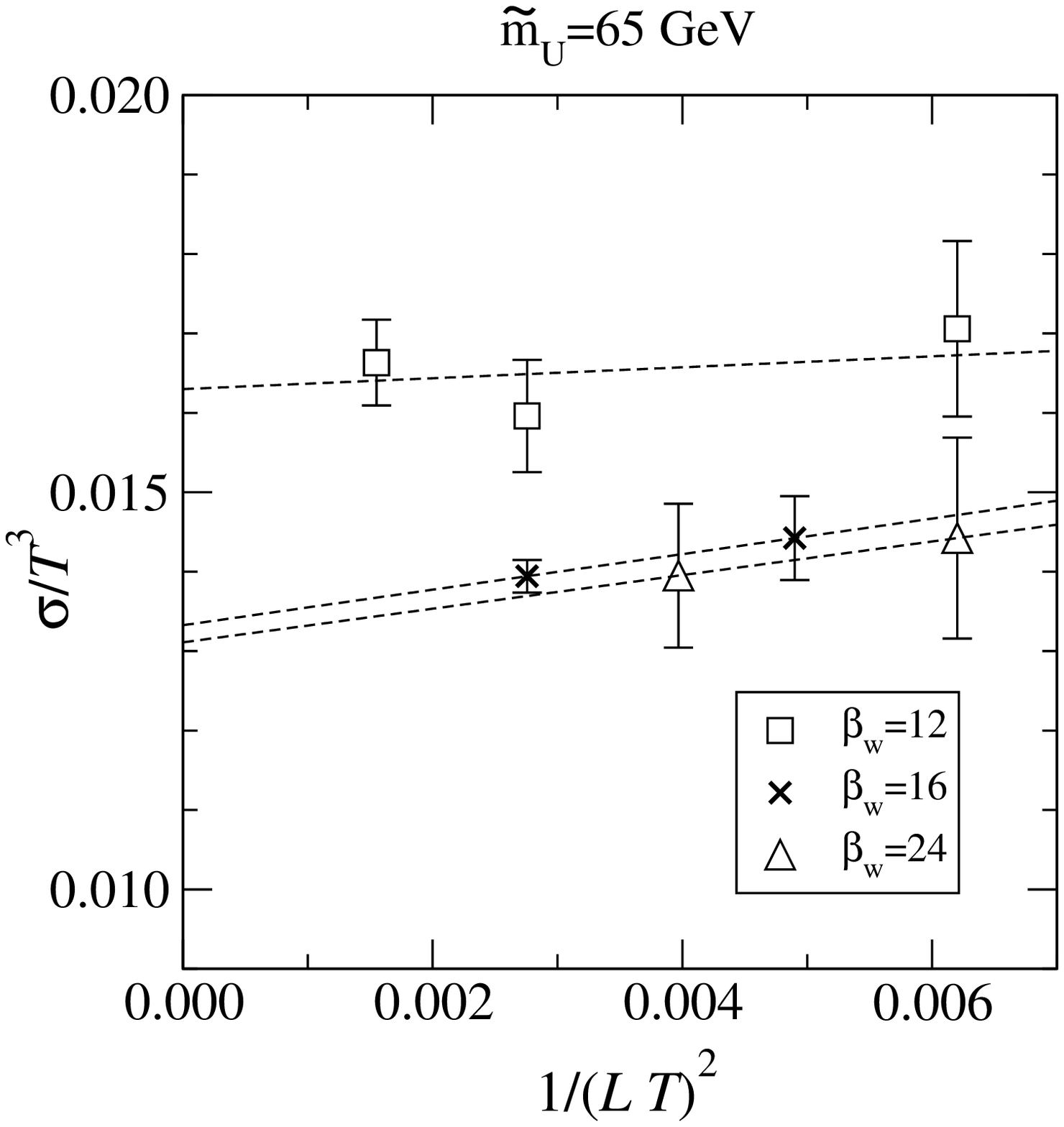}~~
\epsfxsize=6cm\hspace*{1cm}\epsfbox{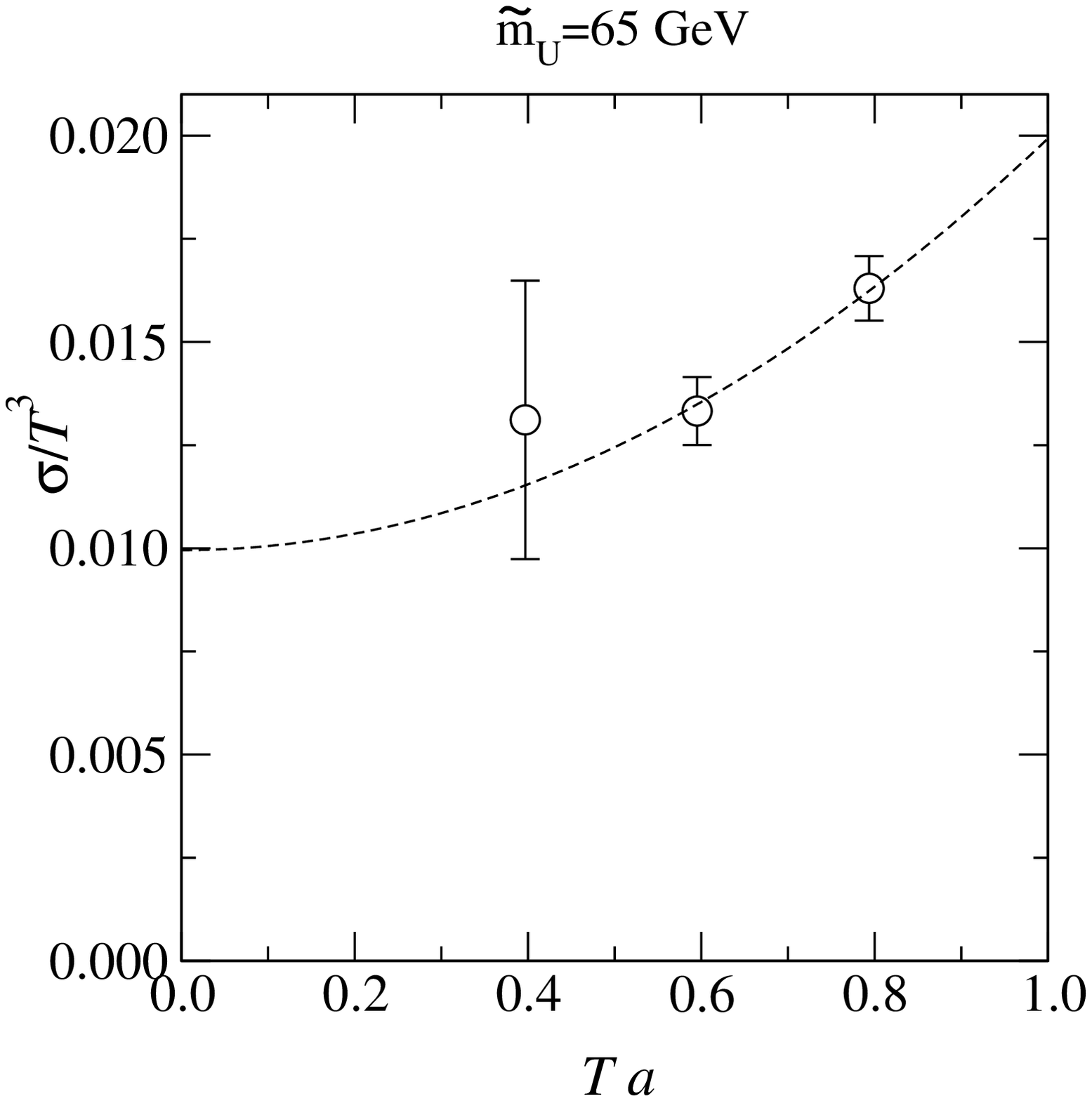}}

\caption[a]{Left: the finite volume dependence of the surface 
tension. Right: a continuum estimate. We should stress 
that our determination of $\sigma/T_c^3$ is not meant to be nearly
as precise as the determination of $(v/T)_c$ (see the text).} 
\la{fig:sigma}
\end{figure}

\paragraph{Surface tension:} 

We measure the tension of the interface between the symmetric and
broken phases (also called the phase boundary)
with the histogram method.  The surface tension
$\sigma$ equals the additional free energy/area 
carried by the interface.  This extra
free energy suppresses mixed phase configurations by a factor
$\propto \exp(-\sigma A/T)$, where $A$ is the area of the interfaces.
This causes a characteristic
``valley'' between the symmetric and broken phase peaks in
probability distributions of order parameter like quantities; see,
for example, \fig\ref{fig:hg}.  

In the histogram method one 
uses the mixed phase suppression to measure the surface tension.
Assuming that the interfaces are perpendicular to the $x_3$-axis
direction, $\sigma$ can be obtained from
\be
  \fr{1}{2L_1 L_2} \ln \fr {P_{\rm max} }{ P_{\rm min} } 
  \rightarrow \fr { \sigma }{ T } \mbox{~~~~~~~ as ~} V\rightarrow\infty\,.
\ee
Here $P_{\rm max,min}$ are the maximum and minimum of the peaks
of the probability distribution, and $L_i = N_i a$ are lattice 
extensions in physical units. The factor $2$ appears in front of 
$L_1 L_2$, because there are two interfaces in a box with 
periodic boundary conditions.

In the actual analysis we measure $\sigma$ from histograms of $H_1^\dagger
H_1 + H_2^\dagger H_2$.  These are reweighted to a temperature where
the peak heights are equal, which simplifies the analysis.  We also
use a finite volume scaling ansatz similar to our earlier work
\cite{mssmsim},
\be
\fr{\sigma}{T_c^3} = \fr{1}{2(LT)^2} \left[ 
  \ln \fr{P_{\rm max} }{ P_{\rm min}} + \fr{1}{2} \ln LT + \mbox{const.}
  \right],
\ee
where we have employed the fact that our lattices are all cubic, 
$L_1 = L_2 = L_3 = L$.  
 
In \fig\ref{fig:sigma} we show the rough infinite volume extrapolations of
$\sigma/T_c^3$ from $\beta_w = 12$, 16 and 24 lattices. (For this
observable, $\beta_w = 8$ gives a surface tension larger by 
about a factor of 5, and we do not include it in the analysis.)  
The continuum limit estimate of $\sigma$ becomes now
\be
  \sigma/T_c^3 \approx 0.010 \pm 0.05.
\ee
The 2-loop perturbative value is 
$\sigma/T_c^3\approx 0.017$, which is {\em larger} 
than the lattice value. However, we stress first of all
that our lattice determination is quite rough here, as 
we have only used cubic volumes, not the cylindrical ones often
employed for getting good infinite volume extrapolations
(see, e.g., \cite{mssmsim}).  
Second, we should also remember that 
in the perturbative estimate no account is taken 
of the effects related to derivative terms
discussed in~\cite{bbfh,baacke,kls,lgy,mr}, 
which would decrease that value.
Nevertheless, the situation is certainly different from the case
studied in~\cite{mssmsim} where even the raw perturbative
value was smaller than the lattice value. On the other hand, 
the situation is qualitatively similar to the case of the 
Standard Model~\cite{nonpert}, where the transition is of 
a relatively weak strength compared with~\cite{mssmsim}, 
as it is here too. 

\paragraph{Correlation lengths:}
\la{sec:masses}

\begin{figure}[t]

\centerline{\epsfxsize=5.8cm\epsfbox{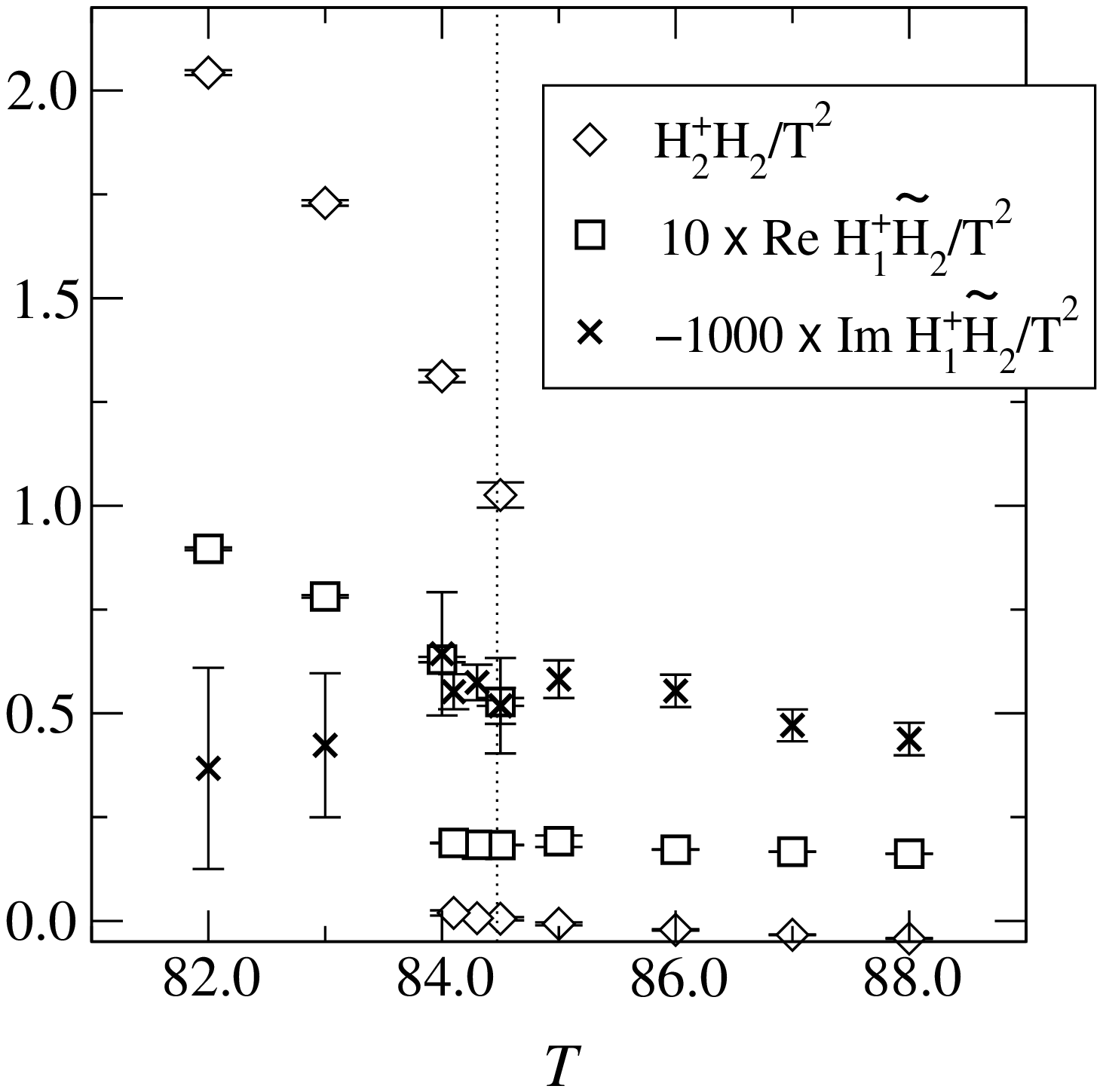}%
~~~ \epsfxsize=6cm\epsfbox{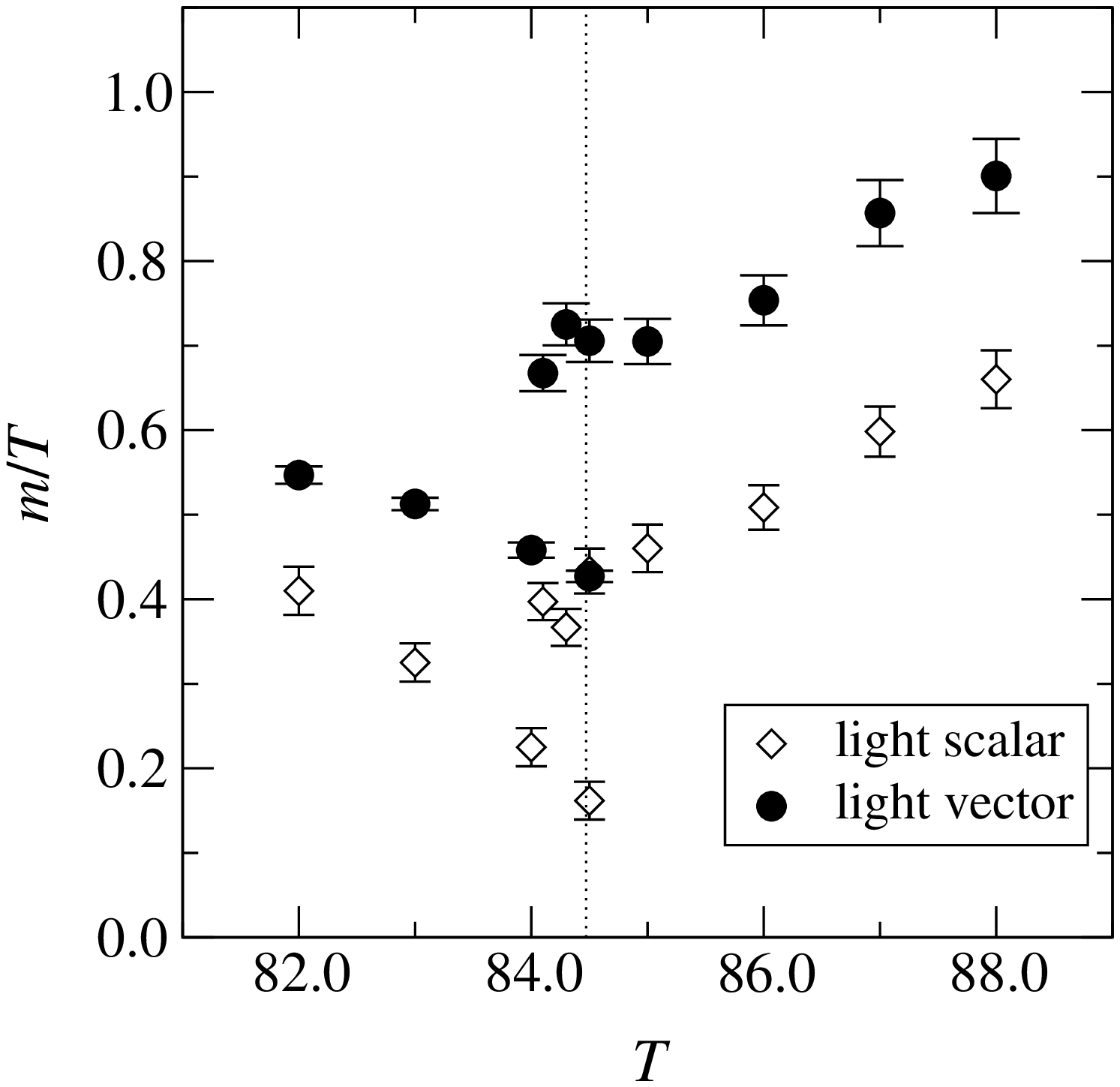}}

\centerline{\epsfxsize=6cm\epsfbox{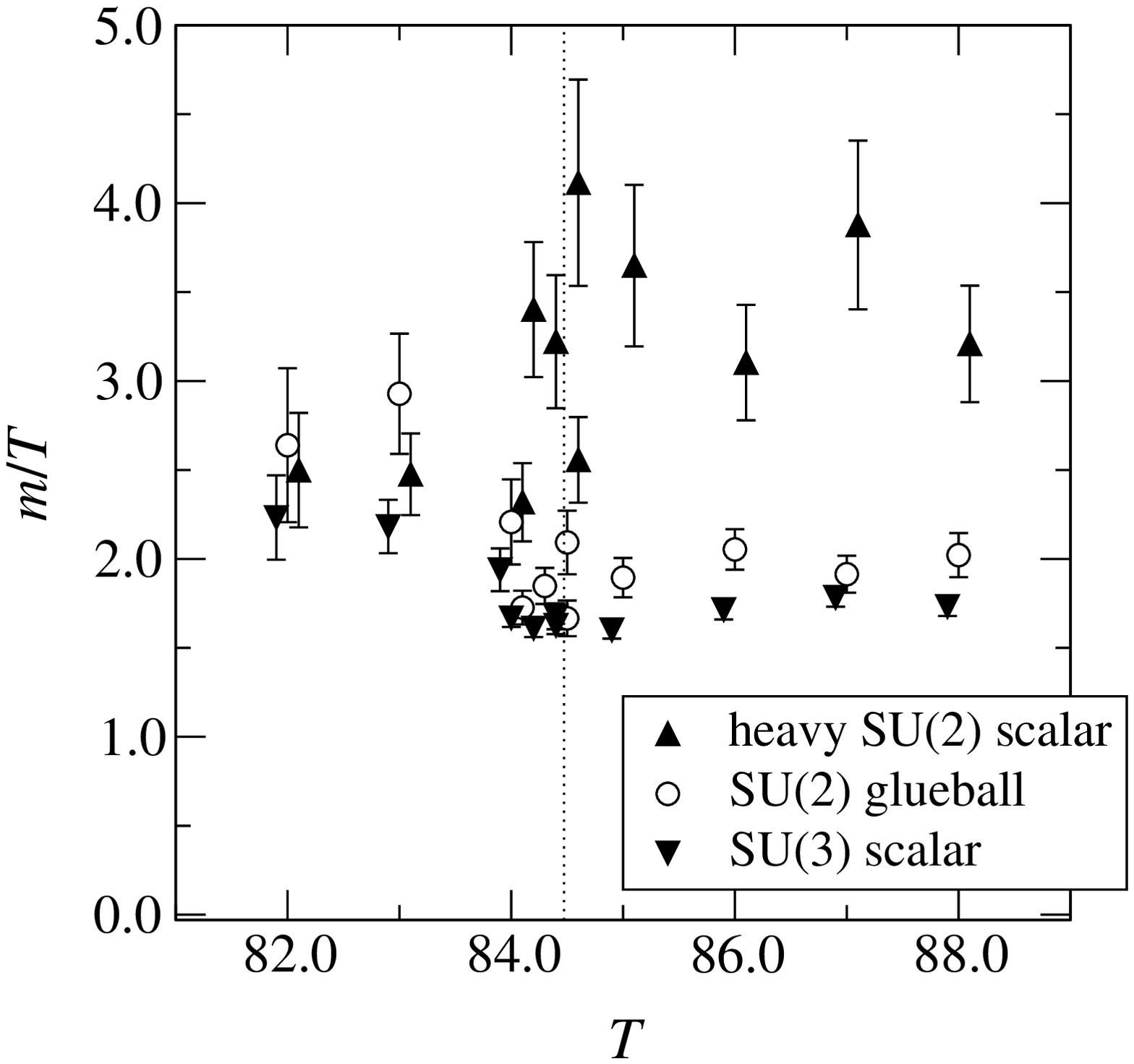}%
~~~ \epsfxsize=6cm\epsfbox{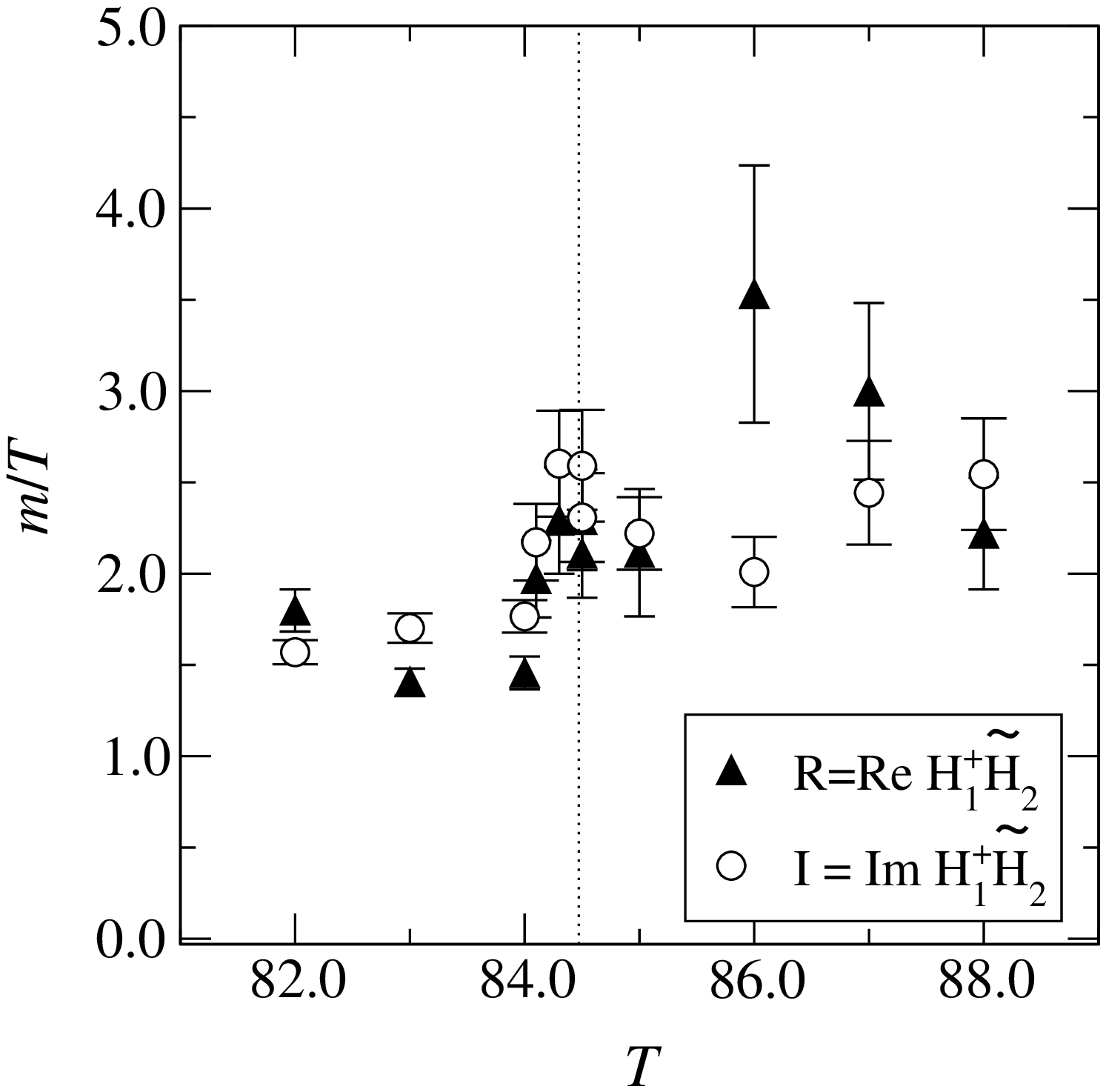}}

\caption[a]{Top left: 
Higgs bilinears, subtracted according to \eq\nr{Higgsub},  
across the transition at
$\beta_w = 16$, volume $=32^3$.  The vertical dotted
line shows the location of $T_c$.
Top right: light SU(2) scalar and vector inverse 
correlation lengths, $m/T$.
Bottom left: heavy SU(2) scalar, SU(2) glueball and SU(3) scalar
inverse correlation lengths.
Bottom right: inverse correlation lengths 
related to the operators $R, I$. All the numbers in the bottom 
panels have relatively large systematic errors (see the text), 
and could almost be compatible with each other.}
\la{fig:masses}
\end{figure}

Next, let us explore spatial correlation lengths
around the transition temperature.  As usual, their
inverses are called ``screening masses''.  The masses
are measured from the 0-momentum correlation functions
\be
  C_a(x_3) = \sum_{\bfx,\bfy} \langle O_a(\bfy,0) O_a(\bfx,x_3) 
              \rangle \propto \exp[ -m_a x_3 ],
\ee
where $\bfx,\bfy$ are in the transverse $(x_1,x_2)$-plane, and 
the gauge invariant operator $O_a(x)$ is one of the following:
\[\begin{array}{ll}
  \mbox{C even SU(2) scalars:}  & 
		S^w_a(x) = H_a^\dagger(x) H_a(x),
	     ~~ R(x)     = \re H_1^\dagger(x)\tilde H_2(x), \\
  \mbox{C odd SU(2) scalar:} & 
                I(x)     =  \im H_1^\dagger(x)\tilde H_2(x), \\
  \mbox{SU(2) vectors:}  & 
                V^w_{a,i}(x) = \im H_a^\dagger(x) U^w_i(x) H_a(x+\hat i), \\
  \mbox{SU(2) $O^{++}$ glueballs:} & 
                G^w(x) = 1-\fr12 \tr P_{12}^w(x), \\ 
  \mbox{SU(3) scalar:}   & 
		S^s(x)   = U^\dagger(x) U(x), \\
  \mbox{SU(3) vector:}   & 
		V^s_i(x) = \im U^\dagger(x) U^s_i(x) U(x+\hat i), \\
  \mbox{SU(3) $O^{++}$ glueballs:} & 
		G^s(x)   = 1-\fr13 \re \tr P_{12}^s(x). \\ 
\end{array}
\]
Here $i=1,2$; $a=1,2$ labels the two SU(2) Higgs fields;  $U_i^w$
and $U_i^s$ are the standard lattice SU(2) and SU(3) gauge links
(denoted by $U^w_{{\bfx},i}, U^s_{{\bfx},i}$
in \eq\nr{lattaction}); 
and $P^w_{12}, P^s_{12}$ 
are $(x_1,x_2)$-plane plaquettes constructed from these links.
In order to reduce statistical noise, we use recursive blocking and
smearing of the gauge and Higgs fields along $(x_1,x_2)$-planes, and
construct operators and measure all of the correlation functions from
the blocked fields at each blocking level.  The blocking is repeated
up to 4 times.  For details of the recursive blocking procedure
we employ here, we refer to \cite{mssmsim} (see also~\cite{owe}).

The measurement of the correlation lengths in an interacting theory
is complicated by the fact that all the operators in a given quantum
number channel in general couple to the same set of physical states.
For example, we can expect that all of the scalar operators
above, including the $O^{++}$ glueballs, will 
in the limit $x_3\rightarrow\infty$  yield
the screening mass of the lightest scalar state.  On the other hand, 
in the real world the correlations may behave at intermediate 
distances in a different way, and to fully resolve this behaviour
one usually measures the
cross-correlation matrix of a large set of operators in a given
quantum number channel, and diagonalises it.  

However, in our case when only rough qualitative accuracy is needed, 
it turns out that this is not necessary: since we use a
large $\tan\beta=12$, the active Higgs component in the transition
projects almost completely to $H_2$, and correspondingly the light
scalar and vector states couple strongly only to
the operators $S_2^w$ and
$V_{2,i}^w$, respectively.  Moreover, the ``coupling'' (or overlap) 
of the SU(2) pure gauge and
the whole SU(3) sector to the SU(2) scalar Higgs sector is weak.  All in all,
this implies that we can obtain the heavier scalar and vector
screening masses
just by measuring the exponential fall-offs of the corresponding
correlation functions at intermediate distances; the accuracy of
this approach is quite sufficient for our conclusions.  

On the second panel of \fig\ref{fig:masses} we show the light SU(2)
scalar and vector masses, and on the third and fourth panel the
heavier masses.  The lightest mass scale at the transition is 
an order of magnitude smaller than the heavy scales.  This makes the condition
in \eq\nr{ineq} very difficult to meet in practice, and in our
case the first part $a \ll \xi_\rmi{min}$ is barely true.
This circumstance also makes the measurement of the heavy masses difficult,
since the correlation functions vanish into noise after a few lattice
units, which explains the large errors in the data.  However,
for our purposes the accuracy obtained is sufficient.

We may now note first of all 
that while the large mass scales make the extrapolation
to the continuum limit delicate
(necessitating several lattice spacings, as we have), 
experience from QCD~\cite{mu} 
indicates that the mass scales $\lsim 3 T$
are still far from being too large for us to have any concerns 
about the applicability of dimensional reduction, used 
in the construction of the 3d effective field theory. The 
integration out of the Matsubara zero modes of the temporal
components of the gauge fields is clearly more critical, but 
as we have discussed in~\cite{cpown}, we expect even that to 
be reasonably under control. In any case, those masses are
still much above the lightest ones in the system ($\ll T$), 
which determine the non-perturbative thermodynamical 
properties of the transition. 

{}Furthermore, the screening mass spectra 
measured add further confidence to two statements we have 
made on other grounds before: (1) The correlation length related
to $I$ is short, and thus not at all ``critical''. This means that 
we are far from the possibility of spontaneous C violation. 
(2) The heavy SU(2) scalar correlation length is much shorter
than the light one, and thus again far from ``critical''.
This means that it is only the ``light'' combination of the 
two Higgs doublets which is really ``dynamical'' at the 
transition point, and any effects related to the other 
one are suppressed.


\section{The properties of the physical phase boundary}
\la{phasewall}

We now turn to the study of the 
properties of the phase boundary.  We have outlined the 
measurements to be carried out, 
as well as some caveats in them, in~\cite{cpown}.

In order to study a phase boundary, we have to make sure 
that there really is one on
the lattice.\footnote{In principle, standard (multicanonical)
simulations at the transition temperature could be used, since 
phase boundaries appear there in the ``tunnelling configurations''
containing regions of both phases.  However,
a lot of effort is wasted, since 
phase boundaries exist only in a small
subset of the total of all configurations.}  This can be achieved by
restricting, say, the volume average of
$h^2 \equiv H_1^\dagger H_1 + H_2^\dagger H_2$ to a
narrow band around $(h^2_{\rm symmetric} + h^2_{\rm broken})/2$.  Provided
that the volume of the system is large enough, this guarantees that
the system will always remain in a broken + symmetric mixed state,
with corresponding phase boundaries, or interfaces.

In the case at hand the interfaces are rather thick, and because
of the periodic boundary conditions, there will be two interfaces
spanning the lattice. This makes it advantageous to use cylindrical
lattices: because of the surface tension, the interfaces will be
preferentially oriented along the smallest cross-sectional area across
the lattice, making them well separated along the longest lattice
direction ($x_3$, say).  This has the further advantage that we always
know the orientation of the interfaces.  Our 
interface simulations were made
on $\beta_w =8$, $12^2\times 96$, and $\beta_w=16$, $24^2 \times 192$
lattices, using up to 450 000 compound update sweeps per lattice.

\subsection{Observables}

We study the interface properties by looking at the Higgs bilinear
operators in \eq\nr{ops} as follows: first, for each configuration, we
average the bilinears across the $(x_1,x_2)$-plane, so that we obtain them
as functions of the $x_3$-coordinate.
Then we average over all configurations in two ways: (1) the
bilinears are measured as functions of the distance from
a certain reference point, 
and (2) one bilinear is measured as a function of 
another.  Let us look at these cases separately.

\paragraph{(1) Interface profiles:}

The Monte Carlo simulation method described does not specify the location
of the interfaces along the $x_3$-direction.  However,
in order to measure the profiles of various observables across the
interface, we have to find a reference point in  
the $x_3$-direction, configuration
by configuration (i.e., we have to remove the zero mode).  
Some care has to be taken here:
for example, one may locate the
$x_3$-value where $h^2$, averaged over $x_1$ and $x_2$, reaches the
value half-way between the symmetric and broken phase ones.  However,
since $h^2$ (as any other local quantity) has large fluctuations, this
kind of a sharply defined location will cause the configuration to be
shifted such that the natural fluctuations across the interface may be
summed ``in phase'', which tends to distort
the profile.  The problem can be avoided by using 
a ``softer'' filter function.  In this
work employ the fact that there are two interfaces on a periodic 
lattice: thus we can find the ``symmetry point'' $x_{3,0}$ of the 
profile by taking a Fourier transform 
($N_3$ is the extent of the lattice in the direction of $x_3$),
\be
C = \sum_{x_3} h^2(x_3) e^{i 2\pi x_3/N_3} \equiv
A \, e^{i 2\pi x_{3,0}/N_3}. 
\ee
Thus $x_{3,0} = N_3/(2\pi) \times (\mbox{polar angle of $C$})$.
Configurations are then superimposed around this point. 

In \fig\ref{fig:profile} we plot the profiles of the
bilinears $H_1^\dagger H_1$, $H_2^\dagger H_2$, $R = \re H_1^\dagger
\tilde H_2$ and $I = \im H_1^\dagger \tilde H_2$, measured from the
$\beta_w=16$ lattice. (We observe no qualitative change compared
with the $\beta_w = 8$ lattice.)  Only half of the profiles (one
interface) are shown.  Besides the obvious one in magnitudes,
it is difficult to see qualitative differences between $H_1^\dagger
H_1$, $H_2^\dagger H_2$ and $R$ across the interface. 

\begin{figure}[t]

\centerline{\epsfxsize=9cm\hspace*{0cm}\epsfbox{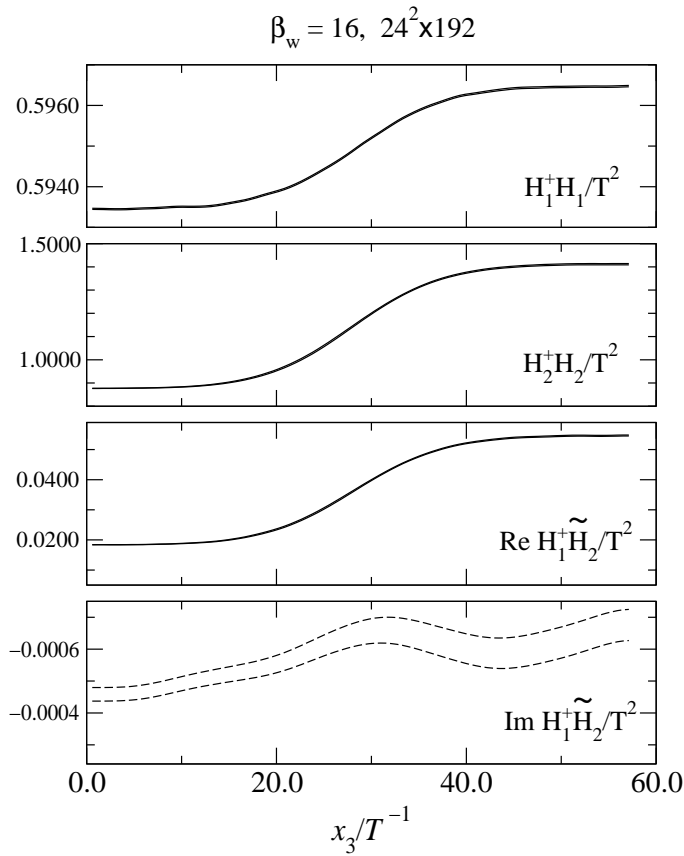}}

\caption[a]{The profiles of the  operators in \eq\nr{ops}
as a function of the spatial coordinate across the wall. 
The cross-sectional area here is $\sim (14/T)^2$. 
The subtraction in \eq\nr{Higgsub} 
has not been carried out.} 
\la{fig:profile}
\end{figure}

As for the C odd condensate $I$, its magnitude is even smaller 
and the errorbars correspondingly larger. 
For clarity, we have smoothed $I(x_3)$ in
\fig\ref{fig:profile} by an approximate Gaussian smearing:
\be
  I(x_3) \leftarrow \fr{1}{3}[I(x_3-a) + I(x_3) + I(x_3+a)]\,,
\ee
and repeated 4 times.  Without this additional smearing it would
be difficult to see any structure in the plot.
After the smearing, we observe that
$I$ increases slightly when the broken phase is entered. 
The overall negative value of $I$ is due to the                
small imaginary part of the $m_{12}^2(T)$ parameter, \eq\nr{m12}
(see Appendix~\ref{suse:redef}). 

{}From \fig\ref{fig:profile} we see that the interface is rather thick:
if we fit $H_i^\dagger H_i$ to a function of the form $a+b\tanh((x_3-c)/L)$, 
we obtain $L \sim 9/T$. 
However, one should bear in mind that in 3 dimensions 
the apparent interface thickness suffers from a logarithmic
divergence as the area is increased (see, e.g., \cite{Binder}).  
A natural way to resolve this arbitrariness is to
consider the interfaces on physically relevant length scales.  For our
case, the longest correlation length $\xi_\rmi{max}$ at the transition 
is $\sim 6/T$ (see \fig\ref{fig:masses}), 
so that the interfaces in \fig\ref{fig:profile} correspond to a 
cross-sectional area $\sim (2 \xi_\rmi{max})^2$ .

\paragraph{(2) $H_1^\dagger H_1$ vs.\ $H_2^\dagger H_2$ and $I$ 
vs.\  $R$ across the interface:}

Instead of plotting the bilinears as functions of $x_3$, it can be
more illustrative to consider the behaviour of one condensate as a
function of another.  This way there is no logarithmic divergence
visible, and no need to locate the interface on the lattice.  

\begin{figure}[p]

\centerline{\epsfxsize=6cm\hspace*{0cm}\epsfbox{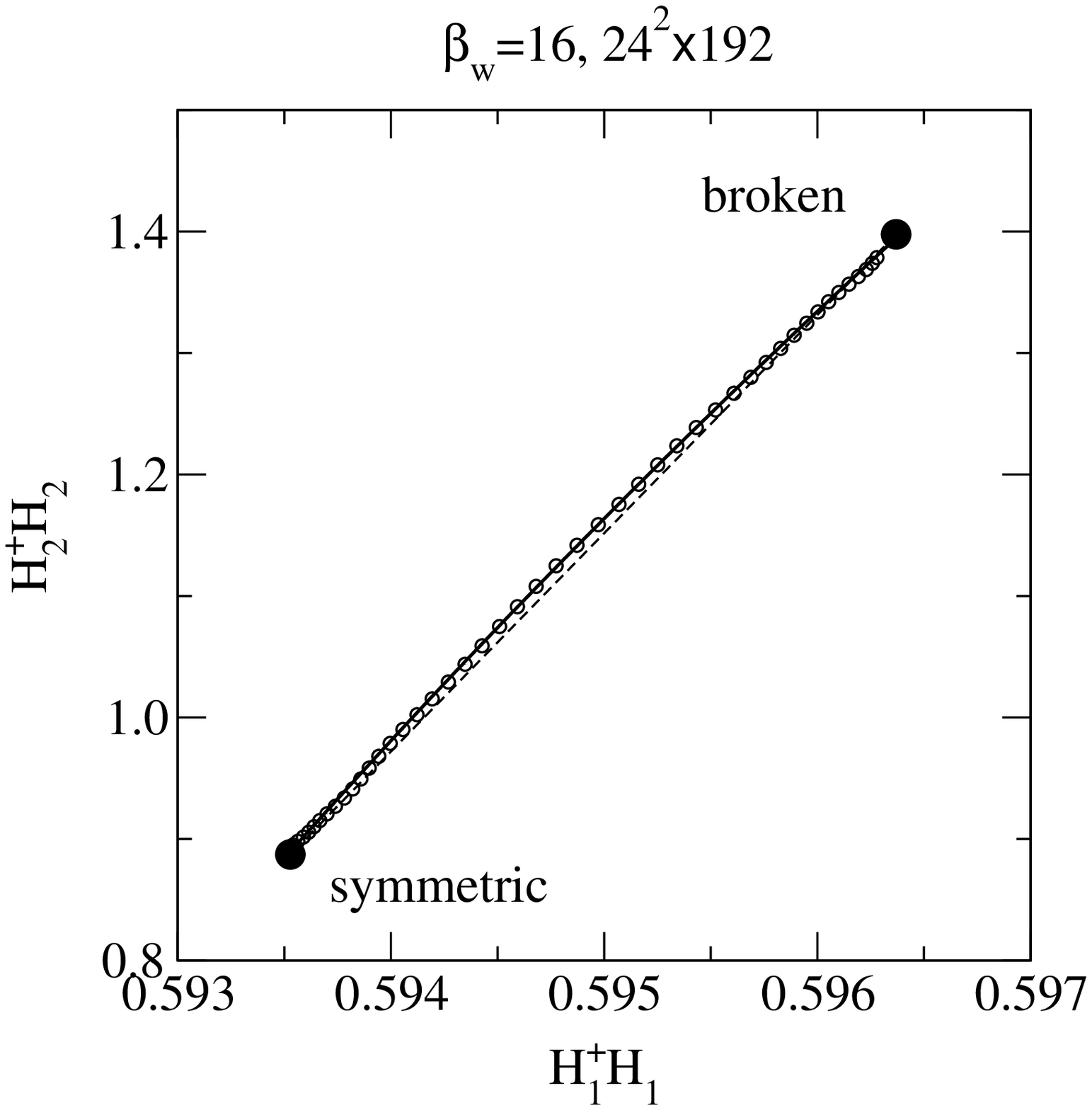}%
\epsfxsize=6cm\hspace*{1cm}\epsfbox{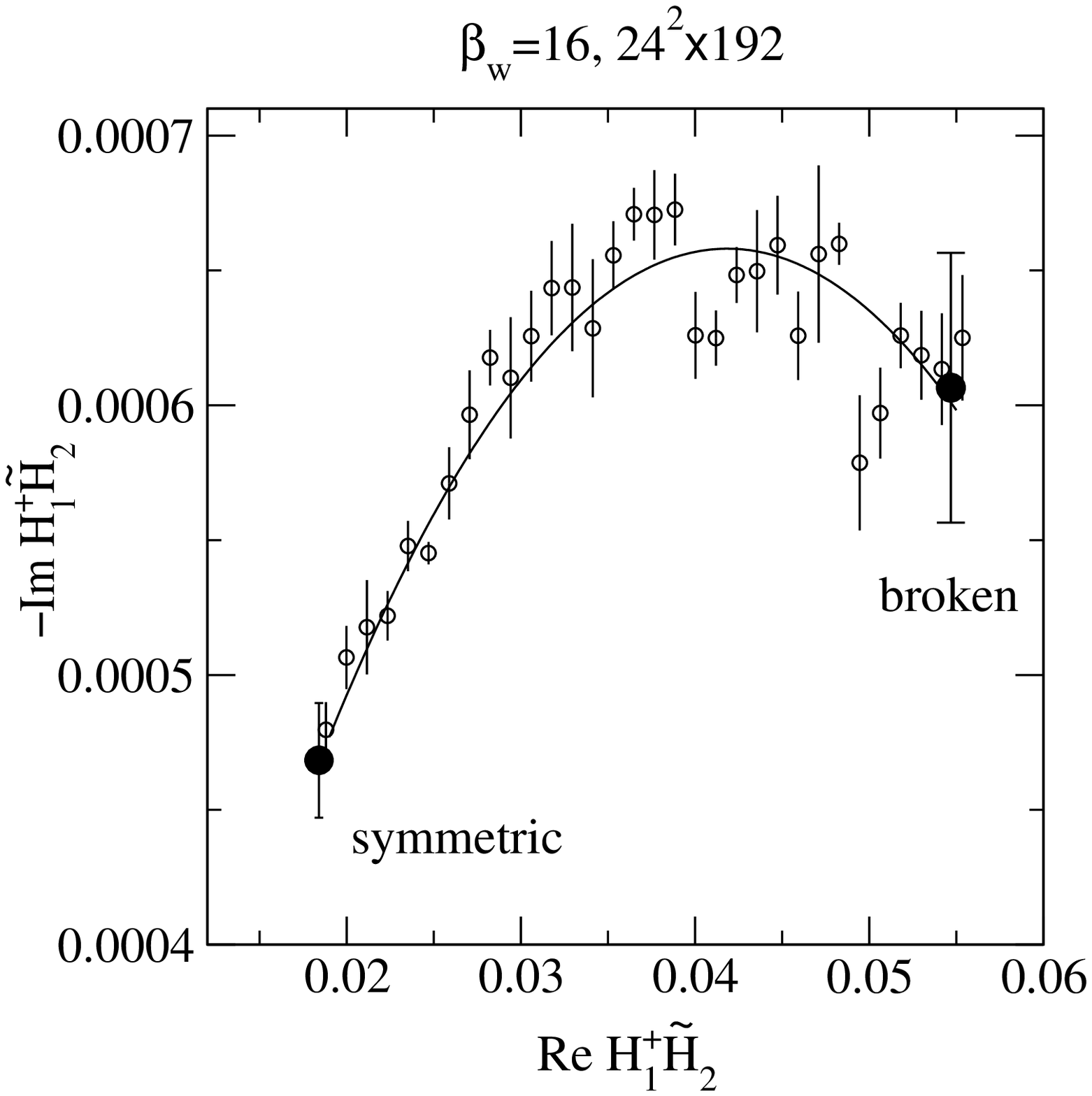}}

\caption[a]{Behaviour of (unsubtracted) 
$H_2^\dagger H_2$, plotted against $H_1^\dagger H_1$,
and $-I= -\im H_1^\dagger \tilde H_2$, plotted  
against $R = \re H_1^\dagger \tilde H_2$, 
across the interface. Values are in units of $T^2$. 
The dashed curve on the left 
is a straight line, and solid curves are fits.} 
\la{fig:ReImb16}
\end{figure}

\begin{figure}[p]

\centerline{\epsfxsize=6cm\hspace*{0cm}\epsfbox{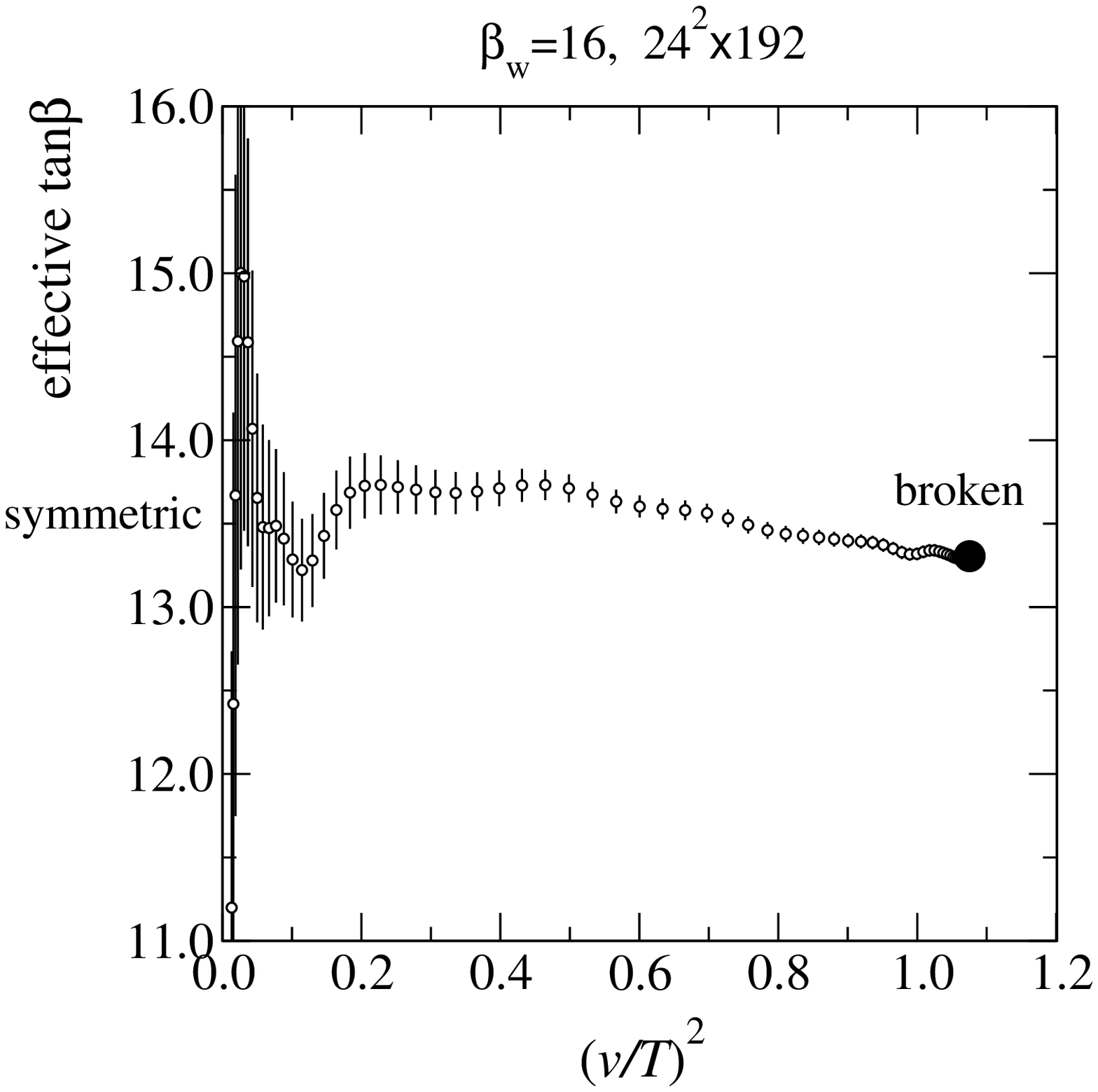}
\epsfxsize=6cm\hspace*{1cm}\epsfbox{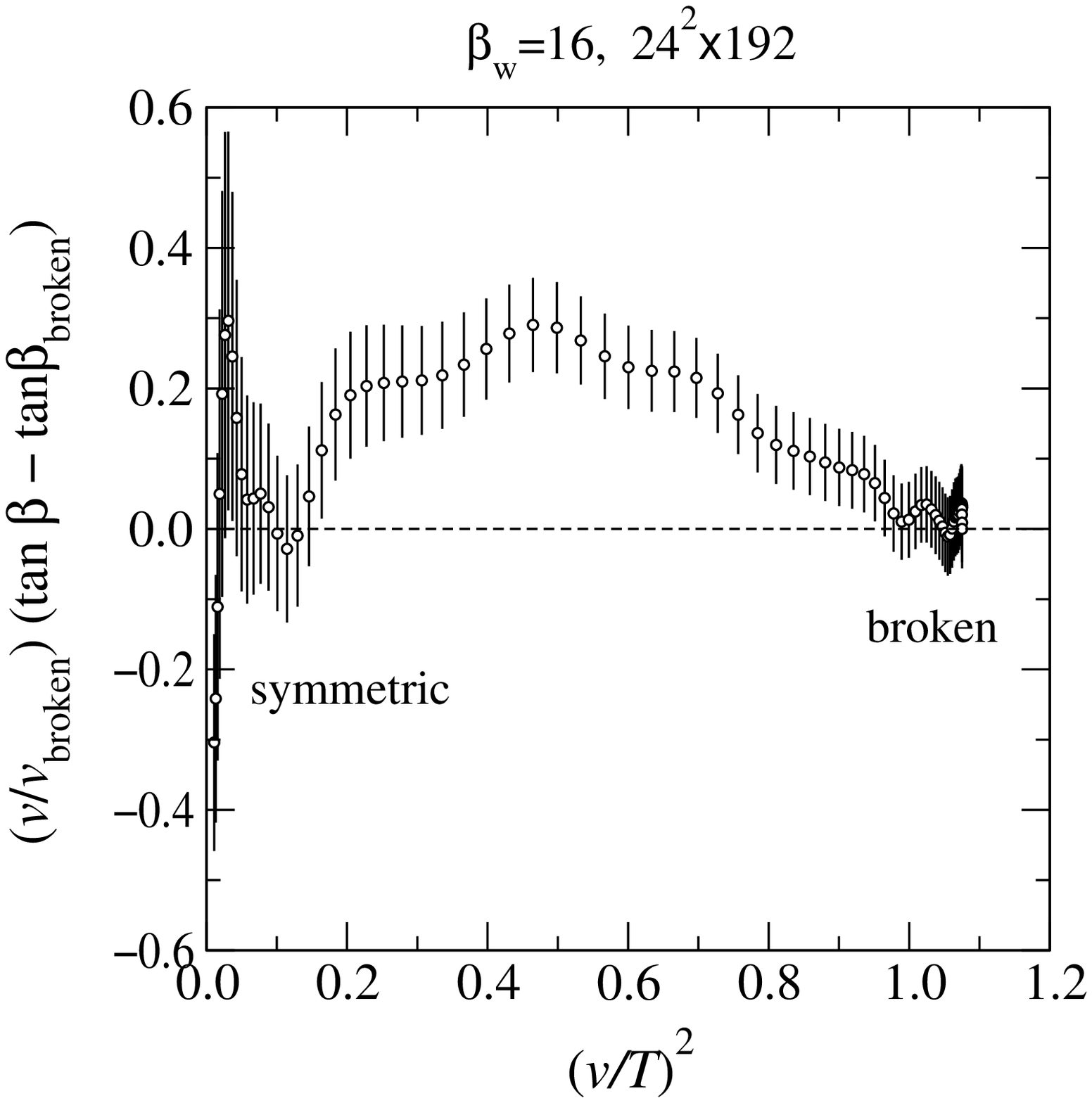}}

\caption[a]{Left: behaviour of the effective $\tb$, 
\eq\nr{lattdef}, across the interface. We note that the 
``dynamical'' value differs only a little from 
the zero temperature input value $\tb\approx 12.0$.
Nevertheless, the effective $\tb$ is not constant 
across the interface, but varies by a relative amount $\lsim 5$\%.
Right: the same data, plotted as 
$(v/v_\rmi{broken})(\tb-\tb_\rmi{broken})$.} 
\la{fig:tanb}
\end{figure}

In \fig\ref{fig:ReImb16} we show $H_2^\dagger H_2$ as a function of
$H_1^\dagger H_1$, and $I$ as a function of $R$.  In the former case,
there is a small but statistically clear deviation from the straight
line between the symmetric and broken phases.  We show the same
data in \fig\ref{fig:tanb} using the definition in \eq\nr{lattdef}.

We observe that 
in terms of $v_i = \sqrt{2\Delta H_i^\dagger H_i}$, the deviation from
a straight line is ${\cal O}(\mbox{a few }\times 10^{-2})$.  
We also observe that
C violation has some structure: the operator 
$I$ tends to saturate close to the broken phase value considerably 
faster than R (or $v/T$).  However, we do not observe any significant 
amplification of $I$ inside the interface, and thus no sign 
of transitional C violation.

The orders of magnitude observed
for the variation in $I$ and $\tb$
agree roughly with perturbation theory~\cite{mqs,cm,pj} and, in the case 
of $\tb$, with previous 4d simulations~\cite{4dmssm}, 
although in most cases the parameter values were not identical to ours.

\section{Summary and Conclusions}
\la{conclusions}

\begin{figure}[p]

\centerline{\epsfxsize=6.6cm\hspace*{0cm}\epsfbox{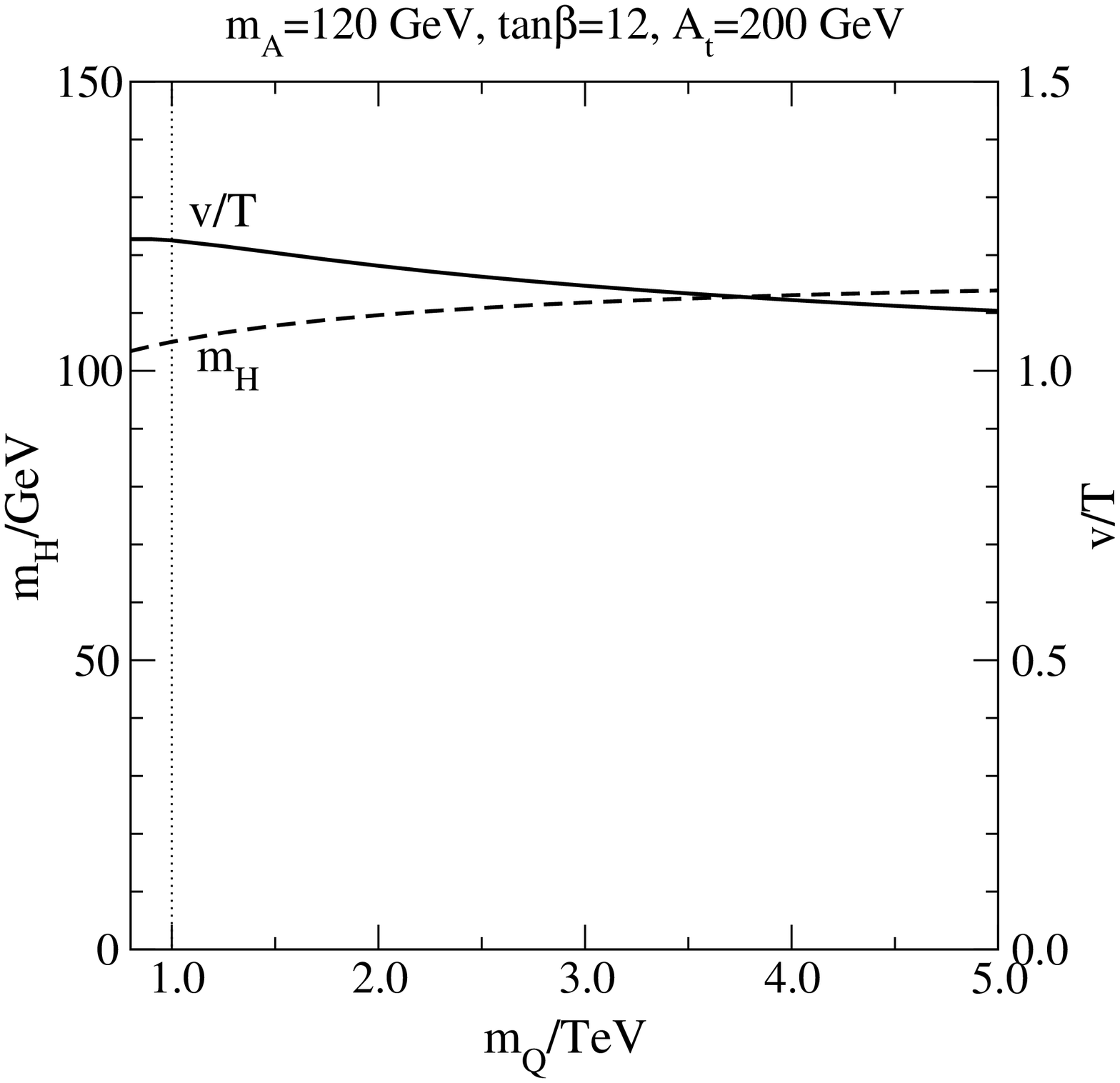}%
\epsfxsize=6.6cm\hspace*{1cm}\epsfbox{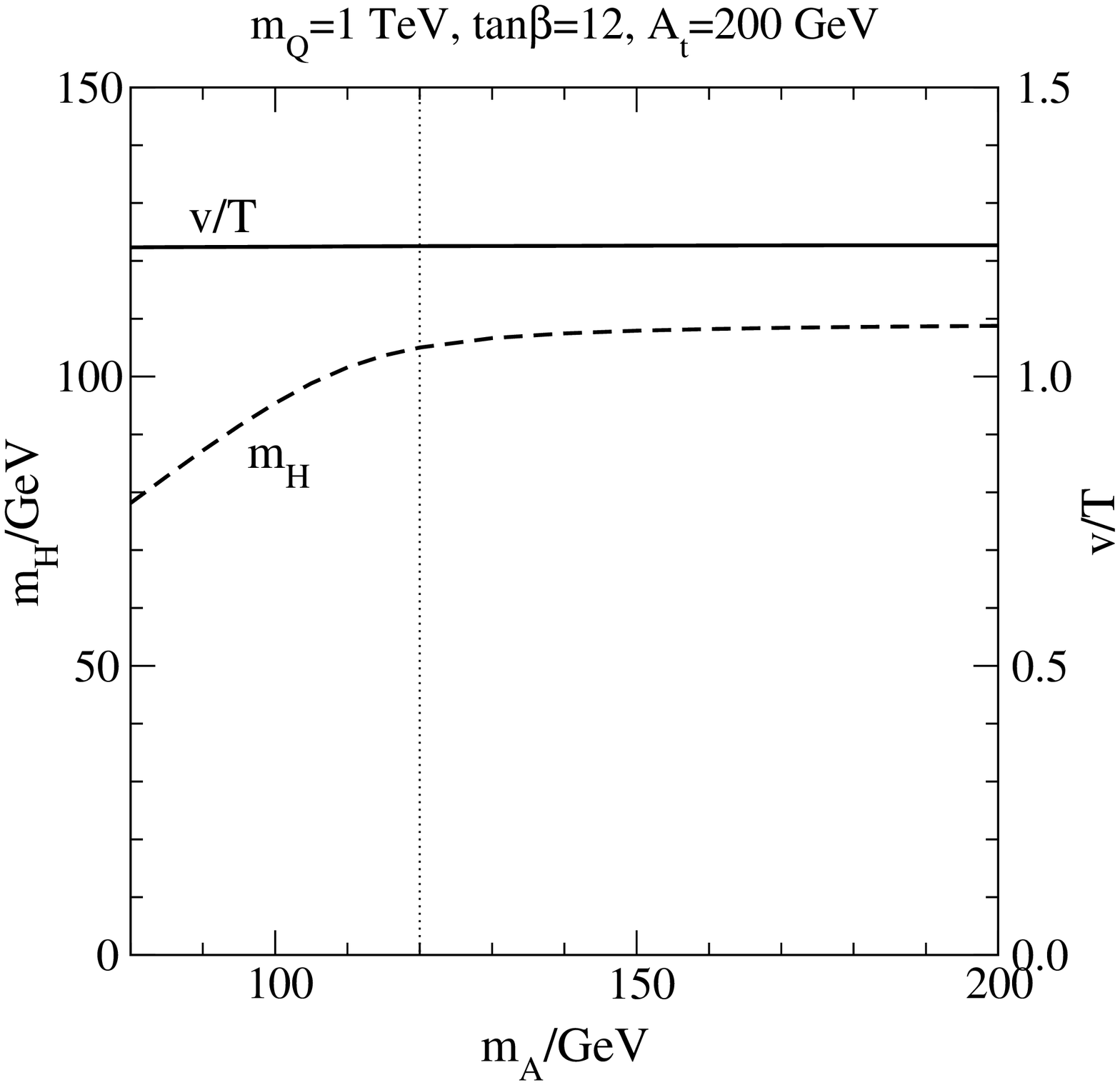}}

\vspace*{1cm}

\centerline{\epsfxsize=6.6cm\hspace*{0cm}\epsfbox{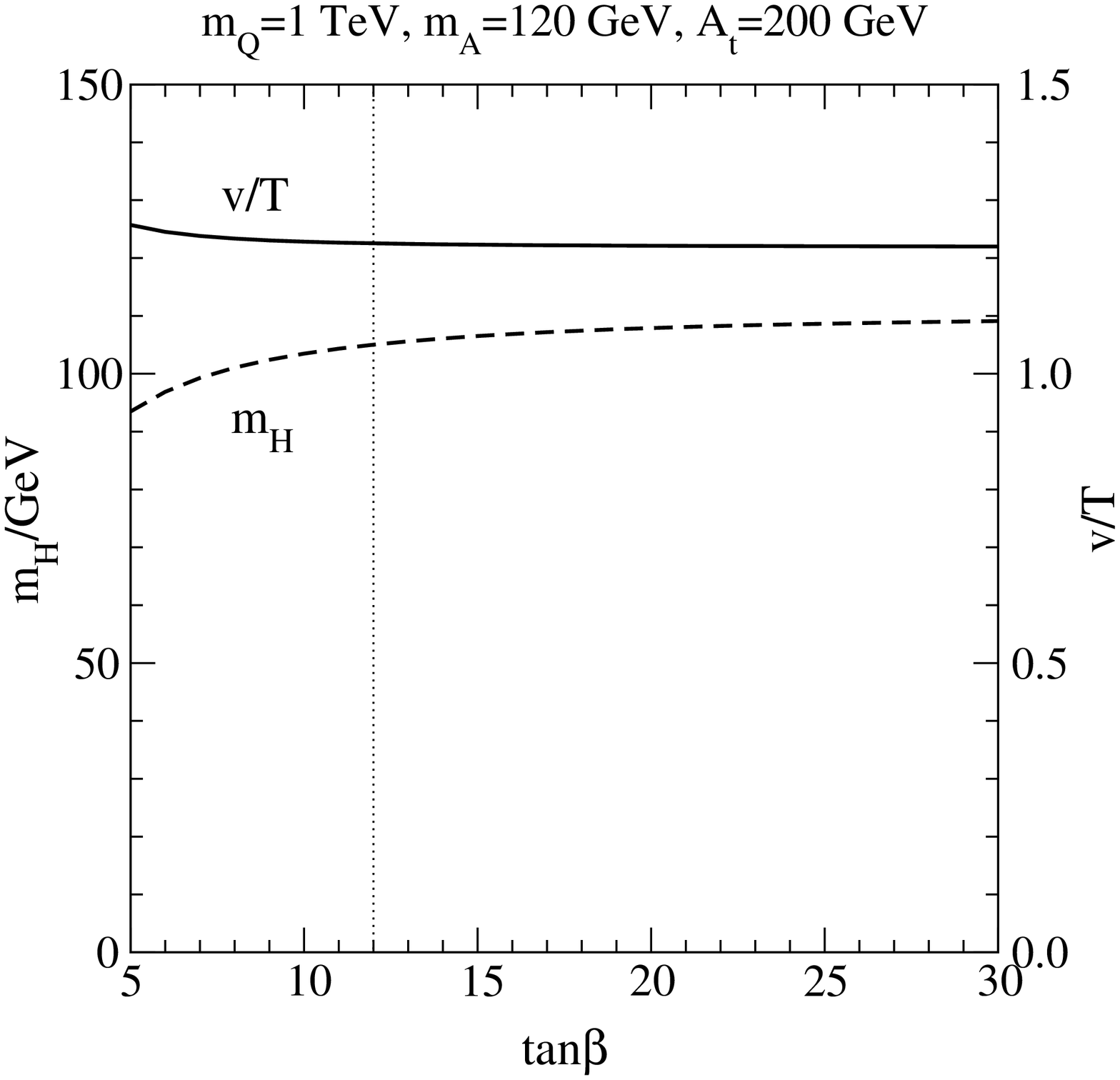}%
\epsfxsize=6.6cm\hspace*{1cm}\epsfbox{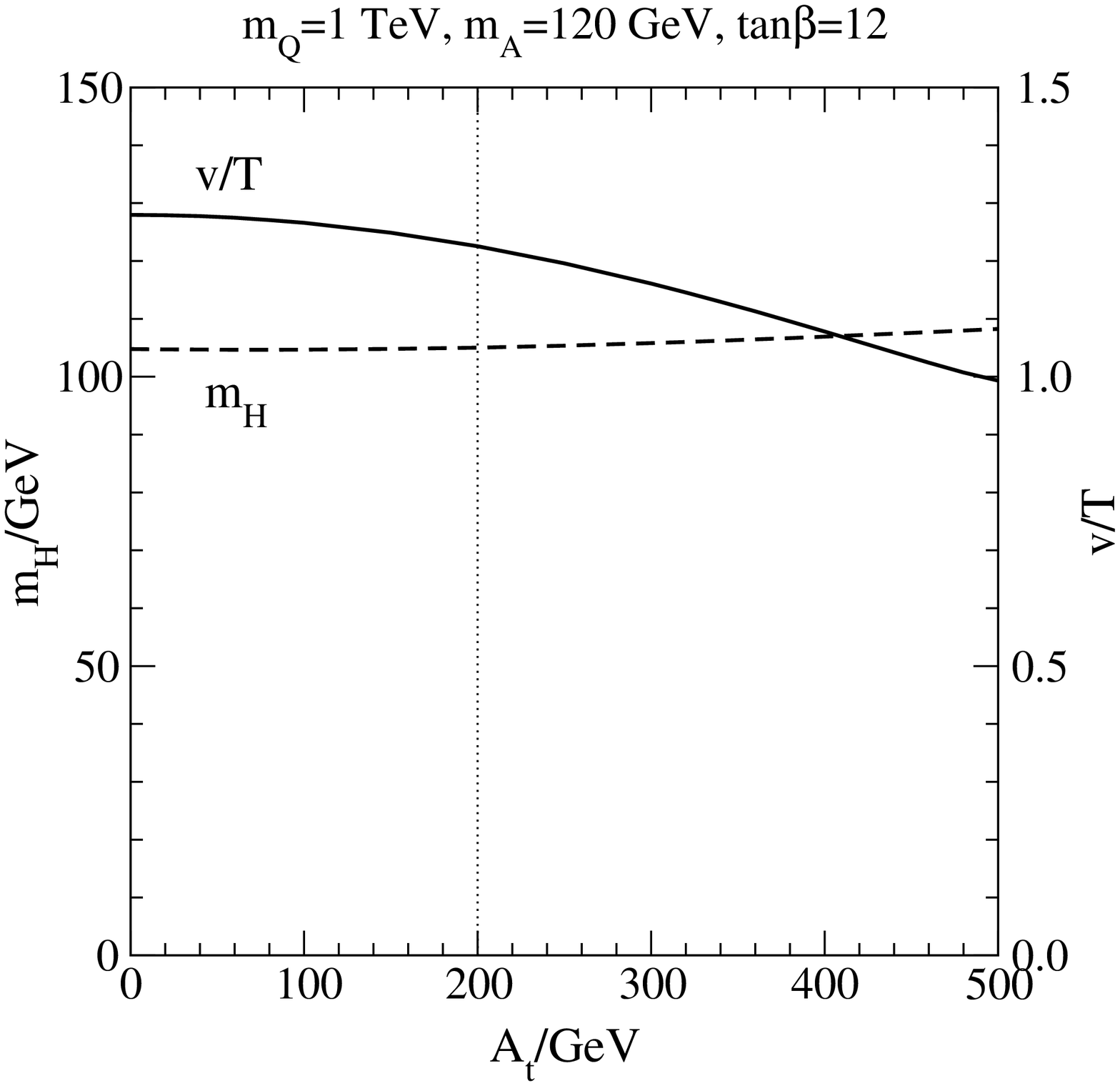}}

\caption[a]{The Higgs mass $m_H$ 
and broken phase expectation value $v/T$
{\em at the triple point} (see \fig\ref{fig:phasediag})
according to 2-loop perturbation theory, 
as a function of various quantities. The vertical lines indicate
the location of our reference parameter values. The Higgs mass
is computed with the 1-loop relations specified in~\cite{cpown}
(corresponding to~\cite{erz}).
Note that the combination relevant for large $m_A$, $A_t-\mu^*\cot\beta$,
is in most cases close to $A_t$, since we keep $\mu = i\, 200$ GeV fixed 
and $\cot\beta = 1/\tan\beta$ is small. In the values of $m_H$ we expect 
an uncertainty of a few GeV, due to unimplemented 
2-loop zero temperature Higgs mass corrections,  2-loop 
dimensional reduction corrections, and perhaps also the fact that
explicit CP violation has not been rigorously treated in the 
zero temperature relations we employ here~\cite{cpown}
(for recent discussions, see~\cite{pw}--\cite{kw}).}  
\la{fig:extensions}
\end{figure}

In this paper, 
we have studied the electroweak phase transition in the MSSM, 
with particular attention on CP violation
in the background (``vacuum'') configuration, 
as well as on the strength of the phase transition. 

The method we have used is based on 3d effective field theories, 
and their non-perturbative study.  
At finite temperatures around the 
electroweak phase transition, the thermodynamics 
of the MSSM can be represented by a theory 
containing two SU(2) Higgs doublets and one SU(3) stop triplet. 
Despite its complexity, we have demonstrated that this theory 
can be studied in a controlled way with lattice simulations.

The phase diagram of this theory is non-trivial, involving a phase where 
CP is (even)
spontaneously violated, as well as a phase where the U(1) symmetry 
corresponding to a massless photon is broken. 
We have studied the phase
with spontaneous CP violation in some detail. 
We have found that for the parameter values 
allowed by the MSSM, one does not end up in this phase close to the 
electroweak phase transition.
In more general two Higgs doublet theories this could happen, 
with potential implications for baryogenesis. 

We have then studied the electroweak phase transition
at physical parameter values, in particular $m_H \approx 105$ GeV, 
not ruled out for the MSSM. 
We observe very clearly the feature familiar from our 
previous MSSM study~\cite{mssmsim} that the transition is significantly
stronger than in 1-loop perturbation theory, 
and even stronger than at 2-loop level, 
due to the fact that the critical temperature $T_c$ is lower.
Let us note that
the situation is different from that studied with 4d simulations 
in~\cite{4dmssm}, where the transition was quite strong 
($m_H \approx 45$ GeV), 
and good agreement even with 1-loop perturbation theory was found.
We do not consider there to be any discrepancy, however, 
since all our previous experience is that perturbation 
theory works the better, the smaller the Higgs mass. 

At the point of our present study,
we observe (\fig\ref{fig:delta}) that the transition 
is strong enough for baryogenesis, since $v/T\approx 1.0$. 
(Based on analytic estimates, one would expect that 
$v/T$ has to be somewhat larger than 1.0~\cite{nonpert}, 
but a dynamical lattice computation suggests that 1.0 should 
be enough~\cite{moore_broken}). However, this 
concerns a particular parameter point, and it is 
important to ask how much room there is around it. 

To get a comprehensive estimate, 
let us now go to the point of the strongest
possible transition, i.e. the triple point shown
in \fig\ref{fig:phasediag}. We then vary various parameters
and use 2-loop perturbation theory to compute $v/T$. 
The results are shown in \fig\ref{fig:extensions}. 
We find the rather remarkable fact that the results are 
almost independent of the Higgs sector parameters $m_A,\tb$. 
This is so because at the triple point the properties of the 
transition are dictated by the stop sector. 

On the other hand, the transition 
weakens rapidly away from the triple point
(see \fig\ref{fig:phasediag}). Our lattice results 
provide a strengthening effect which can partly
compensate for this. Nevertheless, one 
needs to remain close to the triple point in any case,
for instance $\tilde m_U \sim 65...77$ GeV for the parameters
employed in \fig\ref{fig:phasediag}. The perturbative
range would have been $\tilde m_U \sim 69...74$ GeV.

Apart from $v/T$, we have also measured other important 
characteristics of the phase transition. 
The values of the latent heat $L/T_c^4$ 
and surface tension $\sigma/T_c^3$ allow
us to discuss the real time history of the phase transition. 
Estimates such as in~\cite{kk,ikkl,realtime,bjls,mr} lead to the
conclusion that the latent heat is probably large enough to reheat
the system back to $T_c$ after the bubble nucleation period~\cite{mr},
since $L/T_c^4 \gsim 8 (\sigma/T_c^3)^{\fr34}$~\cite{realtime}.
(In fact, the system looks very much like the case B 
studied in~\cite{realtime}, but the physical
friction is orders of magnitude 
larger~\cite{gdm2} than assumed in~\cite{realtime} based on the 
literature available at the time, and the physical
velocities are therefore
smaller than in~\cite{realtime}, $\lsim 0.1$~\cite{gdm2,jsch}.)
The small bubble wall velocities before and particularly after
reheating (when they are $\sim 0.001$) may lead 
to enhanced baryon number production according to the standard 
computations~\cite{non-eq,aso,cjk}. 

Finally we have studied the properties of the phase 
boundary, or bubble wall,  
at the physical transition point. We have determined the 
profiles corresponding to $\tb$ and 
to the C violating phase angle $\cf$ numerically, and 
excluded spontaneous (also called transitional) 
CP violation within the phase boundary, too. 
Explicit CP violating effects in the Higgs background 
are non-vanishing but small, even if the explicit phases 
are of order unity, because they are suppressed by
effective couplings of the type $g_w^2T$ over the heavy mass scale 
$\gsim (m_A^2 + 0.5 T^2)^{\fr12}$. 
The profiles we have determined 
could in principle be used as the 
semiclassical background entering the actual baryogenesis 
computations~\cite{non-eq}--\cite{joyce},\cite{cmqsw}. 

In summary, from the point of view of the non-equilibrium constraint, 
there is some parameter space available
for electroweak baryogenesis in the MSSM.
Our non-perturbative results 
agree well with the ones in~\cite{cm}, based on 2-loop perturbation 
theory and our previous lattice results~\cite{mssmsim}, 
and allow for a strong transition even for a Standard Model 
like Higgs mass $m_H\sim 115...120$ GeV if $m_Q \gsim$ a few TeV 
and $A_t/m_Q\gsim 0.5$ (see \fig\ref{fig:extensions}). 
On the other hand, we find also a small value of $m_A \lsim 120$ GeV
to be acceptable close
enough to the triple point, even though away from it large values
are favoured. Small values of $m_A$ make the MSSM look less like 
the Standard Model and relax the experimental constraint on $m_H$~\cite{lep}, 
\cite{pw}--\cite{kw}, allowing perhaps for somewhat smaller $m_Q$ as well. 
Thus electroweak baryogenesis continues to be a viable scenario, 
besides for instance those based on Majorana type neutrino masses, 
if at the same time quite strongly constrained.

\section*{Acknowledgements}

We thank K.~Kainulainen, A.~Pilaftsis and 
M.~Shaposhnikov for discussions.
Most of the simulations were carried out with a Cray T3E at the Center 
for Scientific Computing, Finland. 
The total amount of computing power used was about 3.7 cpu-years 
of a single node's capacity, corresponding to $\sim 1.4\times10^{16}$ 
floating-point operations. This work was
partly supported by the TMR network {\em Finite Tempera\-ture Phase
Transitions in Particle Physics}, EU contract no.\ FMRX-CT97-0122,
and by the RTN network {\em Supersymmetry and the Early Universe}, 
EU contract no.\ HPRN-CT-2000-00152.

\appendix
\renewcommand{\thesection}{Appendix~\Alph{section}}
\renewcommand{\thesubsection}{\Alph{section}.\arabic{subsection}}
\renewcommand{\theequation}{\Alph{section}.\arabic{equation}}


\section{Integrating out the heavy Higgs direction}
\la{inthiggs}

We review in this Appendix how the effective theory in 
\eq\nr{action} can be simplified
by integrating out a linear combination of the Higgs 
doublets, if we are not interested in C violation but 
only in the strength of the phase transition. 
We have discussed the procedure previously in
Secs.\ 6,7 of~\cite{mssmown} and in Sec.\ 3.1 of~\cite{cpown}. 
We complete those results here by allowing for complex parameters 
(explicit CP violation), as well as by having a 
light dynamical stop. We work at 1-loop level. 

It should be noted that contrary to the case at zero temperature, 
integrating out a linear combination of the Higgs doublets
is reliable even for small values of $m_A$, because 
thermal corrections increase the effective mass of the 
degree of freedom that is integrated out (see below). 

\subsection{Phase redefinition}
\la{suse:redef} 

The starting point is the effective theory in \eq\nr{action}.
We take first a trivial step, removing one extra phase from the
parameters in order to simplify the notation.
Indeed, if $m_{12}^2(T) = |m_{12}^2(T)| \exp(i\phi_{12})$, 
then we can make a field redefinition
\be
H_1 \to H_1 e^{i \phi_{12}}, \quad H_2 \to H_2.
\ee
As a result, the real parameters in \eq\nr{action} remain
unchanged, but the five complex parameters change as
\ba
m_{12}^{2(\rmi{new})}(T) & = & |m_{12}^2(T)|, \\
\gamma_{12}^{(\rmi{new})} & = & \gamma_{12} e^{-i \phi_{12}}, \\
\lambda_{5}^{(\rmi{new})} & = & \lambda_{5} e^{-2 i \phi_{12}}, \\
\lambda_{6}^{(\rmi{new})} & = & \lambda_{6} e^{-i \phi_{12}}, \\
\lambda_{7}^{(\rmi{new})} & = & \lambda_{7} e^{-i \phi_{12}}.
\ea
We leave out the superscripts ``(new)'' in the following, with 
the understanding that after each step, the new parameters are 
denoted with the same symbols as the old ones before it. 

Note that if $\gamma_{12},\lambda_5,\lambda_6,\lambda_7$ are small
and there is no spontaneous CP violation,  then this field redefinition 
directly determines the phase angle of 
$\langle H_1^\dagger \tilde H_2 \rangle$. For
instance, for the parameter value in \eq\nr{m12} with $T\lsim 100$ GeV,
we get $m_{12}^2(T) \sim -1200\, e^{i\, 0.02\pi} \mbox{ GeV}^2$, and 
consequently $H_1^\dagger \tilde H_2 \sim |H_1^\dagger \tilde H_2|
e^{-i\, 0.02\pi}$, and 
$\im H_1^\dagger \tilde H_2 < 0$; see \fig\ref{fig:ReImb16}. 

\subsection{Diagonalising the mass matrix}
 
Next we want to define new fields as linear combinations of 
$H_1, \tilde H_2$, such that the term mixing the two directions,
$\sim m_{12}^2(T) H_1^\dagger \tilde H_2 $, vanishes at tree-level
(1-loop corrections can still induce a mixing and this effect
shows up below). Following~\cite{mssmown,cpown}, we write
\ba
H^1 & = & \cos\!\alpha\, h+\sin\!\alpha\, H, \la{rot1} \\
\tilde{H}^2 & = & -\sin\!\alpha\, h+\cos\!\alpha\, H. \la{rot2}
\ea
The angle $\alpha$ is chosen so that 
\be
\tan\! 2\alpha=\frac{2 m_{12}^2(T)}{m_2^2(T)-m_1^2(T)},\quad
\sin\! 2\alpha=\frac{2 m_{12}^2(T)}
{\sqrt{(m_1^2(T)-m_2^2(T))^2+4 m_{12}^4(T)}}. \la{angle}
\ee
It should be noted that in the practical case considered in 
\se\ref{strength}, the large value of $\tb$ implies that 
$\alpha\approx \pi/2$, which means that the light field $h$ 
is almost in the direction of the original $\tilde H_2$.

After the rotation, the quadratic part of the scalar potential is
\be
{\cal V}_{2} = m_U^2(T) U^\dagger U + m_h^2(T) h^\dagger h + 
m_H^2(T) H^\dagger H, 
\ee
where the new mass parameters are
\ba
m_h^2(T) & = & \fr12
\Bigl[m_1^2(T)+m_2^2(T)-\sqrt{(m_1^2(T)-m_2^2(T))^2+4 m_{12}^4(T)}
\Bigr], \la{d2m1} \\
m_H^2(T) & = & \fr12
\Bigl[m_1^2(T)+m_2^2(T)+\sqrt{(m_1^2(T)-m_2^2(T))^2+4 m_{12}^4(T)}
\Bigr].
\ea
The stop mass parameter $m_U^2(T)$ does not change from the
value in \eq\nr{action}. 

The scalar couplings are modified as follows. The stop self-coupling
$\lambda_U$ does not change. Denoting the quartic 
scalar potential related to the Higgses by
\ba
{\cal V}_{4,\rmi{Higgs}} & = & 
\gamma_1 U^\dagger U h^{\dagger}h+
\gamma_2 U^\dagger U H^{\dagger}H 
+ \Bigl[\gamma_{12} U^\dagger U h^\dagger H +\Hc\Bigr] \nn
& + & 
\lambda_1 (h^\dagger h)^2+\lambda_2 (H^\dagger H)^2+
\lambda_3 h^\dagger h H^\dagger H
+\lambda_4 h^\dagger H H^\dagger h \nn 
& + & \Bigl[ \lambda_5(h^\dagger H)^2 +
\lambda_6h^\dagger h h^\dagger H+
\lambda_7H^\dagger H h^\dagger H + \Hc\Bigr], \la{phithe}
\ea
we get 
\ba
\gamma_1^{(\rmi{new})} & = & 
\gamma_1 \cos^2\alpha + \gamma_2 \sin^2\alpha - \re\gamma_{12}\sin 2\alpha,\\
\gamma_2^{(\rmi{new})} & = & 
\gamma_1 \sin^2\alpha + \gamma_2 \cos^2\alpha +\re\gamma_{12} \sin 2\alpha,\\
\re \gamma_{12}^{(\rmi{new})} & = & \fr12 
(\gamma_1 - \gamma_2)\sin 2\alpha +\re \gamma_{12} \cos 2\alpha ,\\
\im \gamma_{12}^{(\rmi{new})} & = & \im \gamma_{12}, \\
\im \lambda_5^{(\rmi{new})} & = &  \im \lambda_5 \cos 2\alpha +
\fr12  \im(\lambda_6-\lambda_7)\sin 2\alpha, \\ 
\im \lambda_6^{(\rmi{new})} & = &  - \im \lambda_5\sin 2\alpha +
\im\lambda_6 \cos^2\alpha +  \im\lambda_7 \sin^2\alpha, \\ 
\im \lambda_7^{(\rmi{new})} & = &  \im \lambda_5 \sin 2\alpha +
\im\lambda_6 \sin^2\alpha + \im\lambda_7 \cos ^2\alpha .
\ea
The Higgs self-couplings 
$\lambda_1...\lambda_4$, together with the real parts
$\re\lambda_5 ... \re\lambda_7$, 
on the other hand, are related by the matrix in \eq(6.21) of~\cite{mssmown}.

\subsection{Integrating out the heavy direction}

In \eq\nr{angle} the angle $\alpha$ has been chosen such
that the field $h$ is light, as can be seen from~\eq\nr{d2m1}.
Then the heavy field $H$ can be integrated out.
Indeed, the expansion parameters related to this integration are
\be
\frac{g_w^2T}{4\pi m_H(T)},\quad \frac{\lambda_i T}{4\pi m_H(T)},  \la{exppar}
\ee
which are small close to the phase transition.
This is because one of the eigenvalues of the Higgs mass
matrix must be very light at the point of the phase transition, 
$m_h^2(T) \sim (g_w^2 T)^2$, so that the other eigenvalue
is equal to the trace of the mass matrix, given in \eq\nr{sum}:
\be
m_H^2(T) \sim m_1^2(T)+m_2^2(T) \gsim m_A^2+ 0.5 T^2.
\ee

When $H$ is removed, the resulting theory is just
the same as studied in~\cite{bjls,mssmsim}, 
\ba
{\cal L}_\rmi{3d} & = & 
\fr12 \tr G_{ij}^2 + (D_i^s U)^\dagger (D_i^s U) + 
m_U^2(T) U^\dagger U  + \lambda_U (U^\dagger U )^2  \nn
& + & 
\fr12 \tr F_{ij}^2 
+ (D_i^w h)^\dagger(D_i^w h) + 
m_h^2(T) h^{\dagger}h+
\lambda_h (h^\dagger h)^2
+ \gamma\, U^\dagger U h^\dagger h. \la{phi}
\ea
At 1-loop level, the new couplings are:
\ba
g_w^{2(\rmi{new})} \!\! & = &  \!\! 
g_w^2 \biggl( 1-\frac{g_w^2 T}{48\pi m_H(T)}
\biggr), \\
g_s^{2(\rmi{new})} \!\! & = & \!\! g_s^2, \\
\!\! m_h^{2(\rmi{new})}(T) \!\! & = & \!\! m_h^2(T)-
\frac{1}{4\pi}(2\lambda_3+\lambda_4) m_H(T) T, \la{c26} \\
\!\! m_U^{2(\rmi{new})}(T) \!\! & = & \!\! m_U^2(T)-
\frac{1}{2\pi} \gamma_2  m_H(T) T, \la{c27} \\
\lambda_h^{(\rmi{new})} \!\! & = & \!\!  
\lambda_1-\frac{T}{8\pi m_H(T)}
\Bigl(\lambda_3^2+\lambda_3\lambda_4+\fr12 \lambda_4^2+
2|\lambda_5|^2+12\re (\lambda_6 - \lambda_7)\lambda_6^*\Bigr),
\hspace*{0.5cm} \\
\lambda_U^{(\rmi{new})} \!\! & = & \!\!  
\lambda_U-\frac{T}{8\pi m_H(T)}
\Bigl(\gamma_2^2+4 | \gamma_{12}|^2 \Bigr), \\
\gamma^{(\rmi{new})} \!\! & = & \!\!  
\gamma_1-\frac{T}{8\pi m_H(T)}
\Bigl(2 \lambda_3\gamma_2+\lambda_4\gamma_2 +
4 | \gamma_{12}|^2 + 12 \re (\lambda_6 -\lambda_7) \gamma_{12}^*\Bigr). 
\hspace*{0.5cm}
\ea

\subsection{2-loop mass parameters}
\la{Lam}

Finally, let us recall how the results above would change by a 2-loop
integration out of $H$. From the practical point of view, 
the most important effects are in the mass
parameters~\cite{generic}. After the integration, the renormalized
mass parameters in the $\msbar$ scheme can be written as
\ba
m_{h}^{2(\rmi{new})}(T) & =& m_h^2(T) + \pf \Bigl(
\frac{51}{16}g_w^4+9 \lambda_{h}g_w^2-12\lambda_{h}^2-
3\gamma^2+8 g_s^2 \gamma
\Bigr)\ln \frac{\Lambda_{h}}{\bmu}, \hspace*{0.8cm}
\la{mmH3} \\
m_{U}^{2(\rmi{new})}(T) & = & m_U^2(T) + \pf \Bigl(
8 g_s^4+\frac{64}{3} \lambda_Ug_s^2-16\lambda_U^2-
2\gamma^2+3 g_w^2 \gamma
\Bigr)\ln \frac{\Lambda_U}{\bmu},
\la{mmU3}
\ea
where $m_h^2(T), m_U^2(T)$ are 
the 1-loop results in \eqs\nr{c26}, \nr{c27}. Thus
a 2-loop computation amounts to a determination of the expressions for 
$\Lambda_h,\Lambda_U\sim \mbox{a few} 
\times T$~\cite{generic,mssmsim,mlo2,ll};
see \se\ref{se:param} for a discussion of the status 
of such computations.


\section{Lattice counterterms}
\la{cts}

We collect here the lattice counterterms needed in \se\ref{lattact}.
The derivation of the counterterms proceeds as in~\cite{contlatt,ml_cl,lr}, 
and a major part of the results can be extracted from there. However, 
some new parts are needed too, because there are now 
two SU(2) Higgs doublets in contrast to just one. 

The most non-trivial 2-loop changes can be obtained as follows. 
In the contributions proportional to $g^4$, 
we have to replace $T$ by $\sum_i T_i$ in 
\eq(E.4) of \cite{lr}, where $i$ runs over all the 
fields (fundamental or adjoint) interacting with the
SU($N$) gauge fields, and $T_i=1$ in the former case, 
$N$ in the latter. In the present case of two fundamental
doublets, one thus simply needs to put $\sum_i T_i \to 2$
for the SU(2) case $g=g_w$. To obtain
the $g^2\lambda, g^2\gamma$-terms, we replace $m^2 d$ by the 
trace of the scalar mass matrix, computed in the appropriate
Higgs background, in \eq(E.5) of~\cite{lr}. Finally, the 
numerical factors in the terms of types $\lambda^2, \gamma^2$ 
have to be computed by hand. 

The bare parameters appearing in the lattice action 
are then of the form 
\be
m_{i,B}^2 = m_i^2(T) + \delta m_i^2,
\ee
where $m_i^2(T)$ are the $\msbar$ scheme parameters
at a scale $\bmu$.
The results for the counter\-terms $\delta m_i^2$ are:
\ba
\delta m_U^2 \!\! & = & \!\! 
-\frac{\Sigma}{4\pi a} \biggl( 
\fr83 g_s^2 + 8 \lambda_U + 2 \gamma_1 + 2 \gamma_2
\biggr) T \nn
& & 
-\frac{T^2}{16\pi^2} \biggl[
\biggl(
8 g_s^4 + \frac{64}{3}\lambda_U g_s^2 + 3 g_w^2(\gamma_1+\gamma_2) \nn
& & 
- 16 \lambda_U^2 - 2 (\gamma_1^2 + \gamma_2^2 + 2 |\gamma_{12}|^2 )
\biggr)\biggl(
\ln\frac{6}{a\bmu} + 0.08849
\biggr) \nn
& & + 19.633 g_s^4 + 12.362 \lambda_U g_s^2 + 
1.7384 (\gamma_1+\gamma_2) g_w^2
\biggr], \\
\delta m_1^2 \!\! & = & \!\!  
-\frac{\Sigma}{4\pi a} \biggl( 
\fr32 g_w^2 + 6\lambda_1 +2 \lambda_3 + \lambda_4 + 3 \gamma_1
\biggr) T \nn 
& & 
-\frac{T^2}{16\pi^2} \biggl[
\biggl(
\frac{45}{16} g_w^4 + \fr32 (6\lambda_1+2\lambda_3+\lambda_4) g_w^2 + 
8 \gamma_1 g_s^2 - 3 (\gamma_1^2 + |\gamma_{12}|^2 )  \nn
& & 
- (
12\lambda_1^2 +2 \lambda_3^2 + 2\lambda_4^2 + 2 \lambda_3\lambda_4 +
12 |\lambda_5|^2 + 9 |\lambda_6|^2 + 3 |\lambda_7|^2  )
\biggr)\biggl(
\ln\frac{6}{a\bmu} + 0.08849
\biggr) \nn
& & + 5.4650 g_w^4 + 0.86921 (6\lambda_1+2\lambda_3+\lambda_4) g_w^2
+ 4.6358 \gamma_1 g_s^2 
\biggr], \\
\delta m_2^2 \!\! & = & \!\! 
-\frac{\Sigma}{4\pi a} \biggl( 
\fr32 g_w^2 + 6\lambda_2 +2 \lambda_3 + \lambda_4 + 3 \gamma_2
\biggr) T \nn
& & 
-\frac{T^2}{16\pi^2} \biggl[
\biggl(
\frac{45}{16} g_w^4 + \fr32 (6\lambda_2+2\lambda_3+\lambda_4) g_w^2 + 
8 \gamma_2 g_s^2  - 3 (\gamma_2^2 + |\gamma_{12}|^2 ) \nn
& & 
- (
12\lambda_2^2 +2 \lambda_3^2 + 2\lambda_4^2 + 2 \lambda_3\lambda_4 +
12 |\lambda_5|^2 + 3 |\lambda_6|^2 + 9 |\lambda_7|^2  )
\biggr)\biggl(
\ln\frac{6}{a\bmu} + 0.08849
\biggr) \nn
& & + 5.4650 g_w^4 + 0.86921 (6\lambda_2+2\lambda_3+\lambda_4) g_w^2
+ 4.6358 \gamma_2 g_s^2 
\biggr], \\
\delta m_{12}^2 \!\! & = & \!\!  
-\frac{\Sigma}{4\pi a} 3 \Bigl( 
\lambda_6 + \lambda_7 + \gamma_{12} 
\Bigr) T \nn 
& & 
-\frac{T^2}{16\pi^2} \biggl[
\biggl(\fr92 (\lambda_6+\lambda_7) g_w^2 + 
8 \gamma_{12} g_s^2 - 3 \gamma_{12} (\gamma_1+\gamma_2)  \nn 
& & 
- 3 \Bigl(
2 \lambda_1 \lambda_6 + 2 \lambda_2\lambda_7 +
(\lambda_3+\lambda_4)(\lambda_6+\lambda_7) + 
2 \lambda_5 (\lambda_6^*+\lambda_7^*)  
\Bigr) 
\biggr)\biggl(
\ln\frac{6}{a\bmu} + 0.08849
\biggr) \nn
& & + 2.6076 (\lambda_6+\lambda_7) g_w^2 + 
4.6358 \gamma_{12} g_s^2
\biggr].
\ea
Here $\Sigma=3.175911535625$
and $a$ is the lattice spacing.

The continuum operators 
in \eq\nr{ops}, on the other hand, are obtained as 
\ba
\left.\frac{\langle
U ^\dagger U \rangle} {T^2}
\right|_{\msbar, \bmu} & = &  
\left.\langle
\hat U^\dagger \hat U \rangle\right|_\rmi{lattice} - 
\biggl[
\frac{3 \Sigma}{4\pi a T} + 
\frac{8 g_s^2}{(4 \pi)^2} \biggl(
\ln\frac{6}{a\bmu} + 0.66796
\biggr)
\biggr], \\ 
\left.\frac{\langle
H_i^\dagger {\tilde H}_j \rangle} {T^2}
\right|_{\msbar, \bmu} & = &  
\left.\langle
\hat H_i^\dagger {\hat {\tilde H}}_j \rangle\right|_\rmi{lattice} - 
\delta_{ij} 
\biggl[
\frac{\Sigma}{2\pi a T} + 
\frac{3 g_w^2}{(4 \pi)^2} \biggl(
\ln\frac{6}{a\bmu} + 0.66796
\biggr)
\biggr]. \la{Higgsub}
\ea
In practice we choose 
to discuss $\msbar$ parameters with $\bmu=T$, so that 
\be
\frac{6}{a\bmu} = \frac{6}{a T } = \fr32 g_w^2 \beta_w = g_s^2 \beta_s. 
\ee


\section{The C violating phase in perturbation theory}
\la{comppar}

We collect here the details related to the discussion outlined
in \se\ref{se:pdin1loop}. The starting point is the effective
potential in \eq\nr{v1loop}. Note that we are free to choose 
$\beta,\theta \in (0,\frac{\pi}{2}), \phi \in (-\pi,\pi)$. 
We ignore first $\hat A_t,\hat\mu$ and the 1-loop effects
from the SU(2) Higgs masses $m_{S,i}^2$, 
and present a complete parameterization
for the C violating phase in that case. 
We then discuss the effect
of $\hat A_t,\hat\mu\neq 0$ and $m_{S,i}^2\neq 0$. 

\subsection{Minimization with respect to $\theta,\phi$}

Let us assume for the moment that
$v,\beta$, or $v_1,v_2 > 0$, are given.
We denote
\be
M_{12}^2 = m_{12}^2(T) + \fr12 \lambda_6 v_1^2 + \fr12 \lambda_7 v_2^2,
\la{M122}
\ee
and assume first that $M_{12}^2\neq 0$. 
Minimizing \eq\nr{v1loop} with respect to $\phi$, 
we obtain that the region for spontaneous C violation is 
\be
\lambda_4 - 2 \lambda_5 < 0, \quad
\lambda_5 \frac{2 v_1 v_2}{|M_{12}^2|} > 1, 
\la{CPviol}
\ee
and then 
\be
\cos\theta =1, \quad \cos\phi = -  \frac{M_{12}^2}{2\lambda_5 v_1v_2}.
\la{CPmin}
\ee
The region where \uy is broken is 
\be
\lambda_4 - 2 \lambda_5 > 0, \quad
(\lambda_4 - 2 \lambda_5) \frac{v_1 v_2}{2 |M_{12}^2|} > 
1- \lambda_5 \frac{2 v_1 v_2}{ |M_{12}^2|}, 
\ee
and then 
\be
\cos\theta = \frac{2 |M_{12}^2|}{(2\lambda_5 +\lambda_4)v_1v_2}, 
\quad |\cos\phi| = 1.
\ee
For $\lambda_4 -2\lambda_5=0$ and
$\lambda_5 \frac{2 v_1 v_2}{|M_{12}^2|} > 1$, 
$\cos\theta$ and $\cos\phi$ are undetermined but 
\be
\cos\theta\cos\phi = -  \frac{M_{12}^2}{2\lambda_5 v_1v_2}.
\ee
Elsewhere, $\cos\theta = |\cos\phi|=1$.
The special case $M_{12}^2=0$ can be treated 
as a limit of these formulas.


We will in the following 
concentrate on the case in 
\eq\nr{CPviol}.
Then, the value of the effective 
potential at the minimum of \eq\nr{CPmin} is
\ba
\left.V(v_1,v_2) \right|_\rmi{\mbox{\eq\nr{CPmin}}} & = & 
\fr12 m_1^2(T) v_1^2+\fr12 m_2^2(T) v_2^2
+ \fr14 \lambda_1 v_1^4 + \fr14 \lambda_2 v_2^4 + \fr14 \lambda_3 v_1^2 v_2^2 
\nn
& + & \fr14 (\lambda_4 - 2 \lambda_5) v_1^2 v_2^2 -
\frac{1}{4\lambda_5}
\Bigl( 
m_{12}^2(T) + \fr12 \lambda_6 v_1^2 + \fr12 \lambda_7 v_2^2
\Bigr)^2 \nn
& - & \frac{T}{16\pi} g_w^3 (v_1^2+v_2^2)^{\fr32}-
\frac{T}{2\pi} \Bigl(m_U^2(T) + \fr12 h_t^2 v_2^2 \Bigr)^{\fr32}.
\la{vacchange}
\ea

\subsection{Boundedness}
\la{app:bound}

Next, we discuss which values of the couplings naively
leading to spontaneous C violation are actually allowed
from the point of view of the consistency of the theory. 
Let us first of all
recall that according to \eq\nr{CPviol}, 
\be
\lambda_5 > 0, \quad \lambda_4 - 2 \lambda_5 < 0. \la{vc2}
\ee
Furthermore, for the theory to be bounded from below, 
we must clearly also require that 
$\lambda_1,\lambda_2 > 0$ in \eq\nr{v1loop}.
However, this is not enough. It turns out that 
the most critical direction in the field space is where
spontaneous C violation indeed takes place (since this 
means that the 2nd order polynomial in $\cos\phi$, 
\eq\nr{v1loop}, has been successfully minimized).
The value at the minimum is given by \eq\nr{vacchange}.
We observe that the contribution in \eq\nr{vacchange}
effectively normalizes the values of $\lambda_1...\lambda_3$
in \eq\nr{v1loop}. 
It is then easy to see that boundedness requires that
in addition to \eq\nr{vc2}, one has to satisfy
\ba
& & \lambda_6^2 < 4 \lambda_1\lambda_5, \quad
\lambda_7^2 < 4 \lambda_2 \lambda_5, \nn
& & 
\lambda_3+\lambda_4-2 \lambda_5 -\frac{\lambda_6\lambda_7}{2\lambda_5} > 
-2 \sqrt{
\Bigl(\lambda_1-\frac{\lambda_6^2}{4\lambda_5} \Bigr)
\Bigl(\lambda_2-\frac{\lambda_7^2}{4\lambda_5} \Bigr)}.
\la{constraint0}
\ea
These will be replaced by stronger constraints below when 
we restrict ourselves to finding a 
C violating minimum at some finite values of
$v_1,v_2$, but are nevertheless useful as simple relations
involving the quartic couplings only.  

\subsection{Stationary point with respect to $v_1,v_2$}
\la{app:stat}

Next, we should minimize the effective potential
with respect to $v_1,v_2$ in addition to $\theta,\phi$
as has been done before, in order 
to express $v_1,v_2$ in terms of the parameters of the theory. It 
turns out that it is convenient to turn around the question: 
we will use $v_1,v_2$ to parameterise different theories
leading to spontaneous C violation, and express 
$m_1^2(T),m_2^2(T),m_{12}^2(T)$ 
in terms of these. 

Since the potential
has already been minimized with respect to $\theta,\phi$
(c.f.\ \eq\nr{vacchange}), it is sufficient to impose 
$\partial V/\partial v_i = 0, i=1,2$. 
We then find that a stationary 
point at $(v_1,v_2)=v (\cos\beta,\sin\beta)$, with a C violating
angle $\cos\phi$, is obtained for given $\lambda_1...\lambda_7$ 
provided that the mass parameters are
\ba
\frac{m_1^2}{T^2} & = & 
- \biggl[ 
\fr12 \lambda_6 \cos\phi \sin 2\beta +\lambda_1 \cos^2\beta +
\fr12 (\lambda_3+\lambda_4-2 \lambda_5)\sin^2\beta 
\biggr] \frac{v^2}{T^2} + G \frac{v}{T}, \hspace*{1cm} \la{masses1} \\
\frac{m_2^2}{T^2} & = & 
- \biggl[ 
\fr12 \lambda_7 \cos\phi \sin 2\beta +\lambda_2 \sin^2\beta +
\fr12 (\lambda_3+\lambda_4-2 \lambda_5)\cos^2\beta 
\biggr] \frac{v^2}{T^2} + G \frac{v}{T} \nn
&  & +
H \frac{1}{T} \Bigl(M^2 + v^2 \sin^2\!\beta \Bigr)^{1/2}, \la{mm2H} \\
\frac{m_{12}^2}{T^2}  & = & 
-\biggl[ \lambda_5 \cos\phi\sin 2\beta 
+\fr12 \lambda_6 \cos^2\beta +
\fr12 \lambda_7 \sin^2\beta \biggr] \frac{v^2}{T^2}, 
\la{masses3}
\ea
where we have denoted 
\be
G = \frac{3}{16\pi} g_w^3 \approx 0.018, \quad
H = \frac{3}{4\sqrt{2}\pi} h_t^3 \approx 0.169, \quad
M^2 = \frac{2}{h_t^2} m_U^2(T). \la{GHMdef}
\ee

\subsection{Local minimum with respect to $v_1,v_2$}
\la{app:min}

Not all of the stationary points obtained through
\eqs\nr{masses1}--\nr{masses3} are local minima.
The final stage is imposing this condition,
which leads to some further restrictions on the 
parameters (and on the values of $v$ allowed). Of course, 
the requirement of obtaining a {\em global} minimum
in addition to a local one, would lead to
still stronger restrictions, but for the present purpose 
it is enough to consider the local condition. 

As in the previous paragraph, after the minimization with 
respect to $\theta,\phi$ has been carried out, leading to 
\eq\nr{vacchange}, it is enough to consider the potential
as a function of $v_1,v_2$. The constraint is that the 
mass matrix ${\cal M}_{ij}= \partial^2 V/\partial v_i \partial v_j$
have only positive eigenvalues, i.e., 
\be
\det {\cal M} > 0 , \quad
\tr {\cal M} > 0.
\ee
These conditions result in the following constraints:
\ba
& & 
\lambda_1 - \frac{\lambda_6^2}{4\lambda_5}-\frac{GT}{2v} > 0, 
\la{constraint1a} \\
& & 
\lambda_2 - \frac{\lambda_7^2}{4\lambda_6}-\frac{GT}{2v}
-\frac{HT}{2\sqrt{M^2+v_2^2}} > 0, 
\la{constraint1b} \\
& & \biggl|
\lambda_3 + \lambda_4 -2 \lambda_5 -
\frac{\lambda_6 \lambda_7}{2\lambda_5} - G\frac{T}{v} 
\biggr| < \nn
& & \hspace*{2cm}
2 \sqrt{\biggl(
\lambda_1 -\frac{\lambda_6^2}{4\lambda_5}-\frac{GT}{2v}
 \biggr)\biggl(
\lambda_2 -\frac{\lambda_7^2}{4\lambda_5}-\frac{GT}{2v}
-\frac{HT}{2\sqrt{M^2+v_2^2}}
 \biggr)}
\la{constraint2}.
\ea
Note that these equations cannot be satisfied at arbitrarily small 
values of $v/T$, and thus spontaneous C violation can only take
place at sufficiently large $v/T$. 

\subsection{A complete parametrization}

In the previous paragraphs, we have obtained expressions for 
the mass parameters leading to spontaneous C violation, 
\eqs\nr{masses1}--\nr{masses3}, but also a number of 
constraints that have to be satisfied, 
\eqs\nr{constraint1a}--\nr{constraint2}. 
We can now present a complete parametrization for all the
C violating states allowed by the potential in \eq\nr{v1loop}
(with $\hat A_t,\hat \mu, m_{S,i}^2=0$),
such that the constraints are automatically taken care of. 

Suppose we want to have a local minimum where C is 
spontaneously violated \linebreak ($|\cos\phi| < 1$), at a given 
vev $v/T$, with a given $\tan\beta = v_2/v_1$. Take 
arbitrary $\lambda_1, \lambda_2$, $\lambda_5 > 0$, 
and $G,H,M$ as defined in \eq\nr{GHMdef}. Then 
there is a 4-parameter family of possibilities, 
parameterised by 
\be
\alpha_3,\alpha_6,\alpha_7 \in (0,\pi); \quad
p_4 \in (0,\infty),
\ee
provided that 
\be
\frac{v}{T} >  
\mathop{\rm max} 
\biggl( 
\frac{G}{2 \lambda_1\sin^2\alpha_6},\kappa
\biggr), \la{vmin}
\ee
where $\kappa$ is the unique root in the range
\be
\kappa > \frac{G}{2\lambda_2\sin^2 \alpha_7}
\ee
of the equation 
\be
(M^2/T^2 + \kappa^2 \sin^2\!\beta)(2\kappa \lambda_2 \sin^2\!\alpha_7 -G)^2 = 
H^2\kappa^2.
\ee
The remaining couplings have to be chosen as 
\ba
& & \lambda_3 = 2 p_4 + G \frac{T}{v} + 2 \sqrt{\lambda_1\lambda_2}
\biggl[\cos\alpha_6 \cos\alpha_7 \la{lam3} \\
& & + 
\sin\alpha_6 \sin\alpha_7 \cos\alpha_3 
\sqrt{\biggl( 1- \frac{G T}{2 v \lambda_1 \sin^2\alpha_6}\biggr)
\biggl(1 - \frac{G T+HTv/\sqrt{M^2 + v^2\sin^2\!\beta}}
{2 v \lambda_2 \sin^2 \alpha_7} \biggr)} \biggr], 
\nn 
& & \lambda_4 = 2 (\lambda_5 - p_4), \la{lam4} \\
& & \lambda_6 = 2\sqrt{\lambda_1\lambda_5}\cos\alpha_6, \la{lam6} \\
& & \lambda_7 = 2\sqrt{\lambda_2\lambda_5}\cos\alpha_7, \la{lam7}
\ea
and the mass parameters 
according to \eqs\nr{masses1}--\nr{masses3}.

For later purposes, it is also useful to represent the 
parametrization in a slightly different form. Suppose now
that $\lambda_3, \lambda_4$ are given parameters, 
in addition to $\lambda_1, \lambda_2$. For $\lambda_4$, 
this is certainly consistent with the parametrization 
in \eq\nr{lam4} if $\lambda_5 > 0$ and $\lambda_4 < 0$.
The former we assumed to be the case in order to
get C violation, and the latter 
is always true in the MSSM. The 
constraint for $\lambda_3$, \eq\nr{lam3},
then implies that $\lambda_5$ has a maximum 
allowed value for given $\lambda_3,\lambda_4$, 
\ba
& & \!\!\!\! 0 < \lambda_5 < \lambda_{5,\rmi{max}} \equiv
\fr12 \biggl\{ \lambda_3+\lambda_4- G \frac{T}{v}
-  2 \sqrt{\lambda_1\lambda_2}
\biggl[\cos\alpha_6 \cos\alpha_7 \nn 
& & \!\!\!\! - \sin\alpha_6 \sin\alpha_7 
\sqrt{\biggl(1- \frac{G T}{2 v \lambda_1 \sin^2\alpha_6}\biggr)
\biggl(1 - \frac{G T+HTv/\sqrt{M^2 + v^2\sin^2\!\beta}}
{2 v \lambda_2 \sin^2 \alpha_7} \biggr)} \biggr]\biggr\}. \hspace*{0.5cm}
\la{lam5max}
\ea
This holds in the case that 
the expression on the right hand side is positive; 
otherwise no values of $\lambda_5$ are allowed
(this typically happens for small $v/T$ close
to the minimum given by \eq\nr{vmin}). 
The parameters $\lambda_6,\lambda_7$ are still given 
by \eqs\nr{lam6}, \nr{lam7}.

\subsection{The effect of Higgs self-couplings and $\hat A_t, \hat\mu$}
\la{se:Atmu}

In the analysis above, we ignored 1-loop effects from the SU(2) 
Higgses $H_1,H_2$, 
and set $\hat A_t=\hat\mu=0$. Let us discuss here
what happens when these assumptions are relaxed. Because
we consider spontaneous C violation, $\hat A_t,\hat \mu$
are assumed real.  

Clearly the introduction of $\hat A_t, \hat\mu, m_{S,i}^2\neq 0$
does not change the boundedness constraints, \se\ref{app:bound}.
We will also not consider the condition of a local minimum, 
\se\ref{app:min}, since this would be quite tedious. But looking
for a stationary point as in \se\ref{app:stat} leads to useful
observations. 

First, consider the effect of $\hat A_t, \hat\mu\neq 0$. 
Let us look at a local extremum constraint
at some $v_1,v_2,\theta,\phi$, obtained 
with mass parameters $m_1^2(T),m_2^2(T),m_{12}^2(T)$. 
By taking derivatives of \eq\nr{v1loop}, we see that there is an 
extremum at the same point $v_1,v_2,\theta,\phi$
also in the theory without any stop
contribution in~\eq\nr{v1loop}, but at the modified mass parameter 
values $\tilde m_1^2(T),\tilde m_2^2(T),\tilde m_{12}^2(T)$, where
\ba
m_1^2(T) & = &  \tilde m_1^2(T) - 
\frac{3}{4\pi} h_t^2 |\hat\mu|^2\, T 
\left[ A + B \cos\theta \cos\phi \right]^{1/2}, \la{B27} \\
m_2^2(T) & = &  \tilde m_2^2(T) + 
\frac{3}{4\pi} h_t^2 (1- |\hat A_t|^2)\, T 
\left[ A + B \cos\theta \cos\phi \right]^{1/2}, \la{m22T} \\
m_{12}^2(T) & = &  \tilde m_{12}^2(T) + 
\frac{3}{4\pi} h_t^2 \hat A_t \hat\mu \, T 
\left[ A + B \cos\theta \cos\phi \right]^{1/2}, \la{B29}
\ea
and $A,B$ are from \eq\nr{AB}. 

We can now see that the values 
of, say, $m_1^2(T)+m_2^2(T)$ leading to spontaneous C violation, 
differ typically
from those obtained earlier on, $\tilde m_1^2(T)+\tilde m_2^2(T)$, 
by small effects  $\sim - (3/(4\pi)) h_t^2 (|\hat A_t|^2+|\hat\mu|^2) T 
[ m_U^2(T) + (1/2) h_t^2 v_2^2 ]^{\fr12}$. 
(Recall that the dominant term in \eq\nr{m22T}  
which does not depend on $\hat A_t,\hat\mu$,
was already included in our previous discussion.) 
Furthermore, the 
sign is negative, so that the part of the parameter space 
extending to the phenomenologically interesting region 
$m_1^2(T)+m_2^2(T) > 0$ tends to decrease.
The decrease
can be rephrased by noting that $\hat A_t,\hat\mu\neq 0$ tend to 
decrease $v/T$, 
since they effectively decrease the coefficient of the cubic term
which would be obtained from \eq\nr{Atmu1loop} in the limit 
$m_U^2(T)\to 0$, and a smaller $v/T$ makes C violation less likely. 
To summarise, we do not expect the inclusion of $\hat A_t,\hat\mu\neq 0$ 
to change our conclusions. 

Similarly to \eqs\nr{B27}--\nr{B29}, 
the 1-loop effects of the scalars $H_1,H_2$ are expected 
to change $[m_1^2(T)+m_2^2(T)]/T^2$ by terms parametrically of the 
type $\sim \lambda_i m_j/(2 \pi T)$. 
It is hard to dicuss this effect analytically, since in
the general background of \eq\nr{prms}, the scalar 
mass matrix has the dimension $8\times8$. However,  
numerically we observe that the scalar contributions can 
also slightly increase the parameter space leading to spontaneous
C~violation, in contrast to $\hat A_t,\hat \mu$: 
parameters which would otherwise not result
in a C broken minimum, can do so when the last term in 
\eq\nr{s1loop} is included. Nevertheless, the effect is too 
small, numerically of order $\sim \lambda_i m_j/(2 \pi T) \sim 0.1$,   
to have any qualitative significance.


\section{Integrating out the right-handed stop}
\la{intstop}

If the right-handed stop is heavy, it can be integrated out from
the action in \eq\nr{action}. 
In this case the electroweak phase transition is too 
weak for baryogenesis for physical Higgs masses in excess
of 70...80 GeV~\cite{ckold,loold,mssmown}. 
Nevertheless, we summarise
here how the couplings of the  
3d SU(2) + two Higgs doublet model would change 
at 1-loop level
if $U$ is integrated out from~\eq\nr{action}, since
we need the argument in \se\ref{comparison}: 
\ba
\delta m_1^2(T) & = &  
 - \frac{3}{4\pi} \gamma_1 m_U(T) T, \la{dmm1mm2a} \\ 
\delta m_2^2(T) & = &  
-\frac{3}{4\pi} \gamma_2 m_U(T) T,  \la{dmm1mm2b} \\ 
\delta m_{12}^2(T) & = &  
- \frac{3}{4\pi} \gamma_{12} m_U(T) T, \\
\delta \lambda_1 & = & -\frac{3}{16\pi} \frac{T}{m_U(T)} \gamma_1^2, \\ 
\delta \lambda_2 & = &  -\frac{3}{16\pi} \frac{T}{m_U(T)} \gamma_2^2, \\ 
\delta \lambda_3 & = & -\frac{3}{8\pi} \frac{T}{m_U(T)} \gamma_1\gamma_2, \\  
\delta \lambda_4  & = &  -\frac{3}{8\pi} \frac{T}{m_U(T)} |\gamma_{12}|^2, \\  
\delta \lambda_5 & = & -\frac{3}{16\pi} \frac{T}{m_U(T)} \gamma_{12}^2, \\  
\delta \lambda_6 & = & -\frac{3}{8\pi} \frac{T}{m_U(T)} \gamma_1\gamma_{12},\\
\delta \lambda_7 & = &  -\frac{3}{8\pi} \frac{T}{m_U(T)} \gamma_2\gamma_{12}.  
\la{eq:intstop}
\ea
This integration
is reliable in the limit that $\gamma_i T, \lambda_U T, 
g_s^2 T \ll m_U(T)$. As to the numerical magnitudes of the 1-loop corrections, 
recall from \se\ref{sec:ac}
that typically $\gamma_1\sim |\gamma_{12}| \ll \gamma_2\sim 1$.



\begin{thebibliography}{99}

\bibitem{krs}
V.A.~Kuzmin, V.A.~Rubakov and M.E.~Shaposhnikov,
Phys.\ Lett.\ {B 155} (1985) 36.

\bibitem{asy}
P.~Arnold, D.T.~Son and L.G.~Yaffe,
Phys.\ Rev.\  {D 55} (1997) 6264 [hep-ph/9609481];
%
{\em ibid.}\  {59} (1999) 105020 [hep-ph/9810216];
%
{\em ibid.}\  {60} (1999) 025007 [hep-ph/9901304];
%
P.~Arnold and L.G.~Yaffe,
hep-ph/9912305;
%
hep-ph/9912306.

\bibitem{gdm}
G.D.~Moore, C.~Hu and B.~M\"uller,
Phys.\ Rev.\  {D 58} (1998) 045001
[hep-ph/9710436];
%
G.D.~Moore,
Nucl.\ Phys.\  {B 568} (2000) 367 [hep-ph/9810313];
%
G.D.~Moore and K.~Rummukainen,
Phys.\ Rev.\  {D 61} (2000) 105008
[hep-ph/9906259];
%
D.~B\"odeker, G.D.~Moore and K.~Rummukainen,
Phys.\ Rev.\  {D 61} (2000) 056003
[hep-ph/9907545];
%
G.D.~Moore,
Phys.\ Rev.\  {D 62} (2000) 085011
[hep-ph/0001216].

\bibitem{db}
D.~B\"odeker,
Phys.\ Lett.\  {B 426} (1998) 351 [hep-ph/9801430];
%
Nucl.\ Phys.\  {B 566} (2000) 402 [hep-ph/9903478];
%
{\em ibid.}\  {559} (1999) 502 [hep-ph/9905239].

\bibitem{bp}
W.~Buchm\"uller and M.~Pl\"umacher,
hep-ph/0007176.

\bibitem{isthere}
K. Kajantie, M. Laine, K. Rummukainen and M. Shaposhnikov,
Phys.\ Rev.\ Lett.\  {77} (1996) 2887
[hep-ph/9605288];
%
K.~Rummukainen, M.~Tsypin, K.~Kajantie, M.~Laine and M.~Shaposhnikov,
Nucl.\ Phys.\  {B 532} (1998) 283
[hep-lat/9805013].

\bibitem{gis}
M.~G\"urtler, E.M.~Ilgenfritz and A.~Schiller,
Phys.\ Rev.\ {D 56} (1997) 3888 [hep-lat/9704013].

\bibitem{cfh}
F.~Csikor, Z.~Fodor and J.~Heitger,
Phys.\ Rev.\ Lett.\  {82} (1999) 21
[hep-ph/9809291].

\bibitem{bext}
K.~Kajantie, M.~Laine, J.~Peisa, K.~Rummukainen and M.~Shaposhnikov,
Nucl.\ Phys.\  {B 544} (1999) 357 [hep-lat/9809004];
%
M.~Laine,
hep-ph/0001292.

\bibitem{lep}
ALEPH, DELPHI, L3 and OPAL Collaborations, 
presentation at \newline 
``Rencontres de Moriond'', Les Arcs, France, March 11--25, 2000 \newline
[http://alephwww.cern.ch/ALPUB/oldconf/oldconf00/29/moriond.ps]; \newline
P. Igo-Kemenes, presentation at CERN, November 3, 2000 \newline
[http://lephiggs.web.cern.ch/LEPHIGGS/talks/pik\_lepc\_nov3\_2000.ps].


\bibitem{cqw1}
M.~Carena, M.~Quir\'os and C.E.M.~Wagner,
Phys.\ Lett.\  {B 380} (1996) 81 [hep-ph/9603420].

\bibitem{dggw}
D.~Delepine, J.-M.~G\'erard, R.~Gonzalez Felipe and J.~Weyers,
Phys.\ Lett.\  {B 386} (1996) 183
[hep-ph/9604440].

\bibitem{e}
J.R. Espinosa, 
Nucl.\ Phys.\  {B 475} (1996) 273 [hep-ph/9604320];
%
B. de Carlos and J.R. Espinosa,
Nucl.\ Phys.\  {B 503} (1997) 24
[hep-ph/9703212].

\bibitem{ckold}
J.M.~Cline and K.~Kainulainen,
Nucl.\ Phys.\  {B 482} (1996) 73 [hep-ph/9605235];
%
{\em ibid.}\  {510} (1998) 88 [hep-ph/9705201].

\bibitem{loold}
M.~Losada,
Phys.\ Rev.\  {D 56} (1997) 2893 [hep-ph/9605266];
%
G.R.~Farrar and M.~Losada,
Phys.\ Lett.\  {B 406} (1997) 60
[hep-ph/9612346].

\bibitem{mssmown}
M.~Laine,
Nucl.\ Phys.\  {B 481} (1996) 43
[hep-ph/9605283]; {\em ibid.}\ 548 (1999) 637 (E).

\bibitem{bjls}
D. B\"odeker, P. John, M. Laine and M.G. Schmidt, 
Nucl.\ Phys.\  {B 497} (1997) 387
[hep-ph/9612364].

\bibitem{cqw}
M.~Carena, M.~Quir\'os and C.E.M.~Wagner,
Nucl.\ Phys.\  {B 524} (1998) 3 [hep-ph/9710401].

\bibitem{mssmsim}
M. Laine and K. Rummukainen, 
Phys.\ Rev.\ Lett.\  {80} (1998) 5259 [hep-ph/9804255];
%
Nucl.\ Phys.\  {B 535} (1998) 423
[hep-lat/9804019].

\bibitem{cm}
J.M.~Cline and G.D.~Moore,
Phys.\ Rev.\ Lett.\  {81} (1998) 3315 [hep-ph/9806354].

\bibitem{mlo2}
M.~Losada,
Nucl.\ Phys.\  {B 537} (1999) 3 [hep-ph/9806519].

\bibitem{cms}
J.M.~Cline, G.D.~Moore and G.~Servant,
Phys.\ Rev.\  {D 60} (1999) 105035 [hep-ph/9902220].

\bibitem{mlo3}
M. Losada, 
Nucl.\ Phys.\  {B 569} (2000) 125 [hep-ph/9905441];
%
S.~Davidson, T.~Falk and M.~Losada,
Phys.\ Lett.\  {B 463} (1999) 214 [hep-ph/9907365].

\bibitem{non-eq}
M.~Carena, M.~Quir\'os, A.~Riotto, I.~Vilja and C.E.M.~Wagner,
Nucl.\ Phys.\  {B 503} (1997) 387 [hep-ph/9702409].

\bibitem{aso}
M.~Aoki, A.~Sugamoto and N.~Oshimo,
Prog.\ Theor.\ Phys.\  {98} (1997) 1325
[hep-ph/9706287].

\bibitem{cjk}
J.M. Cline, M. Joyce and K. Kainulainen,
Phys.\ Lett.\  {B 417} (1998) 79 [hep-ph/9708393];
%
JHEP {0007} (2000) 018 [hep-ph/0006119];
%
J.M.~Cline and K.~Kainulainen,
hep-ph/0002272.

\bibitem{rio}
A.~Riotto,
Phys.\ Rev.\  {D 58} (1998) 095009
[hep-ph/9803357].

\bibitem{risa}
N.~Rius and V.~Sanz,
Nucl.\ Phys.\  {B 570} (2000) 155 [hep-ph/9907460].

\bibitem{hsch}
S.J.~Huber and M.G.~Schmidt,
Eur.\ Phys.\ J.\  {C 10} (1999) 473 [hep-ph/9809506];
%
hep-ph/0003122.

\bibitem{pj}
P.~John,
Phys.\ Lett.\  {B 452} (1999) 221 [hep-ph/9810499];
%
S.J.~Huber, P.~John, M.~Laine and M.G.~Schmidt,
Phys.\ Lett.\  {B 475} (2000) 104 [hep-ph/9912278].

\bibitem{joyce}
M.~Joyce, K.~Kainulainen and T.~Prokopec,
Phys.\ Lett.\  {B 468} (1999) 128 [hep-ph/9906411];
%
{\em ibid.}\   {474} (2000) 402 [hep-ph/9910535];
%
JHEP {0010} (2000) 029
[hep-ph/0002239].

\bibitem{Brhlik:1999qr}
M.~Brhlik, G.J.~Good and G.L.~Kane,
hep-ph/9911243.

\bibitem{gdm2}
G.D.~Moore,
JHEP {0003} (2000) 006 [hep-ph/0001274].

\bibitem{jsch}
P.~John and M.G.~Schmidt,
hep-ph/0002050.

\bibitem{mr}
G.D. Moore and K. Rummukainen, 
hep-ph/0009132. 

\bibitem{cmqsw}
M.~Carena, J.M.~Moreno, M.~Quir\'os, M.~Seco and C.E.M.~Wagner,
hep-ph/0011055.

\bibitem{lee}
T.D. Lee, 
Phys.\ Rev.\  {D 8} (1973) 1226.

\bibitem{nm}
N.~Maekawa, 
Phys.\ Lett.\  {B 282} (1992) 387.

\bibitem{apo}
A.~Pomarol, 
Phys.\ Lett.\  {B 287} (1992) 331
[hep-ph/9205247].

\bibitem{cp}
D. Comelli and M. Pietroni,
Phys.\ Lett.\  {B 306} (1993) 67 [hep-ph/9302207].

\bibitem{emq}
J.R. Espinosa, J.M. Moreno and M. Quir\'os,
Phys.\ Lett.\  {B 319} (1993) 505
[hep-ph/9308315].

\bibitem{cpr}
D. Comelli, M. Pietroni and A. Riotto,
Nucl.\ Phys.\  {B 412} (1994) 441
[hep-ph/9304267].

\bibitem{fkot}
K.~Funakubo, A.~Kakuto, S.~Otsuki and F.~Toyoda,
Prog.\ Theor.\ Phys.\  {99} (1998) 1045
[hep-ph/9802276];
%
K.~Funakubo, S.~Otsuki and F.~Toyoda, 
Prog.\ Theor.\ Phys.\  {102} (1999) 389
[hep-ph/9903276].


\bibitem{mstv}
L. McLerran, M. Shaposhnikov, N. Turok and M. Voloshin, 
Phys.\ Lett.\  {B 256} (1991) 451.


\bibitem{nck}
A.E. Nelson, D.B. Kaplan and A.G. Cohen, 
Nucl.\ Phys.\  {B 373} (1992) 453.

\bibitem{jpt}
M. Joyce, T. Prokopec and N. Turok, 
Phys.\ Rev.\  {D 53} (1996) 2958
[hep-ph/9410282].

\bibitem{hn}
P. Huet and A.E. Nelson, 
Phys.\ Rev.\  {D 53} (1996) 4578
[hep-ph/9506477].

\bibitem{ckv}
J.M. Cline, K. Kainulainen and A.P. Vischer, 
Phys.\ Rev.\  {D 54} (1996) 2451
[hep-ph/9506284].

\bibitem{cpown}
M. Laine and K. Rummukainen, 
Nucl.\ Phys.\  {B 545} (1999) 141
[hep-ph/9811369].

\bibitem{su2u1}
K.~Kajantie, M.~Laine, K.~Rummukainen and M.~Shaposhnikov,
Nucl.\ Phys.\  {B 493} (1997) 413
[hep-lat/9612006].

\bibitem{allor}
M.~Laine and M.~Shaposhnikov,
Phys.\ Lett.\  {B 463} (1999) 280 [hep-th/9907194].

\bibitem{generic}
K.~Kajantie, M.~Laine, K.~Rummukainen and M.~Shaposhnikov,
Nucl.\ Phys.\  {B 458} (1996) 90
[hep-ph/9508379].

\bibitem{contlatt}
K.~Farakos, K.~Kajantie, K.~Rummukainen, and M.~Shaposhnikov,
Nucl.\ Phys.\  {B 442} (1995) 317 [hep-lat/9412091].

\bibitem{moore_a}
G.D.~Moore,
Nucl.\ Phys.\  {B 493} (1997) 439
[hep-lat/9610013];
%
{\em ibid.}\   {523} (1998) 569 [hep-lat/9709053].

\bibitem{nonpert}
K. Kajantie, M. Laine, K. Rummukainen and M. Shaposhnikov,
Nucl.\ Phys.\  {B 466} (1996) 189
[hep-lat/9510020].

\bibitem{parity}
K. Kajantie, M. Laine, K. Rummukainen and M. Shaposhnikov,
Phys.\ Lett.\  {B 423} (1998) 137
[hep-ph/9710538].

\bibitem{Ambj3}
J.~Ambj{\o}rn, K.~Farakos and M.E.~Shaposhnikov,
Mod.\ Phys.\ Lett.\ {A 6} (1991) 3099;
%
Nucl.\ Phys.\ {B 393} (1993) 633 [hep-lat/9205022].

\bibitem{hart}
A.~Hart, O.~Philipsen, J.D.~Stack and M.~Teper,
Phys.\ Lett.\  {B 396} (1997) 217
[hep-lat/9612021].

\bibitem{su3adj}
K.~Kajantie, M.~Laine, A.~Rajantie, K.~Rummukainen and M.~Tsypin,
JHEP {9811} (1998) 011 [hep-lat/9811004].

\bibitem{pa}
P.~Arnold,
Phys.\ Rev.\  {D 46} (1992) 2628 [hep-ph/9204228].

\bibitem{2hdcp}
M.~Laine and K.~Rummukainen,
Nucl.\ Phys.\ B (Proc.\ Suppl.)\  {83} (2000) 577 [hep-lat/9908045].

\bibitem{4dmssm}
F.~Csikor, Z.~Fodor, P.~Heged\"us, 
A.~Jakov\'ac, S.D.~Katz and A.~Pir\'oth,
Phys.\ Rev.\ Lett.\  {85} (2000) 932 [hep-ph/0001087].

\bibitem{pw}
A.~Pilaftsis,
Phys.\ Rev.\  {D 58} (1998) 096010 [hep-ph/9803297];
%
A.~Pilaftsis and C.E.M.~Wagner,
Nucl.\ Phys.\  {B 553} (1999) 3 [hep-ph/9902371];
%
M.~Carena, J.~Ellis, A.~Pilaftsis and C.E.M.~Wagner,
Nucl.\ Phys.\ B 586 (2000) 92 [hep-ph/0003180];
%
hep-ph/0009212.

\bibitem{cpmasses}
D.A.~Demir,
Phys.\ Rev.\  {D 60} (1999) 055006 [hep-ph/9901389].

\bibitem{cdl}
S.Y.~Choi, M.~Drees and J.S.~Lee,
Phys.\ Lett.\  {B 481} (2000) 57 [hep-ph/0002287].

\bibitem{kw}
G.L.~Kane and L.~Wang,
Phys.\ Lett.\  {B 488} (2000) 383
[hep-ph/0003198].

\bibitem{mqs}
J.M.~Moreno, M.~Quir\'os and M.~Seco,
Nucl.\ Phys.\  {B 526} (1998) 489 [hep-ph/9801272].

\bibitem{edm}
P.~Nath,
Phys.\ Rev.\ Lett.\  {66} (1991) 2565.

\bibitem{ko}
Y.~Kizukuri and N.~Oshimo,
Phys.\ Rev.\  {D 46} (1992) 3025.

\bibitem{prs}
S.~Pokorski, J.~Rosiek and C.A.~Savoy,
Nucl.\ Phys.\  {B 570} (2000) 81 [hep-ph/9906206].

\bibitem{ckn2}
A.G.~Cohen, D.B.~Kaplan and A.E.~Nelson,
Phys.\ Lett.\  {B 388} (1996) 588 [hep-ph/9607394].


\bibitem{ckp}
D.~Chang, W.~Keung and A.~Pilaftsis,
Phys.\ Rev.\ Lett.\  {82} (1999) 900
[hep-ph/9811202]; {\em ibid.} 83 (1999) 3972 (E);
%
A.~Pilaftsis,
Phys.\ Lett.\  {B 471} (1999) 174 [hep-ph/9909485];
%
Phys.\ Rev.\  {D 62} (2000) 016007 [hep-ph/9912253];
%
D.~Chang, W.~Chang and W.~Keung,
Phys.\ Lett.\  {B 478} (2000) 239 [hep-ph/9910465].

\bibitem{ll}
M.~Laine and M.~Losada,
Nucl.\ Phys.\  {B 582} (2000) 277 [hep-ph/0003111].

\bibitem{mcdo}
J.~McDonald,
Phys.\ Lett.\  {B 413} (1997) 30 [hep-ph/9707290].

\bibitem{grhi}
J.~Grant and M.~Hindmarsh,
Phys.\ Rev.\  {D 59} (1999) 116014 [hep-ph/9811289].

\bibitem{bbfh}
D.~B\"odeker, W.~Buchm\"uller, Z.~Fodor and T.~Helbig,
Nucl.\ Phys.\  {B 423} (1994) 171
[hep-ph/9311346].

\bibitem{baacke}
J.~Baacke,
Phys.\ Rev.\  {D 52} (1995) 6760 [hep-ph/9503350].

\bibitem{kls}
J.~Kripfganz, A.~Laser and M.G.~Schmidt,
Z.\ Phys.\  {C 73} (1997) 353 [hep-ph/9512340].

\bibitem{lgy}
A.~Parnachev and L.G.~Yaffe,
Phys.\ Rev.\  {D 62} (2000) 105034
[hep-th/0005269].

\bibitem{owe}
O.~Philipsen, M.~Teper and H.~Wittig,
Nucl.\ Phys.\  {B 469} (1996) 445 [hep-lat/9602006].

\bibitem{mu}
M. Laine and O. Philipsen, 
Nucl.\ Phys.\ B 523 (1998) 267 [hep-lat/9711022];
Phys.\ Lett.\ B 459 (1999) 259 [hep-lat/9905004];
A. Hart and O. Philipsen, 
Nucl.\ Phys.\ B 572 (2000) 243 [hep-lat/9908041];
A.~Hart, M.~Laine and O.~Philipsen,
Nucl.\ Phys.\  {B 586} (2000) 443
[hep-ph/0004060].

\bibitem{Binder}
K. Binder, in {\em Phase Transitions and Critical Phenomena\,}, 
eds.\ C.~Domb and M.S.~Green, Vol.\ {2} 
(Academic Press, New York, 1972).

\bibitem{moore_broken}
G.D.~Moore,
Phys.\ Lett.\  {B 439} (1998) 357 [hep-ph/9801204];
%
Phys.\ Rev.\  {D 59} (1999) 014503
[hep-ph/9805264].

\bibitem{erz}
J.~Ellis, G.~Ridolfi and F.~Zwirner,
Phys.\ Lett.\  {B 262} (1991) 477.

\bibitem{kk}
K.~Kajantie,
Phys.\ Lett.\ {B 285} (1992) 331.

\bibitem{ikkl}
J.~Ignatius, K.~Kajantie, H.~Kurki-Suonio and M.~Laine,
Phys.\ Rev.\  {D 50} (1994) 3738
[hep-ph/9405336].

\bibitem{realtime}
H.~Kurki-Suonio and M.~Laine,
Phys.\ Rev.\ Lett.\  {77} (1996) 3951 [hep-ph/9607382].

\bibitem{ml_cl}
M.~Laine,
Nucl.\ Phys.\  {B 451} (1995) 484
[hep-lat/9504001].

\bibitem{lr}
M. Laine and A. Rajantie,
Nucl.\ Phys.\  {B 513} (1998) 471
[hep-lat/9705003].

\end{thebibliography}
\end{document}